\documentclass{aa}  

\usepackage{graphicx}
\usepackage{txfonts}
\usepackage{natbib}
\usepackage{enumitem} 
\usepackage{color}
\usepackage[colorlinks=true, linkcolor = blue,
            urlcolor  = blue,
            citecolor = blue,
            anchorcolor = blue]{hyperref}

\bibpunct{(}{)}{;}{f}{}{,} 

\newcommand{\teffa}{$T_{\mathrm{eff}}^{\mathrm{H}\alpha}$}

\newcommand{\teff}{$T_{\mathrm{eff}}$}
\newcommand{\teffI}{$T_{\mathrm{eff}}^{IRFM}$}
\newcommand{\logg}{\mbox{log~\textit{g}}}

\begin{document}

  \title{Accurate effective temperature from H$\alpha$ profiles
  \thanks{Based on observations collected at Observat\'{o}rio do Pico dos Dias
  (OPD), operated by the Laboratório Nacional de Astrofísica, CNPq,
  Brazil and on data from the ESO Science Archive Facility.}}

     \author{R. E. Giribaldi\inst{1,2}
          \and 
          M. L. Ubaldo-Melo\inst{2}   
          \and
	  G. F. Porto de Mello\inst{2}          
          \and
          L. Pasquini\inst{1} 
          \and
          H.-G. Ludwig\inst{3}
          \and\\
          S. Ulmer-Moll\inst{4,5}          
          \and   
          D. Lorenzo-Oliveira\inst{6}
          }

   \institute{ESO - European Southern Observatory, Karl-Schwarzchild-Strasse 2, 
             85748 Garching bei M\"{u}nchen, Germany. \\
             \email{rescateg@eso.org, riano@astro.ufrj.br}
             \and  Observat\'{o}rio do Valongo, Universidade Federal de Rio de Janeiro, 
	     Ladeira Pedro Ant\^onio 43, 20.080-090 Rio de Janeiro RJ, Brazil
	     \and  Zentrum f\"ur Astronomie der Universit\"at Heidelberg, Landessternwarte, K\"onigstuhl 12, D-69117 Heidelberg, Germany
             \and  Instituto\,de\,Astrof\'{i}sica\,e\,Ci\^{e}ncias\,do\,Espa\c{c}o,\,Universidade\,do\,Porto,\,CAUP,\,Rua\,das\,Estrelas,\,4150-762\,Porto,\,Portugal
             \and  Departamento de F\'{i}sica e Astronomia, Faculdade de Ci\^{e}ncias, 
             Universidade do Porto, Rua do Campo Alegre 687, PT4169-007 Porto, Portugal.
	     \and  Universidade de S\~ao Paulo, Departamento de Astronomia do IAG/USP, Rua do Mat\~ao 1226, 
             Cidade Universit\'aria, 05508-900 S\~ao Paulo, SP, Brazil}
             

 
  \abstract
   {The determination of stellar effective temperature (\teff)   
   in \mbox{F, G, and K stars} using H$\alpha$ profile fitting is a quite remarkable and powerful tool, 
   because it practically does not depend on other atmospheric parameters and reddening. 
   Nevertheless, this technique is not frequently used because of the complex 
   procedure to recover the profile of broad lines 
   in echelle spectra.  
   As a consequence, tests performed on different models have sometimes provided ambiguous results.}
   {The main aim of this work is to test the H$\alpha$ profile fitting technique 
   to derive stellar effective temperature. 
   To improve its applicability to echelle spectra and to test how well 1D + LTE
   models perform on a variety of F-K stars.
   We also apply the technique to HARPS spectra and 
   test with the Sun the reliability and the stability of the HARPS response over several years.} 
   {We have therefore developed a normalization method for recovering undistorted H$\alpha$ profiles
   and we have first applied it to spectra acquired with the single order coud\'{e} instrument 
   (resolution $R = 45~000$) at \textit{do Pico dos Dias Observatory}  
   to avoid the problem of blaze correction.
   The continuum location around H$\alpha$ is optimized using an iterative procedure,
   where the identification of minute telluric features is performed.
   A set of spectra was acquired with the MUSICOS echelle spectrograph ($R = 40~000$)  
   to independently validate the normalization method. 
   The accuracy of the method and of the 1D + LTE model is determined
   using coud\'{e}/HARPS/MUSICOS spectra of the Sun and only coud\'{e} spectra of 
   a sample of 10 \textit{Gaia Benchmark Stars} 
   with effective temperature determined from
   interferometric measurements. 
   HARPS spectra ($R = 100~000$) are used to determine the effective temperature of
   26 stars in common with the coud\'{e} data set by the same procedure.}
   {We find that a proper choice of spectral windows of fits plus the identification of
   telluric features allow a very careful normalization of the spectra and
   produce  reliable H$\alpha$ profiles. 
   We also find that the most used solar atlases cannot be used as templates for H$\alpha$ temperature diagnostics
   without renormalization.
   The comparison with the Sun 
   shows that the effective temperatures derived by us with H$\alpha$ profiles from 1D + LTE models 
   underestimate the solar effective temperature by 28 K.   
   A very good agreement is found with the interferometric benchmarks and with the 
   Infrared Flux Method determination, 
   that shows a shallow dependency on metallicity according to the relation 
   \mbox{$T_{\mathrm{eff}} = T_{\mathrm{eff}}^{H\alpha}$ $-159$[Fe/H] + 28 K}
   within the metallicity range $-0.7$ to $+0.45$ dex.
   The comparison with Infrared Flux Method show a 59 K scatter
   dominated by photometric errors (52 K).
   In order to investigate the origin of this dependency, 
   we analyzed in the same way spectra generated by 3D models and found that  
   they  produce hotter temperatures, and that their use largely improve the agreement with the 
   interferometric and Infrared Flux Method measurements. 
   Finally, we find HARPS spectra to be fully suitable for H$\alpha$ profiles temperature diagnostics, 
   they are perfectly compatible with the coud\'{e} spectra,  and the same effective temperature for the 
   Sun is found analyzing HARPS spectra
   over a time span of more than 7 years.}
   {} 


   \maketitle
%

\section{Introduction}

Effective temperature 
is a fundamental stellar parameter because
it defines the physical conditions of the stellar atmosphere and 
it directly relates to the physical properties of the star: mass, radius and luminosity. 
Its measurement is essential to determine the evolutionary state of the stars, to perform detailed chemical 
abundance analysis, and to characterize exoplanets.

Among a variety of model-dependent techniques used to derive $T_{\mathrm{eff}}$ in F, G, and K type stars,
fitting  Balmer lines offers two important advantages: it
is not sensitive to reddening and is
very little sensitive to other stellar parameters, such as metallicity 
([Fe/H]\footnote{[A/B] = log $\text{N(A)/N(B)}_{\text{star}} -$ log $\text{N(A)/N(B)}_{\text{Sun}}$, 
where N denotes the number abundance of a given element.})
and surface gravity (\logg) \citep{Fu1993,Fu1994,BPO2000,BPO2002}.
For instance, variations of about 0.1 dex in either of these parameters induce 
3 to 35 K variations in $T_{\mathrm{eff}}$,
depending on the metallicity of the star (see Table 4 in \citet{BPO2002}, hereafter BPO02).
Thanks to this, the degeneracy between $T_{\mathrm{eff}}$ and [Fe/H]
when both parameters are simultaneously constrained with the 
excitation and ionization balance of iron lines 
(the parameters measured with this technique will be referred as ``spectroscopic'' hereafter) 
can be reduced by fixing the first to subsequently derive the second.
Thus, it is possible to distinguish minute differences in chemical abundances, as done e.g. 
by \citet{porto2008} and \citet{ram2011}.

In spite of  these advantages, the use of Balmer profiles fitting 
remains sporadic because:
\begin{enumerate}[label=(\roman*)]
   \item The complex normalization of wide line-profiles,
   especially in cross-dispersed echelle spectra because of 
   the instrumental blaze and of the fragmentation of the spectrum into multiple orders.
   \item The accuracy of the models of Balmer lines is not well established, 
   which is partially a consequence of (i).
   A clear example are the two ranges of $T_{\mathrm{eff}}$
   derived for the Sun using the model of 
   BPO02 and spectra from different instruments including two versions of the
   Kitt Peak National Observatory solar atlas \citet{kurucz1984} and \citet{kurucz2005} 
   (hereafter KPNO1984 and KPNO2005, respectively). 
   A ``cool'' value of $\sim\!\!5670$ K is found by 
   \citet{pereira2013}\footnote{The authors used a different 
   implementation of self-broadening with a later model atmospheres and different input physics.} 
   and \citet{onehag2014} from KPNO2005 and KPNO1984, respectively,
   while a ``hot'' value of $\sim\!\!5730$ K is found by 
   BPO02, \citet{ram2011,ram2014} and \citet{cor2012} from other 
   spectra; precise values are listed in Table~\ref{zero-point}. 
\end{enumerate} 

The problem of normalizing H$\alpha$ in echelle spectra has been approached 
making use of fiber-fed spectra, whose blaze function is efficiently 
removed by the flat field procedure
\citep[e.g.][]{fuhrmann1997,korn2003,korn2006,korn2007,lind2008,onehag2014}.
Also, a complex normalization method explained by BPO02 (hereafter 2D-normalization)
has been applied by some authors to remove the blaze 
\citep[e.g.][]{fuhrmann1997,allende2004,ram2011,ram2014,matsu2017b,matsu2017}.
Briefly, the method consist on interpolating the blaze function for the echelle orders contiguous 
to that containing H$\alpha$. 

It is recognized 
that the introduction of the self-broadening theory of hydrogen atoms by
BPO02 constitutes a significant advance to the completeness of the physics of the Balmer lines formation,
however the tests on the Sun performed by the authors quoted above
indicate that the model, or its application, is not accurate enough.
As a consequence, subsequent works concentrated on improving the  model
by adding more transitions in the self-broadening  
\citep{allard2008,cayrel2011}, and replacing
LTE and 1D by non-LTE and 3D model atmospheres
\citep{barklem2007, ludwig2009, pereira2013, amarsi2018}
but the solar $T_{\mathrm{eff}}$ has not yet been recovered.
The large discrepancies in the solar temperatures derived using the same model and
different instruments suggest that 
the treatment of observational spectra is the dominant source 
of uncertainty;
H$\alpha$ profiles are so sensitive that a minute error in the continuum location 
may significantly vary the derived temperature.
The continuum location problem was already identified by BPO02, 
who also estimated the errors induced by this process in the derived temperature.
In this work we aim to minimize these errors by a meticulous analysis of 
spectra of F, G, and K stars.

We first eliminate instrumental blaze and spectral fragmentation inherent to echelle spectra by 
using a long-slit single order spectrograph.
The continuum location is  then optimized by a normalization-fitting iterative procedure,
and it is also fine tuned during the process by identifying telluric features that
contaminate the spectra.

As a first step of our program of chemical tagging, 
mainly based in HARPS spectra, we establish the methodology 
to derive $T_{\mathrm{eff}}$ from H$\alpha$ profiles.
We determine the accuracy of the temperature diagnostics 
with H$\alpha$ profiles from 1D + LTE model atmospheres
and the self-broadening theory of BPO02 
(these profiles will be referred henceforth as profiles from 1D model atmospheres
and their temperatures will be represented by \teffa)
by comparing them with the accurate \teff's of the \textit{Gaia Benchmark Stars} derived by interferometry.
The method we present is further validated by comparing the temperatures of the same stars
from \mbox{MUSICOS} spectra normalized by the \mbox{2D-normalization}, which is an independent method.
Finally, we prove the absence of residual blaze features in HARPS spectra
by processing them in the same way we performed with coud\'{e},  
and obtaining compatible \teffa's.

This paper is organized as follows. In section~\ref{data} the selection of the sample is described
together with the characteristics of the spectroscopic observations.
In section~\ref{norm} we describe the normalization method. 
In section~\ref{fitting} we describe the fitting procedure.
In section~\ref{2DN} we validate the normalization method.
The results are presented from section~\ref{accuracy} on.
In this section the accuracy of H$\alpha$ profiles from 1D models is determined.
In section~\ref{consistency} \teffa~is compared against temperature diagnostics from other frequent techniques.
In section~\ref{otherHa} we compare our H$\alpha$ temperature scale with others from the same and different models.
In section~\ref{3D} the effect of replacing 3D by 1D models is tested.
In section~\ref{Reliability} the suitability of HARPS for the use of this technique is tested.
Finally, in section~\ref{resume} we summarize our results and conclusions.

\section{Data}
\label{data}
\subsection{Sample selection}
\begin{figure}
   \centering
   {\includegraphics[width=.4\textwidth]{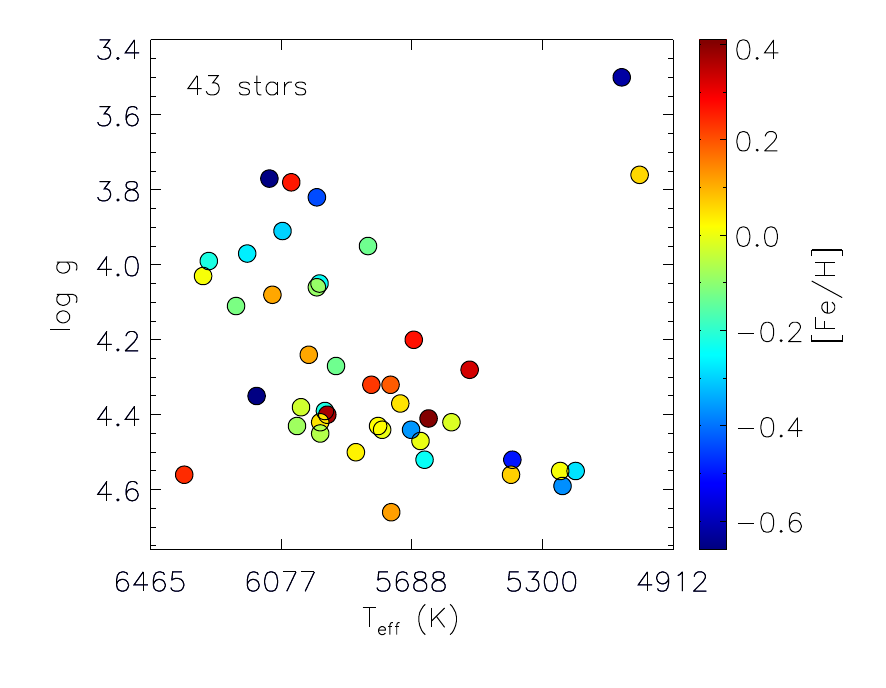}}
   \caption{Parameter space covered by the sample stars. The values are listed in Table~\ref{objects}.}
   \label{sample_space}
   \end{figure}

The sample stars are presented in Table~\ref{objects}.
These are 43 F, G, and K type stars including the Sun observed by means of the proxies Ganymede, Ceres, Calisto and Moon.
They were selected from the HARPS/ESO archive of reduced and calibrated data, 
brighter than $V = 7$ to obtain spectra of good quality with
the MUSICOS and coud\'{e} instruments.
Thus, three samples of spectra were collected
(named according to the spectrograph of acquisition).
More stars were observed with coud\'{e} in order to cover 
as much as possible the \mbox{\teff--[Fe/H]--\logg} parameter space.
Therefore, every object in the HARPS and MUSICOS subsamples has associated coud\'{e} spectra. 
The parameter space covered by the sample stars is presented in Fig.~\ref{sample_space}.
Stellar parameters were extracted from a compilation of catalogs from the literature coded henceforth as follows:
(Sousa08) \citet{sousa2008}, 
(Ghezzi10) \citet{gh2010}, 
(Tsantaki13) \citet{tsa2013}, 
(Ramirez13) \citet{ram2013}, 
(Bensby14) \citet{bensby2014}, 
(Ramirez14a) \citet{ram_2014}, 
(Ramirez14b) \citet{ram2014}, 
(Maldonado15) \citet{maldo2015}, 
(Heiter15) \citet{heiter2015}. 
In order to compare literature $T_{\mathrm{eff}}$ scales 
with ours, we selected works that derived $T_{\mathrm{eff}}$ with three different techniques: 
excitation and ionization of Fe lines (Sousa08, Ghezzi10, Tsantaki13, Bensby14, 
Ramirez14a, Ramirez14b, Maldonado15),
photometric calibrations based in the Infrared flux method (Ramirez13)
and interferometry (Heiter15). 
Most of the parameters in Table~\ref{objects} belong to Ramirez13 because our selection started 
with this catalog, which has a large number of stars from H{\scriptsize IPPARCOS}
observable in the southern telescopes.

We added Ceres to the HARPS sample 
to expand the data in time in order to 
check the temporal stability of the instrument. 
The solar proxies analyzed are listed in Table~\ref{proxies} together with their date of observation, S/N ratio, and 
the temperatures derived in this work. We extracted 10 random spectra of the same object per day/year.
The only 6 spectra available of 2010/10 were complemented with 
spectra of the close date 2010/12, and for 2007 and 2009 only the available spectra were used.

\begin{table*}
\small
\centering
\caption{ \small Sample stars.
The column 4 specifies the spectrograph of acquisition: coud\'{e} (Co), HARPS (HA) and MUSICOS (MU).
The columns 5, 6 and 7 list the atmospheric parameters used to select the sample.
The last column indicate the catalogs 
that provide parameters of the star, with which we compare our results in Sect.~\ref{accuracy} and \ref{consistency}.
The identification code is: 
(1) \citet{sousa2008}, (2) \citet{gh2010},
(3) \citet{tsa2013}, (4) \citet{ram2013}, (5) \citet{bensby2014}, (6) \citet{ram_2014},
(7) \citet{ram2014}, (8) \citet{maldo2015}, (9) \citet{heiter2015}.
The catalog from which the parameters in columns 5, 6 and 7 were taken is highlighted in bold.}
\label{objects}
\begin{tabular}{l c c c c c c l}
\hline\hline \\
Name & HD & HIP & spectrum & $T_{\mathrm{eff}}$ (K) & \logg & [Fe/H] & \;\;\;ctlg  \\
\hline \\
Moon    & & & Co/HA/MU & 5771 & 4.44 & 0.00 & \\
Ganymede & & & Co/HA/MU & 5771 & 4.44 & 0.00 & \\
Calisto  & & & Co & 5771 & 4.44 & 0.00 & \\
Ceres & & & HA & 5771 & 4.44 & 0.00 & \\
$\zeta$ Tuc & 1581	& 1599 & Co/HA & 5947 & 4.39 & $-0.22$ & 1,2,3,\textbf{4},5,8 \\ 
$\beta$ Hyi & 2151	& 2021 & Co & 5819 & 3.95 & $-0.13$ & 3,\textbf{4},9 \\
& 3823	& 3170 & Co/HA & 5963 & 4.05 & $-0.24$ & 1,2,3,\textbf{5},8 \\
$\tau$ Cet & 10700	& 8102 & Co/HA & 5390 & 4.52 & $-0.50$ & 1,2,3,\textbf{4},8,9 \\
$\epsilon$ For & 18907	& 14086 & Co/HA & 5065 & 3.50 & $-0.62$ & \textbf{4},9 \\
$\alpha$ For & 20010	& 14879 & Co & 6073 & 3.91 & $-0.30$ & \textbf{4},5 \\
$\kappa$ Cet & 20630	& 15457 & Co & 5663 & 4.47 & 0.00 & 2,\textbf{4},8 \\
$10$ Tau & 22484	& 16852 & Co & 5971 & 4.06 & $-0.09$ & 2,\textbf{4},5,8 \\
$\delta$ Eri & 23249	& 17378 & Co/HA & 5012 & 3.76 & 0.06 & 1,3,\textbf{4},8,9 \\
40 Eri & 26965	& 19849 & Co/HA & 5202 & 4.55 & $-0.28$ & 1,3,\textbf{4},8 \\
& 100623	& 56452 & Co/HA & 5241 & 4.59 & $-0.37$ & \textbf{4},5,8 \\
$\beta$ Vir & 102870	& 57757 & Co/MU & 6103 & 4.08 & 0.11 & 2,\textbf{4},9 \\
& 114174	& 64150 & Co & 5723 & 4.37 & 0.05 & \textbf{4},7 \\
$59$ Vir & 115383	& 64792 & Co & 5995 & 4.24 & 0.11 & 2,\textbf{4},5,8 \\
$61$ Vir & 115617	& 64924 & Co/HA/MU & 5571 & 4.42 & $-0.02$ & 1,2,3,\textbf{4},5,8 \\
$\eta$ Boo & 121370	& 67927 & Co/HA & 6047 & 3.78 & 0.26 & \textbf{4},9 \\
& 126053	& 70319 & Co & 5691 & 4.44 & $-0.36$ & 2,\textbf{4},8 \\
$\alpha$ Cen A & 128620	& 71683 & Co/HA & 5809 & 4.32 & 0.23 & \textbf{4},8,9 \\
$\psi$ Ser & 140538	& 77052 & Co/HA & 5750 & 4.66 & 0.12 & 7,\textbf{8} \\
& 144585	& 78955 & Co/HA & 5940 & 4.40 & 0.37 & 1,3,\textbf{5},6 \\
18 Sco & 146233	& 79672 & Co/HA/MU & 5789 & 4.43 & 0.02 & 1,3,\textbf{4},7,8,9 \\
& 147513	& 80337 & Co & 5855 & 4.50 & 0.03 & 1,2,3,\textbf{4},5,8 \\
$\zeta$ TrA & 147584	& 80686 & Co/HA & 6030 & 4.43 & $-0.08$ & \textbf{4},5 \\
12 Oph & 149661	& 81300 & Co/HA & 5248 & 4.55 & 0.01 & \textbf{4},5,8 \\
& 150177	& 81580 & Co/HA & 6112 & 3.77 & $-0.66$ & \textbf{5} \\
& 154417	& 83601 & Co/HA & 6018 & 4.38 & $-0.03$ & \textbf{4},5 \\
$\mu$ Ara & 160691	& 86796 & Co/HA/MU & 5683 & 4.20 & 0.27 & 2,\textbf{4},6,9 \\
70 Oph & 165341	& 88601 & Co & 5394 & 4.56 & 0.07 & \textbf{4},8 \\
$\iota$ Pav & 165499	& 89042 & Co & 5914 & 4.27 & $-0.13$ & \textbf{8} \\
& 172051	& 91438 & Co & 5651 & 4.52 & $-0.24$ & \textbf{4},5,8 \\
& 179949	& 94645 & Co/HA & 6365 & 4.56 & 0.24 & 1,2,3,\textbf{5},6 \\
31 Aql & 182572	& 95447 & Co/MU & 5639 & 4.41 & 0.41 & \textbf{5} \\
& 184985	& 96536 & Co/HA & 6309 & 4.03 & 0.01 & \textbf{2},5 \\
$\delta$ Pav & 190248	& 99240 & Co/HA & 5517 & 4.28 & 0.33 & 1,2,3,\textbf{4},8 \\
15 Sge& 190406	& 98819 & Co & 5961 & 4.42 & 0.05 & 2,\textbf{4},8 \\
$\phi^2$ Pav & 196378	& 101983 & Co & 5971 & 3.82 & $-0.44$ & \textbf{8} \\
$\gamma$ Pav & 203608	& 105858 & Co/HA/MU & 6150 & 4.35 & $-0.66$ & \textbf{4},5,8 \\
& 206860	& 107350 & Co/HA & 5961 & 4.45 & $-0.06$ & \textbf{4},8 \\
$\xi$ Peg & 215648	& 112447 & Co/MU & 6178 & 3.97 & $-0.27$ & \textbf{2} \\
49 Peg & 216385	& 112935 & Co/HA & 6292 & 3.99 & $-0.22$ & \textbf{4} \\
51 Peg & 217014	& 113357 & Co/HA/MU & 5752 & 4.32 & 0.19 & \textbf{4},5,8 \\
$\iota$ Psc & 222368	& 116771 & Co/HA & 6211 & 4.11 & $-0.12$ & 2,\textbf{4},8 \\
\hline \\
\end{tabular}
\end{table*}

\subsection{do Pico dos Dias observations}
We used coud\'{e} and MUSICOS in 2016 and 2017.
Both spectrographs are fed by the 1.60 m Perkin-Elmer telescope of 
do Pico dos Dias Observatory (OPD, Braz\'{o}polis, Brazil), 
operated by Laboratório Nacional de Astrofısica (LNA/CNPq).
In the coud\'{e} spectrograph the slit width was adjusted to give a two-pixel resolving power 
$R = \lambda/ \Delta \lambda = 45\,000$.
A 1800 l/mm diffraction grating was employed in the first order, 
projecting onto a 13.5 $\mu$m, 2048 pixels CCD.
The spectral region is centered on the H$\alpha$ line \mbox{$\lambda = 6562.797$~\AA},
with a spectral coverage of $155$~\AA.

MUSICOS is a fiber-fed echelle spectrograph \citep[e.g.][]{1992A&A...259..711B} 
(on loan from Pic du Midi Observatory since 2012) 
available for the OPD/LNA. We employed the red channel, 
covering $\lambda$5400-8900~\AA ~approximately, comprising about 50 spectral orders, 
at \mbox{R $\sim 40~000$} and 0.05~\AA/pix dispersion in the H$\alpha$ wavelength range.

The exposure times were chosen to obtain S/N ratios of at least 250 for the faintest stars ($V \sim 7$)
and 300 in average for the other stars.

\section{Normalization}
\label{norm}
The challenge in normalizing H$\alpha$ profiles 
arises from the uncertainty of the continuum location,
that is estimated defining ``continuum windows''.
Thus, the success of the normalization
resides in the capability of identifying many wide
windows that allow to determine the shape of the spectrograph response.
   
Frequently, the continuum windows are determined 
using  automatic or semiautomatic procedures, as the 
IRAF\footnote{\textit{Image Reduction and Analysis Facility} (IRAF) is distributed by 
the National Optical Astronomical Observatories (NOAO), which is operated by the
Association of Universities for Research in Astronomy (AURA), Inc., under contract to 
the National Science Foundation (NSF).} task \textit{``continuum''}, 
selecting the wavelength bins with the highest fluxes by applying clipping.
We improve this procedure by iterating on the normalization and fitting processes, 
in this way the compatibility at the extremes of the wings are checked after every fit. 
This check is fundamental for consistent temperature measurements because, 
although the spectrograph response may be well described by a low order polynomial 
(as is the case of coud\'{e}),
the normalization by interpolation may be highly imprecise close to the line-core.
It occurs because the continuum regions available to interpolate the polynomial are short compared to the 
fitted region, 
thus small errors in the outer profile wings 
trigger larger errors close to the line-core, where the H$\alpha$ profile is more sensitive to the temperature.
With this method, explained below in detail, we minimize the main source of uncertainty, 
as demonstrated by the very low dispersion of \teffa~values obtained with many solar 
spectra in Sect.~\ref{zero-p} and \ref{Reliability}.

Normalization is more complex in echelle spectra,
because of the correction of the blaze 
and order merging.
As discussed by \citet{skoda2004}, distortions in the spectra, such as
discontinuities of the orders and ripple-like patterns 
\citep[see][Fig.~11]{skoda2008}
are often produced in slit echelle spectrographs but possibly also in fiber-fed instruments.
When this occurs, the spectra are useless and a new reduction from raw data should be applied
following the recipe recommended by \citet{skoda2008}.
Of course, empirical corrections on the reduced spectra could recover the profiles, 
but their goodness must be tested by recovering the \teff~accuracy obtained with non-distorted profiles.
On the other hand, also spectra with no obvious distortions need to be tested, 
because subtle residual blaze features may remain and systematically impact the 
\teff~estimate.
Residual blaze features distort the profiles making them shallower 
(more strongly close to the center of the spectral order), 
thus the distorted spectra mimic profiles of cooler temperatures.
In order to investigate this effect in HARPS, the 1D pipeline-reduced HARPS spectra were analyzed 
in the same way as the coud\'{e} ones, and the derived \teffa's were compared. 
The results of this analysis are presented in Sect.~\ref{Reliability}.

The normalization method applied to coud\'{e} and HARPS is  independently validated by 
deriving \teffa's from MUSICOS spectra normalized with the 2D-normalization. 
These results are presented in Sect.~\ref{2DN}.

   \begin{figure*}
   \centering
   {\includegraphics[width=14cm]{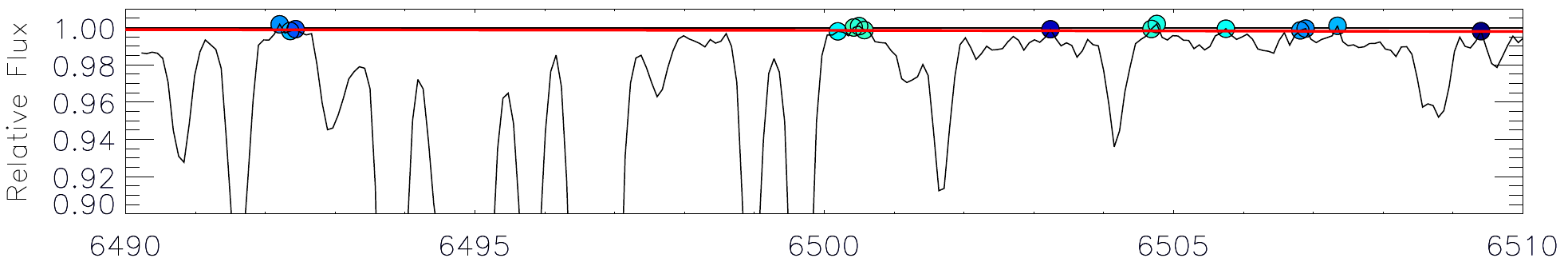}}
   {\includegraphics[width=14cm]{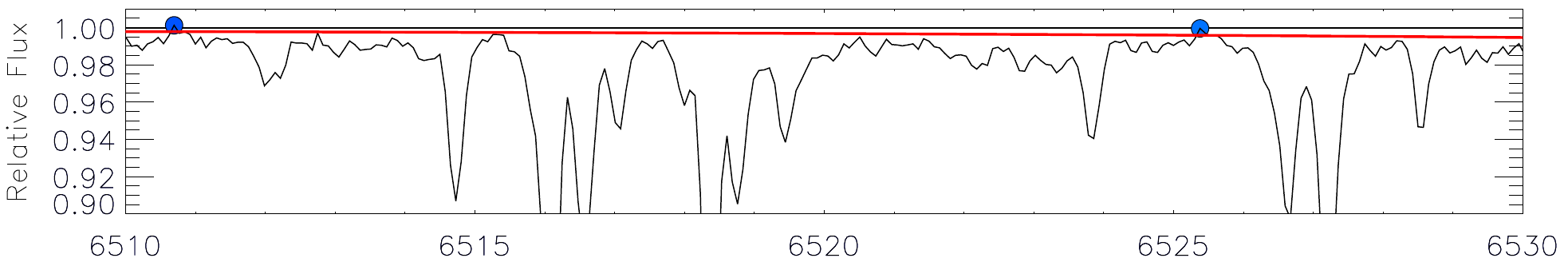}}
   {\includegraphics[width=14cm]{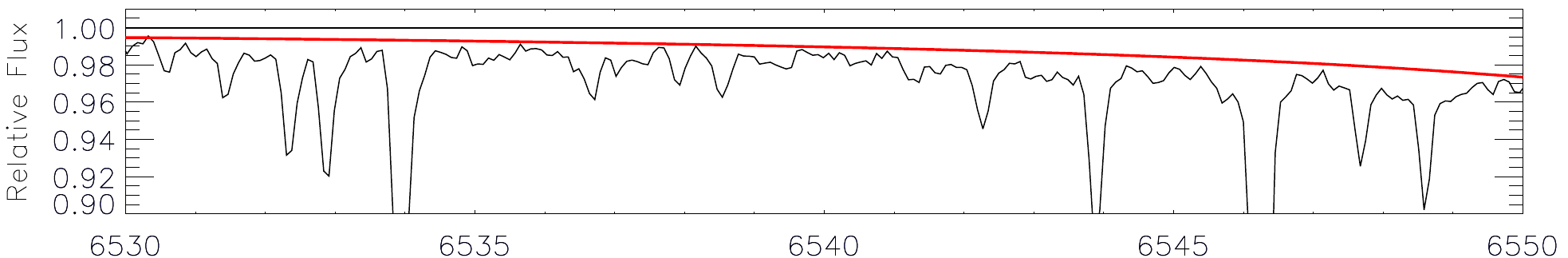}}
   {\includegraphics[width=14cm]{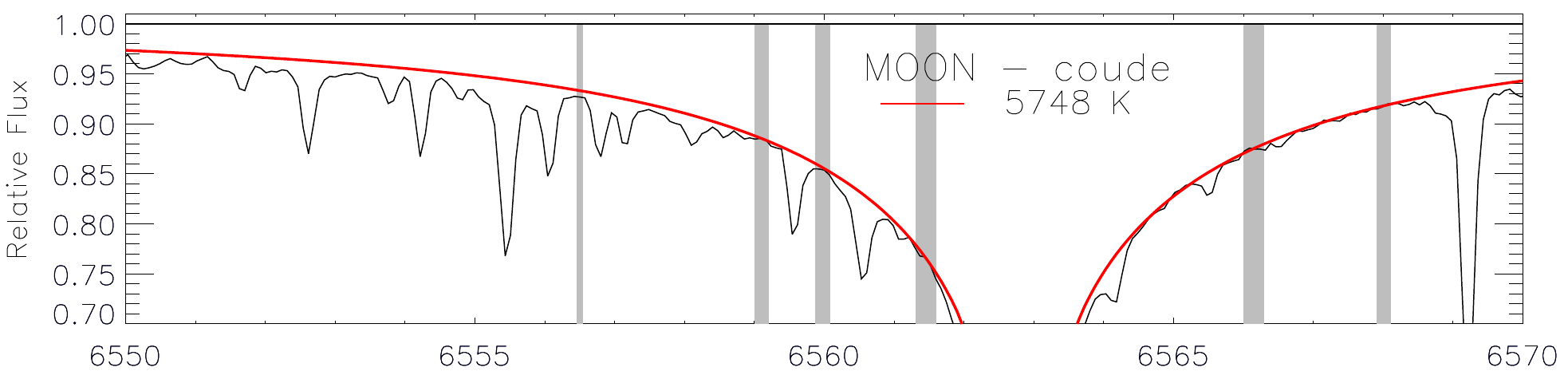}}
   {\includegraphics[width=14cm]{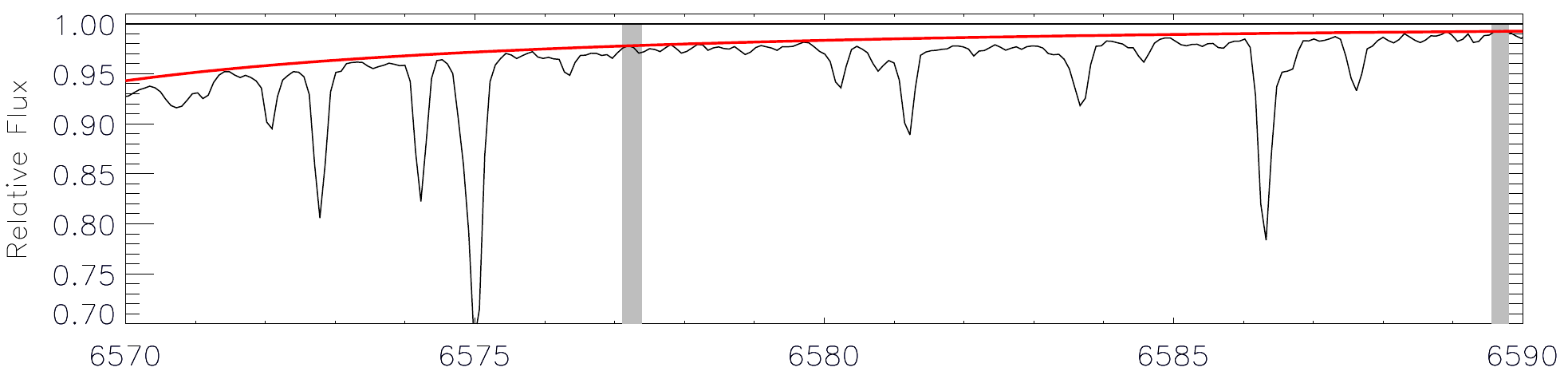}}
   {\includegraphics[width=14cm]{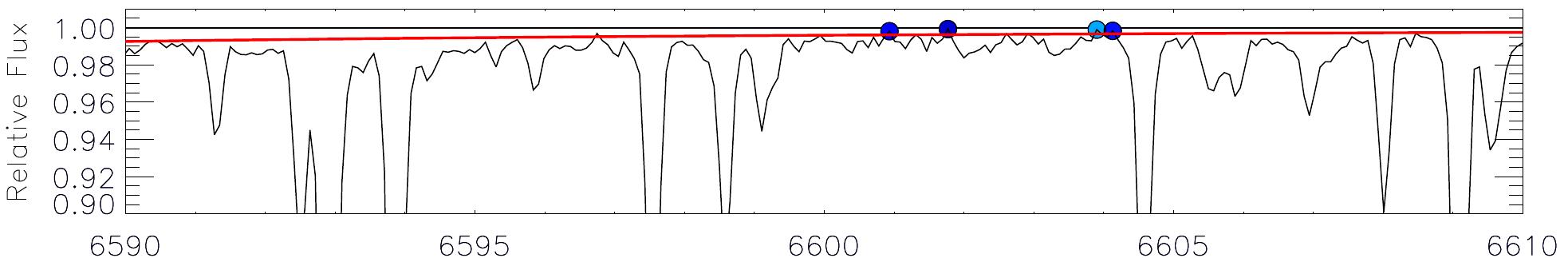}}
   {\includegraphics[width=14cm]{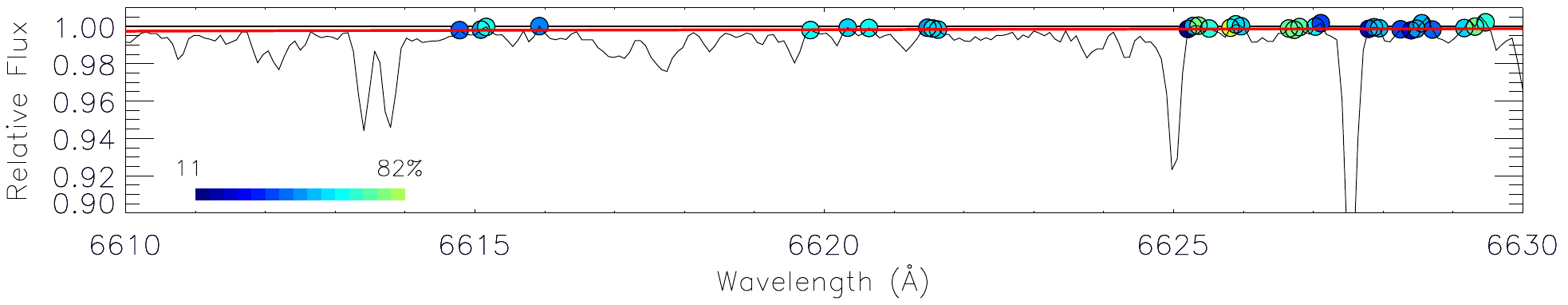}}
   {\includegraphics[width=.4\textwidth]{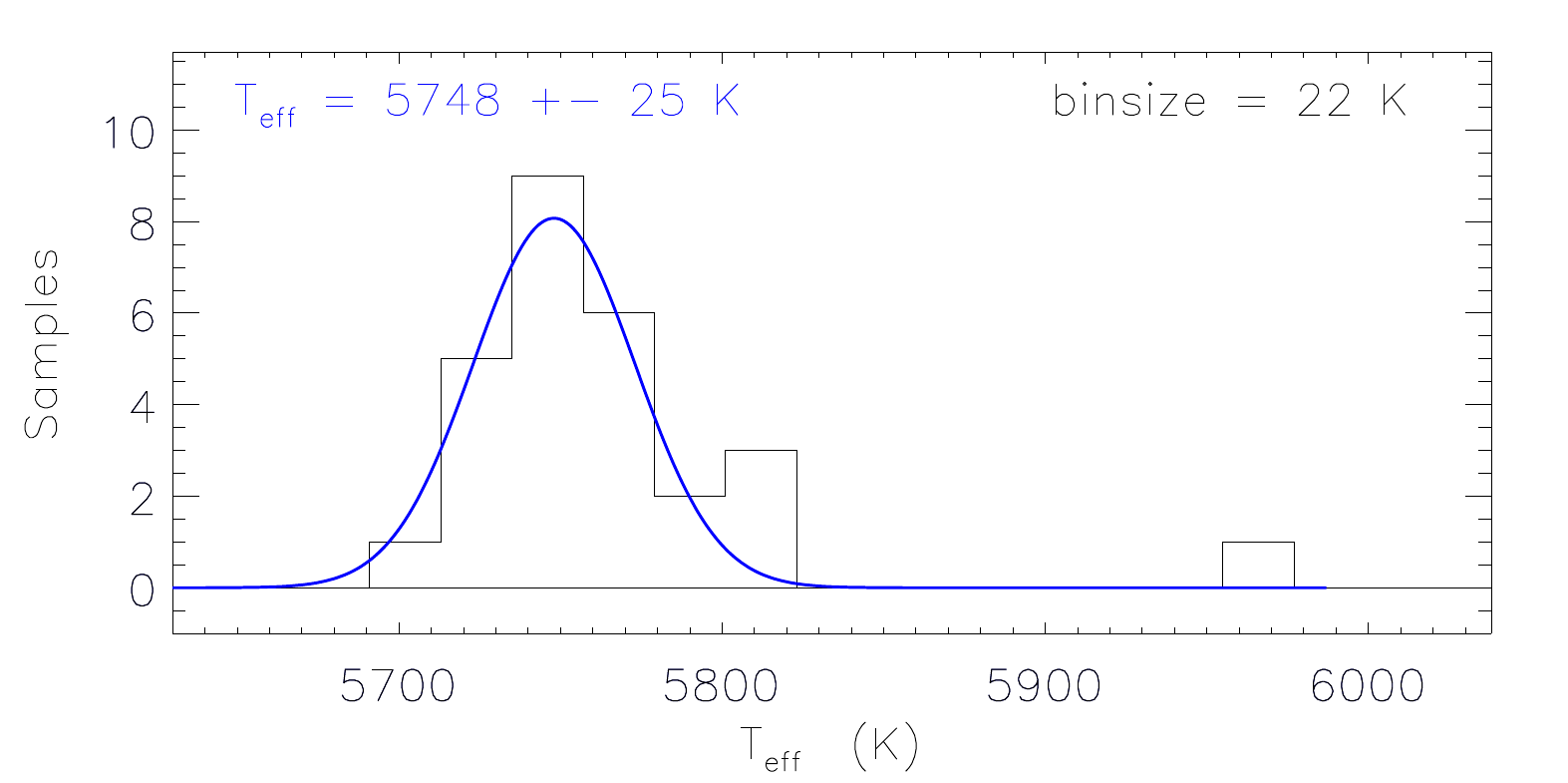}}
   \caption{Coud\'{e} H$\alpha$ profile of one of the solar proxies in Table~\ref{proxies}. 
   The red and black lines represent the synthetic and observed profiles.
   The shaded regions are the windows of fits and
   the circles represent the continuum bins 
   color-coded according to their frequency of appearance 
   in all coud\'{e} spectra.
   The most frequent continuum windows are observed at [6500.25, 6500.50], [6504.50, 6505.00], [6619.70, 6620.50],
   [6625.60, 6625.80] and [6626.50, 6626.80].
   \textit{Bottom panel:} Histogram of temperatures related to the wavelength bins within the windows of fits.
   A Gaussian is fitted to its median and robust standard deviation.}
   \label{sun_normal}
   \end{figure*}   
   
   \begin{figure*}[!]
   \centering
   {\includegraphics[width=14cm]{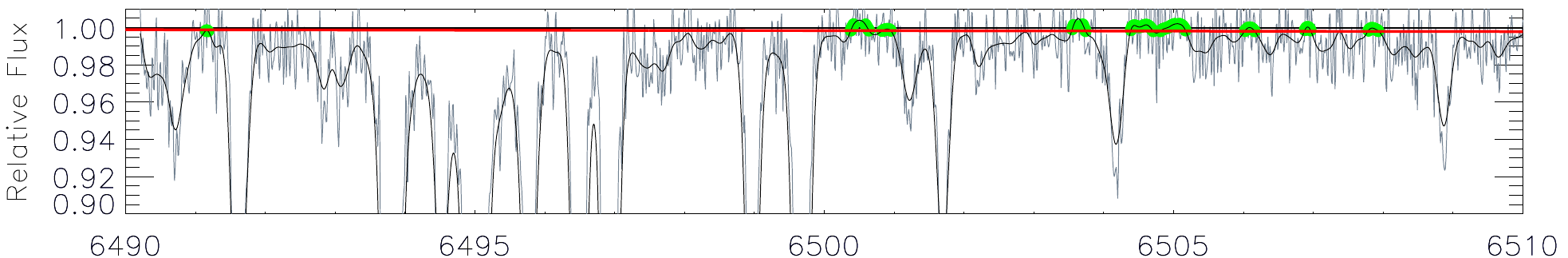}}
   {\includegraphics[width=14cm]{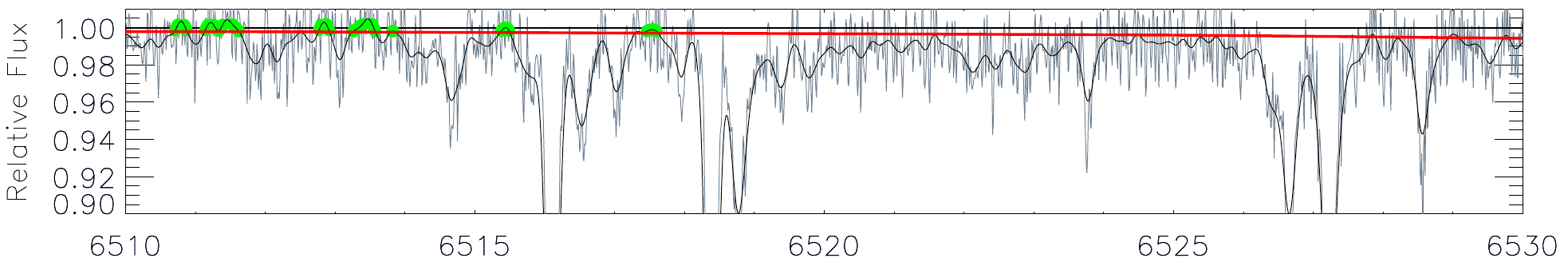}}
   {\includegraphics[width=14cm]{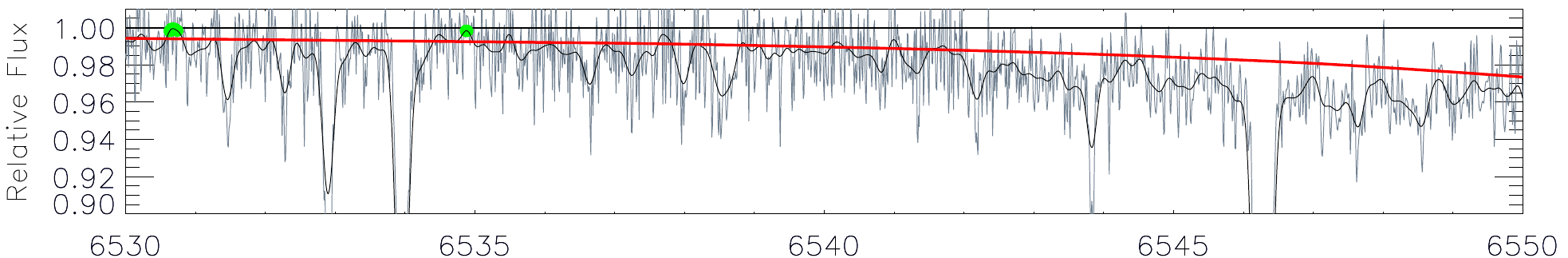}}
   {\includegraphics[width=14cm]{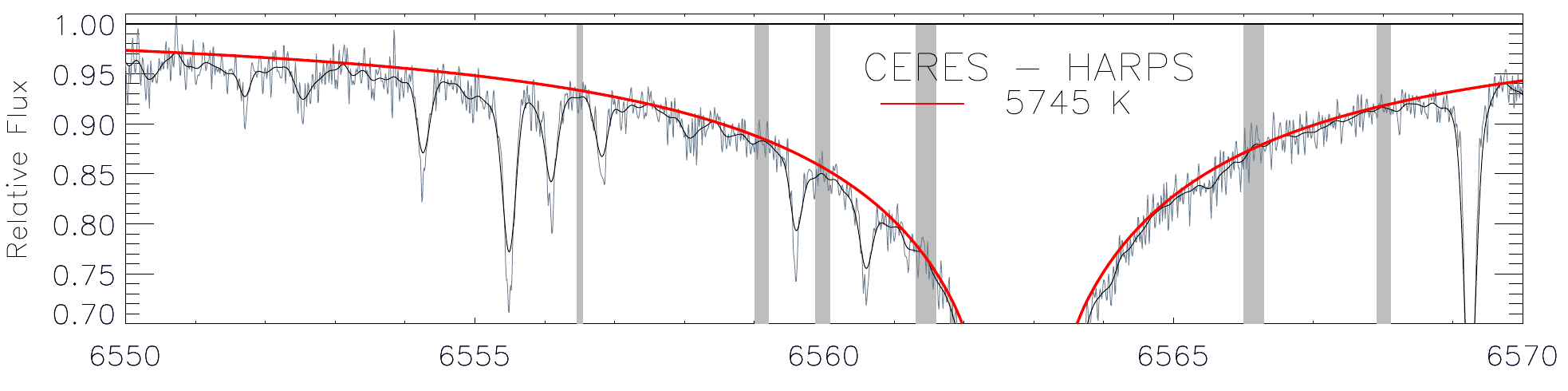}}
   {\includegraphics[width=14cm]{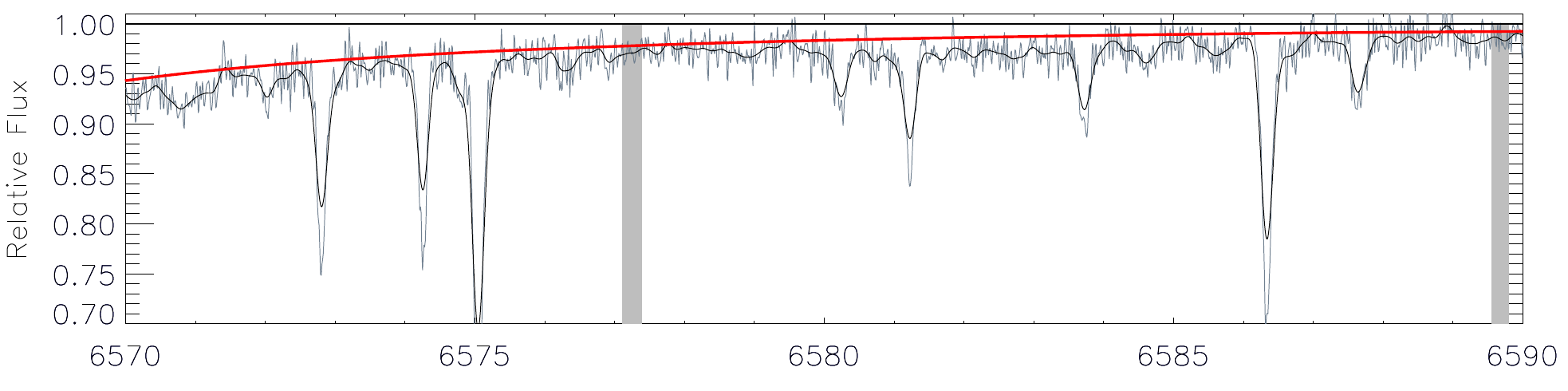}}
   {\includegraphics[width=14cm]{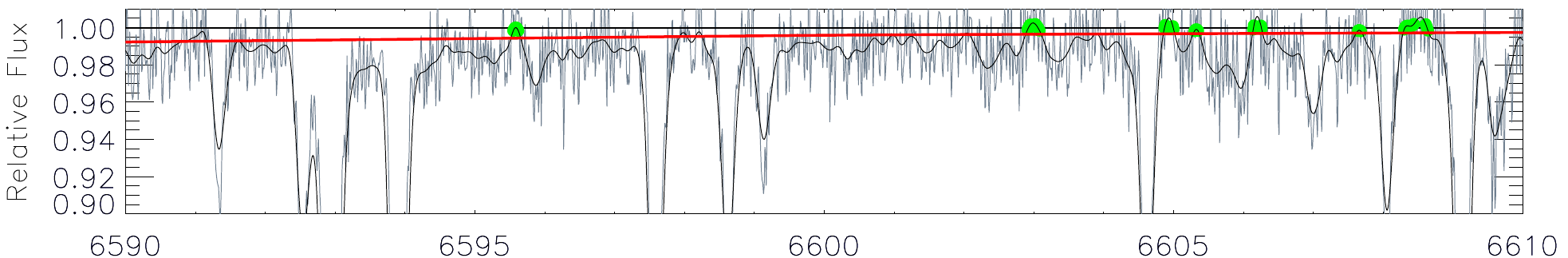}}
   {\includegraphics[width=14cm]{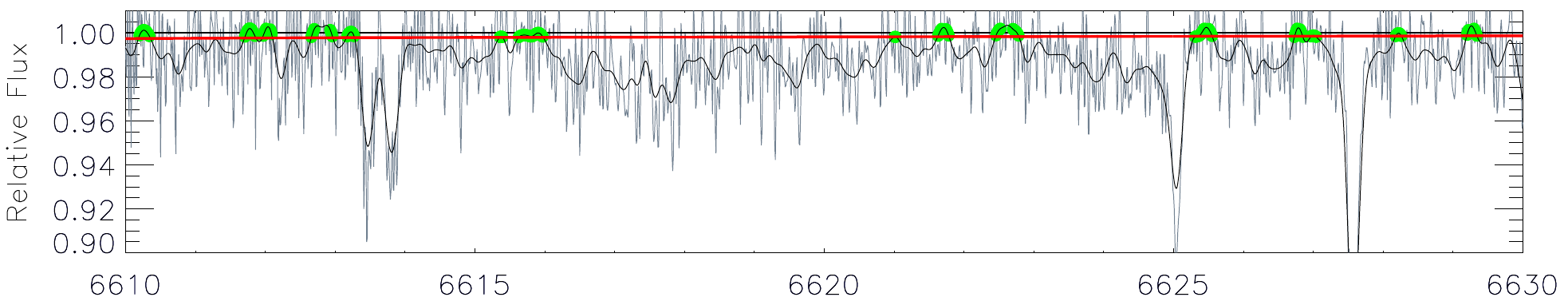}}
   {\includegraphics[width=.4\textwidth]{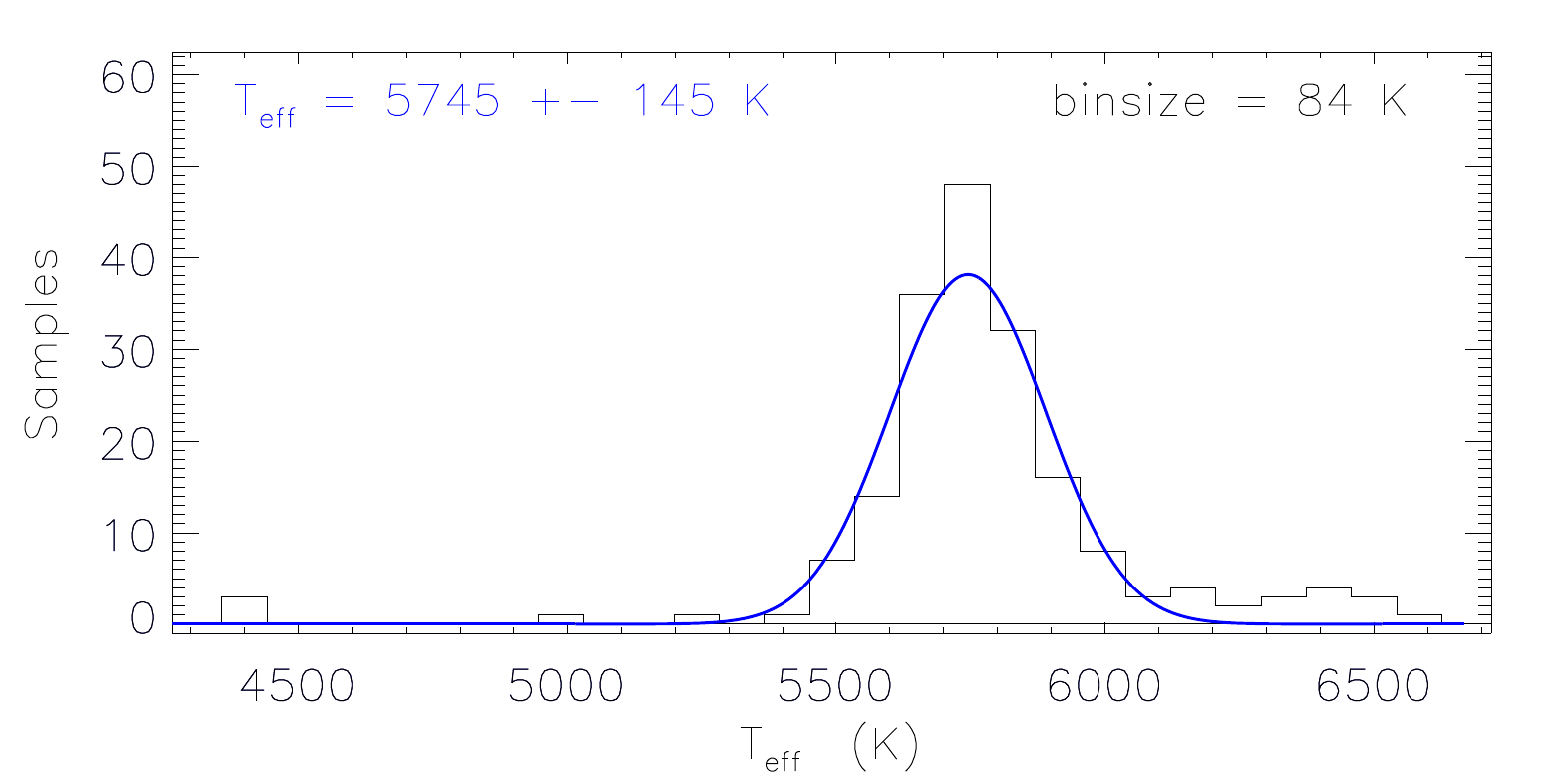}}
   \caption{Analogous to Fig.~\ref{sun_normal} with a HARPS spectrum of one of the solar proxies
   from Table~\ref{proxies}.
   The gray line represents the spectrum in its original resolution and the black line represents the spectrum degraded to the 
   resolution of coud\'{e}.  
   Continuum bins in the degraded spectrum are highlighted in green; notice that they mostly match those of 
   Fig.~\ref{sun_normal}.}
   \label{sun_normal_HARPS}
   \end{figure*}  
   
   \begin{figure*}
   \centering
   {\includegraphics[width=.8\textwidth]{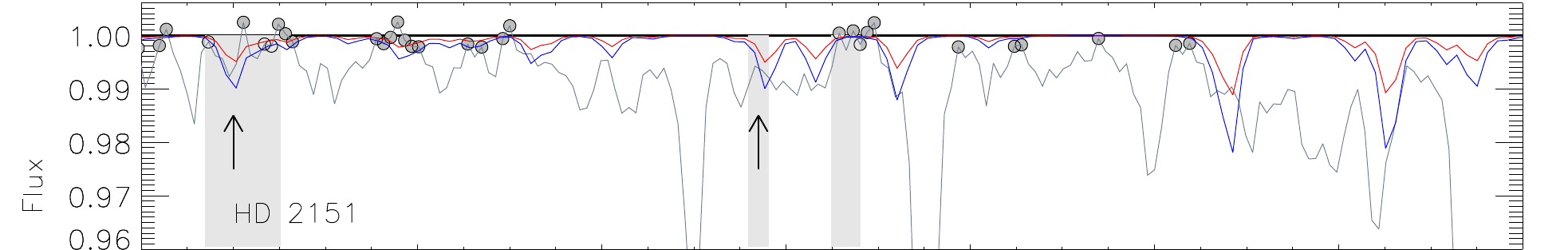}}
   {\includegraphics[width=.1265\textwidth]{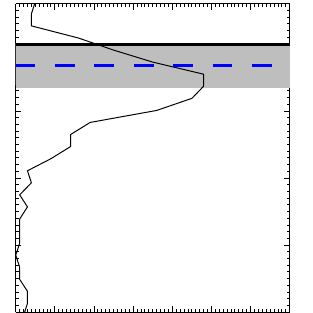}}
   {\includegraphics[width=.8\textwidth]{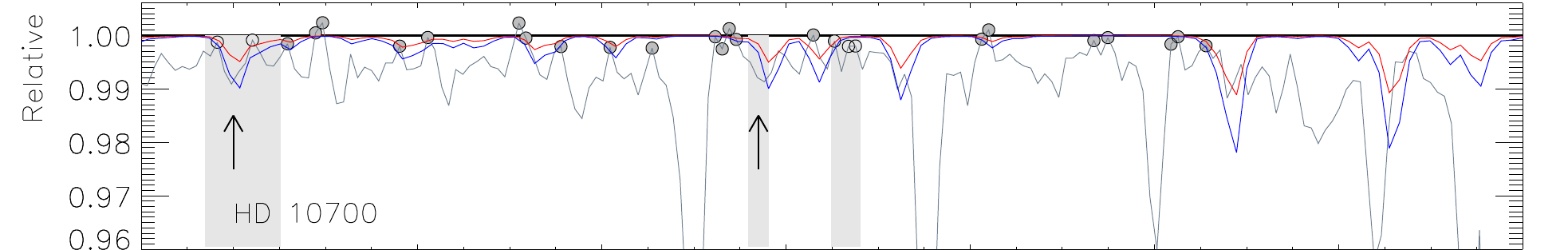}}
   {\includegraphics[width=.1265\textwidth]{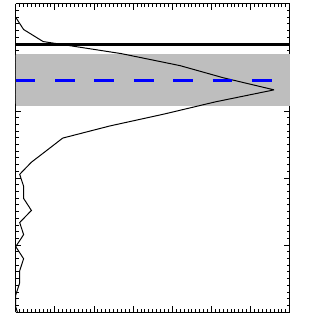}}
   {\includegraphics[width=.8\textwidth]{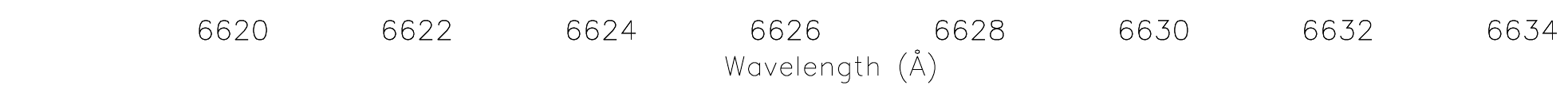}}
   {\includegraphics[width=.1265\textwidth]{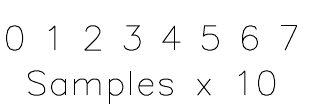}}
   \caption{\textit{Left panels:} Fitting of two coud\'{e} spectra (gray line) 
   with synthetic spectra of PWV with concentrations of 10 and 20 mm (red and blue lines).
   The circles are the continuum wavelength bins on 1 $\pm~\sigma$(noise).
   The shades represent 3 of the 5 continuum windows selected in Fig.~\ref{sun_normal}.
   The arrows point the windows contaminated by telluric features.
   \textit{Right panels:} Flux histograms of the spectra on the left panels 
   with the same flux scale. The black horizontal line points the continuum, 
   the dashed line is the average flux of the 5 continuum windows of Fig.~\ref{sun_normal}
   and the shades are the spread.}
   \label{telluric}
   \end{figure*}  
   
   \subsection{Normalization of coud\'{e} and HARPS spectra} 
   \label{coude}
   The normalization is applied by interpolating low order polynomials with 
   the IRAF task \textit{``continuum''}, integrated with the fitting code described in Sect.~\ref{fitting} in an iterative procedure:
   
   \begin{enumerate}
   \item A first gross normalization is performed neglecting 
   the region $6514-6610$~\AA~in the interpolation.
   Although the extension of the H$\alpha$ wings is variable,
   this region is kept the same for all the sample stars
   with the purpose of keeping enough room to apply weights in nearby regions 
   to modulate the normalizing curve.
   
   \item The obtained profile is used to fit a precipitable water vapor (PWV) spectrum 
   that will be used to verify the continuum level after every iteration, 
   see Sect.~\ref{Continuum_tuning}.
   
   \item The same normalized profile is compared with the grid of synthetic profiles 
   using the fitting code described in Sect.~\ref{fitting} to find the most compatible one.
   
   \item The compatibility between the normalized and synthetic profiles
   must be visually checked at the
   ``transition regions'' ($\lambda < 6536$~\AA~and $6590$~\AA~< $\lambda$)
   in which  the continuum turns into line wings.
   The regions of the  line interior  are very sensitive to 
   temperature, hence they are predominant in the fittings.  
   For this reason, if distortions are artificially introduced in the profile during the normalization,
   they become more evident in the transition regions.
   This procedure makes our normalizations dependent on the model but very weakly, 
   because metallicity and
   surface gravity (the parameters set beforehand) do not greatly influence the shape of the line, 
   especially in the transition regions. 
   We verified that changes as large as \mbox{$\sim\!\!\pm0.3$ dex} do not modify significantly 
   the shape of the normalized profiles, while larger changes may truncate the procedure.
   For consistency, HARPS spectra were degraded to the resolution of coud\'{e} in this step 
   (only for this step, not for the fitting procedure), see Fig.~\ref{sun_normal_HARPS}. 
   In Fig.~\ref{transition_regions} examples of transition regions at the red wing of 
   H$\alpha$ in solar spectra normalized by different authors are provided.
   In it, the fit of the coud\'{e} spectrum of Fig.~\ref{sun_normal} 
   is compared with fits of KPNO2005, and the solar atlas of \citet{wall2011} (KPNO2011) 
   to show how this method improves the normalization. 

   \item Usually the first normalization is deficient, in this case  
   a second one is performed \textit{from scratch} applying weights to the wings around 6514 and $6610$~\AA~to 
   make the profile deeper or shallower as required to match the flux of the synthetic profile. 
   Then, another fit is applied and the matching check 
   described in step 4 is repeated. 
   The procedure finishes when the observed and synthetic profiles are 
   compatible in the transition regions, as shown in
   Fig.~\ref{sun_normal} and \ref{sun_normal_HARPS}. 
   An example of the difference between the first gross normalization and the final normalization 
   is shown in Fig.~\ref{gross_norm}.
   \end{enumerate}
   
   \subsection{Continuum fine-tune}
   \label{Continuum_tuning}
   The solar KPNO2005 atlas and the lines catalog of
   \citet{moore} were used to select windows free from metallic lines to check the continuum 
   during the normalization procedure.
   However, the availability of these windows diminish progressively in cool and metal-rich stars
   and because of the presence of telluric lines.
   Since the humidity at do Pico dos Dias Observatory often exceeded $90\%$ during our observations,
   the contribution of many minute telluric lines is relevant in the coud\'{e} spectra.
   To fine-tune the continuum level, as part of the procedure described in Sect.~\ref{coude},
   we separated telluric features from noise fitting the observed spectra with synthetic telluric spectra
   as shown in Fig.~\ref{telluric}.
   
   Attempts for the fittings were performed with the \textit{Molecfit} software package 
   described in detail in Sect.~\ref{telluric_correction} 
   and with the PWV library of \citet{Moehler2014}\footnote{\url{ftp://ftp.eso.org/pub/dfs/pipelines/skytools/telluric\_libs}}.
   The first demonstrated to be precise for fitting strong features
   but many more weak features are present in the second, which makes it more suitable for this
   analysis.
   The PWV library is available at resolutions R = 300~000 and R = 60~000, 
   for the air-masses 1.0, 1.5, 2.0, 2.5, 3.0
   and water content of 0.5, 1.0, 1.5, 2.5, 3.5, 5.0, 7.5, 10.0, and 20.0 mm.
   The fitting is performed degrading the resolution of the original PWV spectra to match 
   those of the spectrograph used, and selecting the set of PWV spectra with the air-mass closest to that of the observation.
   
   We quantified the displacement of the continuum due to the presence of telluric features as follows.
   After normalized all coud\'{e} spectra, continuum wavelength bins were identified in the 
   solar spectrum of Fig.~\ref{sun_normal} applying \mbox{$\sigma$-clipping}.
   The fluxes of these wavelength bins were then checked in all other normalized coud\'{e} spectra, 
   and none of them was found to remain as continuum in all the sample.
   The color-code of the plot in the figure represents the percentage rate,
   being the windows at [6500.25, 6500.50], [6504.50, 6505.00], [6619.70, 6620.50], [6625.60, 6625.80], 
   [6626.50, 6626.80]~\AA~the most frequent.
   Fig.~\ref{telluric} shows two cases where two of these windows are affected by the presence of minute telluric lines,
   and how much the average flux of the five mentioned windows decreases.
   Analyzing all the sample spectra, we find that when the content of PWV is high, say between 7.5 and 20.0 mm,
   minute telluric features are almost omnipresent and displace the continuum flux by about $0.5\%$.
   In our experience, this issue may induce to underestimate the stellar temperature 
   between 30 and 100 K.
   It is however difficult to provide a precise estimate because
   the displacement produced is often not homogeneous, but a distortion of the continuum shape. 
   We stress that no correction is applied during this procedure, only a visual check.
   The correction of strong features is done later, and it is explained Sect.~\ref{telluric_correction}.

\section{Profiles fitting}
      \begin{figure*}[t]
   \centering
   \includegraphics[width=14cm]{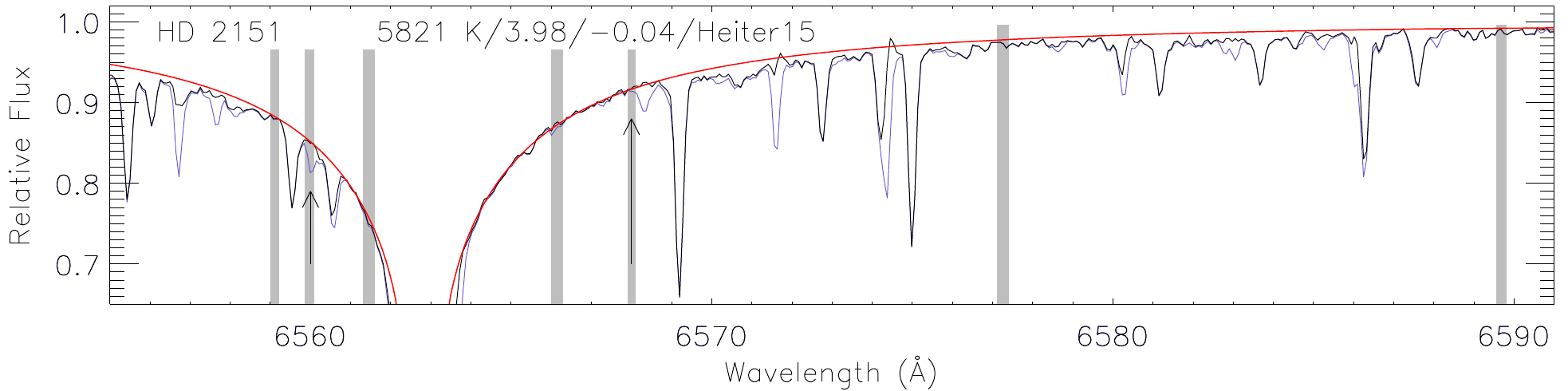}
   \includegraphics[width=3.5cm]{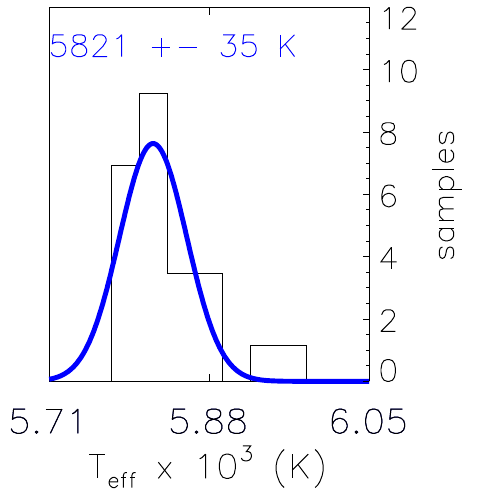}\\
   \caption{Telluric correction and profile fitting of the coud\'{e} spectrum of \mbox{HD 2151}. 
           \textit{Left panel: }Corrected and non corrected spectra are represented by 
           the black and blue lines, respectively. The windows of fits are represented by the shades,
           and the arrows point those where the relative flux was perfectly recovered.
           The red line represent the synthetic profile fitted.
           \textit{Right panel: }
           Histogram of temperatures related to the wavelength bins inside the windows of fits.
           The most probable \teffa~is shown in the top part of the plots, also log \logg~and [Fe/H] values used for the fittings 
           along with their source in the literature are shown.}
   \label{t_correction}
   \end{figure*}

\label{fitting}
This study is based on the grid of synthetic profiles of BPO02
computed using the self-broadening theory developed in \citet{BPO2000}
and the 1D LTE plane-parallel model atmospheres from the MARCS code 
\citep{asplund1997}.
The atmospheric parameters of the grid are
$T_{\mathrm{eff}}$: 4400 to 7500 K with steps of 100 K,
[Fe/H]: $-3.0$ to $+0.5$ dex with steps of 0.5 dex,
\logg: 3.4 to 5.0 dex with steps of 0.5 dex and 
microturbulence velocity of 1.5 Km$/$s.
In order to derive very precise $T_{\mathrm{eff}}$'s around solar parameters, 
a more detailed grid from the same theoretical recipe used by \citet{ram2011}
(kindly provided by the first author by private communication) is also used here, 
its parameters are 
$T_{\mathrm{eff}}$: 5500 to 6100 K with steps of 10 K,
[Fe/H]: $-3.0$ to $+0.3$ dex with steps of 0.05 dex,
\logg: 4.2 to 4.65 dex with steps of 0.05 dex and 
microturbulence velocity of 1.5 Km$/$s.
The fitting between the observed and synthetic profiles is performed using 
the ``windows of fits'' free from metallic lines: [6556.45, 6556.55], [6559.00, 6559.20], [6559.86, 6560.08],
[6561.30, 6561.60], [6566.00, 6566.30], [6567.90, 6568.10], [6577.10, 6577.40],
[6589.55, 6589.80]\footnote{No more windows in the blue wing of the profile were included 
because our spectra appear systematically contaminated by telluric features in this region}.

A program in IDL\footnote{Interactive Data Language, version 7.0} was written to perform the fits 
eliminating the influence of contaminated wavelength bins.  
It first interpolates the resolution of the grids to 1 K, 0.01 dex, 0.01 dex  
in $T_{\mathrm{eff}}$, [Fe/H], \logg. 
Then, for each wavelength bin, the temperature related to the interpolated synthetic profile 
with the closest flux value is chosen, [Fe/H] and \logg~previously fixed by the user.
The most probable temperature and its uncertainty are determined by the median and the robust 
standard deviation (1.4826 times the median absolute deviation) of the histogram, 
see e.g. Fig.~\ref{sun_normal} and \ref{sun_normal_HARPS}.

   \subsection{Telluric correction}
   \label{telluric_correction}   
   The resolution and sampling of the coud\'{e} spectra allow a total of 26 to 27 wavelength bins 
   inside the windows of fit, enough to perform the fitting procedure described in Sect~\ref{fitting}.
   In order to optimize \teffa~and its error determination when windows of fits are contaminated 
   and to provide a spectral library clean from telluric features, 
   we corrected the normalized coud\'{e} spectra with the \textit{Molecfit} software package
   \citep{smette_molecfit_2015, kausch_molecfit:_2015}.   
   This software computes the transmission of the Earth's atmosphere at the time of the observations 
   with the radiative transfer code LBLRTM \citep{clough_atmospheric_2005}, 
   taking into account spectroscopic parameters from the HITRAN database 
   \citep{rothman_hitran2012_2013} and an atmospheric profile.
   The atmospheric transmission is fitted to the observed spectrum, 
   and the telluric correction is done dividing the observed 
   spectrum by the atmospheric transmission. 
   We used the average equatorial atmospheric profile, which is \textit{Molecfit}'s 
   default profile. We chose to fit H$_2$O (the main absorber in this wavelength region), 
   O$_2$, and O$_3$. The line shape is fitted by a boxcar profile; as starting value for 
   the boxcar FWHM we used 0.36 times the slit width. 
   The wavelength solution of the atmospheric transmission is adjusted with a first degree polynomial.
   First, we ran \textit{Molecfit} automatically on all spectra, 
   avoiding the center of the H$\alpha$ line from 6560 to 6566~\AA. 
   If the residuals of this first telluric correction were larger than 2\% of the continuum, 
   we adapted the starting value of the water abundance and performed a second fit. 
   This telluric correction allowed us 
   to recover with precision the stellar flux inside the contaminated windows of fits in most cases. 
   An example is shown in Fig~\ref{t_correction} where the corrected 
   and non-corrected spectra of \mbox{HD 2151} are over-plotted.
   
   The telluric corrected and non-corrected normalized coud\'{e} spectra of the sample stars in 
   Table~\ref{objects} can be accessed at an on-line repository\footnote{\url{https://github.com/RGiribaldi/Halpha-FGKstars}},
   or by contacting the first author.

\section{Validation of the normalization method}
\label{2DN}

BPO02 found the 2D-normalization efficient in removing the spectral blaze,
the method is described in detail in their paper.
It is referred as 2D-normalization because it depends on the two spacial dimensions of the CCD detector.
Namely, the normalization curve of the spectral order of
interest is found by interpolating the normalization curves of the adjacent orders in the pixel domain.

\begin{figure}
   \centering
   {\includegraphics[width=.45\textwidth]{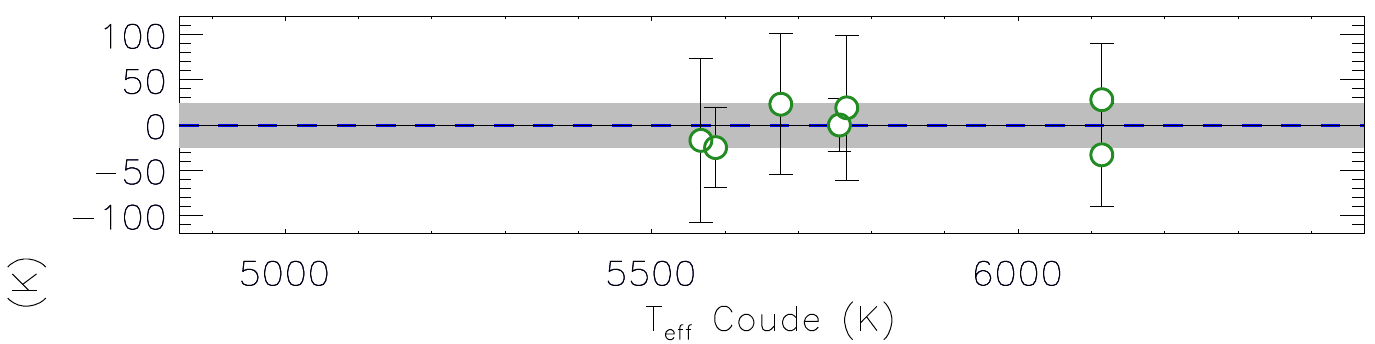}}
   {\includegraphics[width=.45\textwidth]{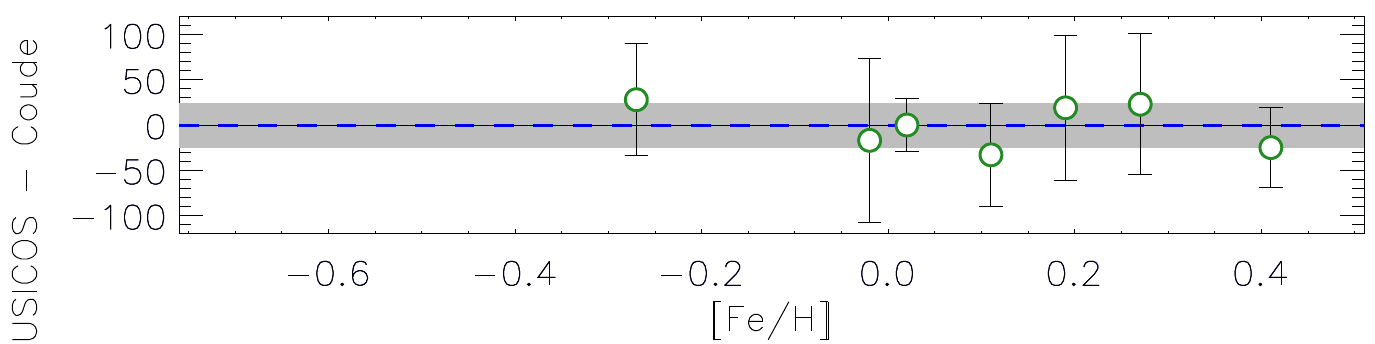}}
   {\includegraphics[width=.45\textwidth]{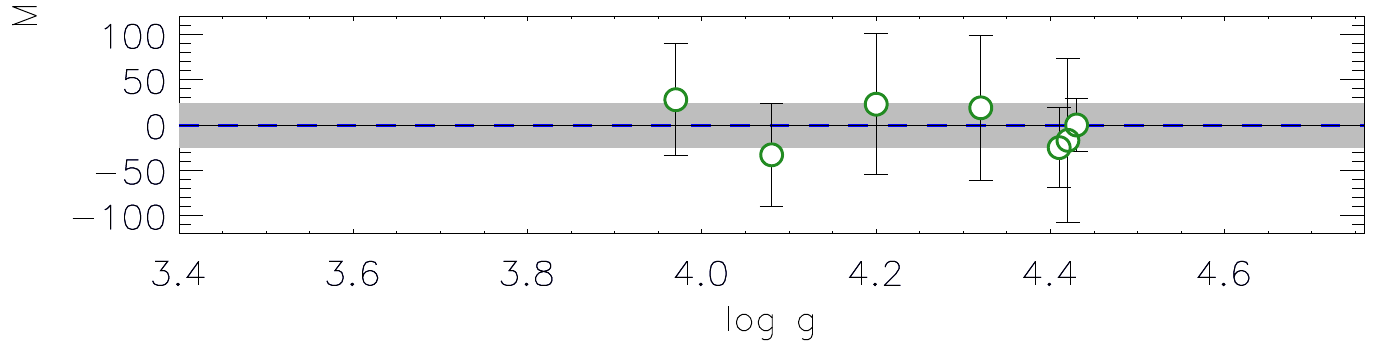}}
   \caption{Temperature diagnostics from MUSICOS with respect to those of coud\'{e} vs.
   atmospheric parameters.
   [Fe/H] and \logg~values from Table~\ref{objects} were used here.
   The \mbox{$-1$ K} offset and its 25 K scatter are represented by the dashed lines and the shades, respectively.}
   \label{coude_MUSICOS}
   \end{figure}

We validate the normalization method described in Sect.~\ref{coude} 
used on coud\'{e} and HARPS spectra,
deriving \teffa~with MUSICOS spectra normalized by the 2D-normalization.
The comparison in Fig.~\ref{coude_MUSICOS} shows that 
\teffa~derived with coud\'{e} and MUSICOS 
are compatible for all stars.
We find no trend with respect to the
atmospheric parameters, a negligible offset of $-1$ K and a 
low scatter of 25 K. 
Solar spectra reflected in the Moon and Ganymede
were also normalized with this method, from which we derive the average value $5745 \pm 16$ K 
(see comparative values in Table~\ref{proxies}, the profile fits are shown in Fig.~\ref{MUSICOS_fittings})
consistent with \teffa's listed in Table~\ref{zero-point}
derived from coud\'{e} and HARPS spectra.

\section{Accuracy of 1D model atmospheres} 
\label{accuracy}
\subsection{The zero-point}
\label{zero-p}

   \begin{figure*}[t]
   \centering
   {\includegraphics[width=0.8\textwidth]{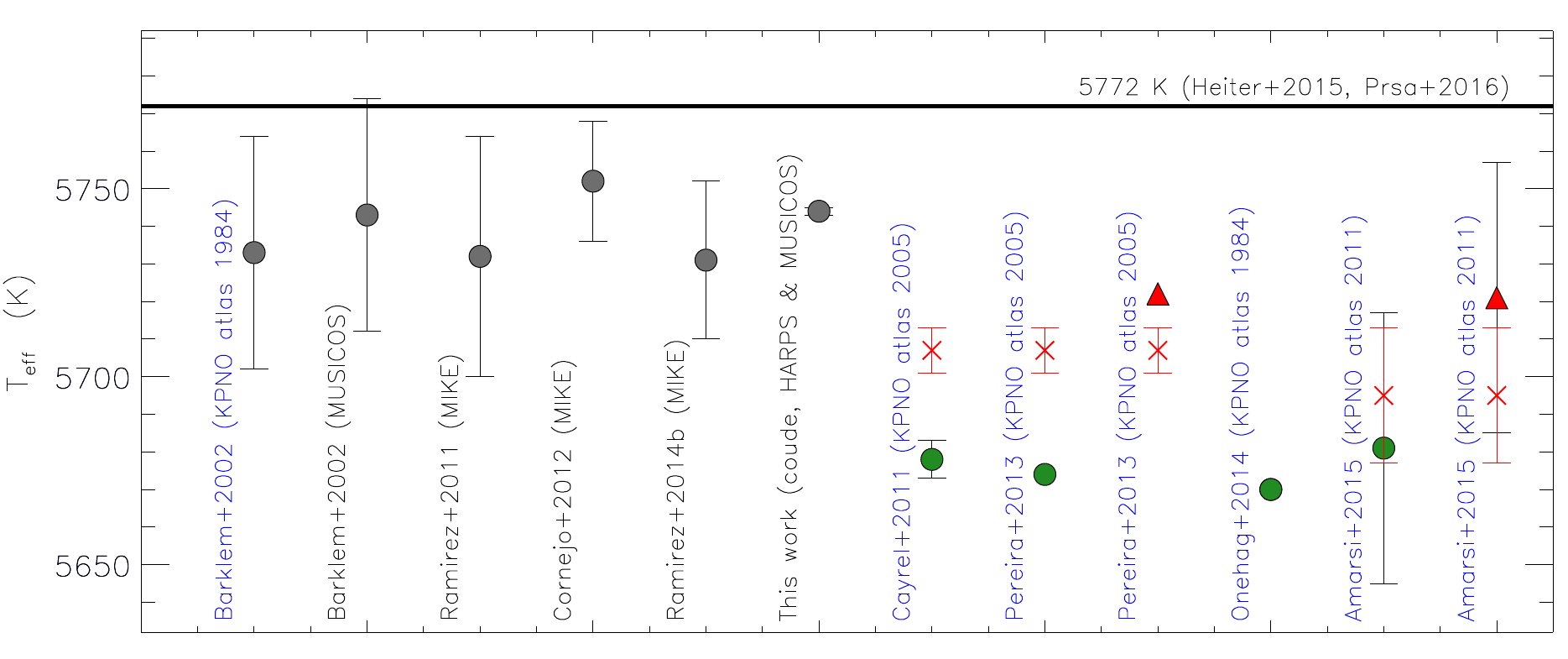}}
   \caption{Graphic representation of solar \teff~values in Table~\ref{zero-point}.
   The horizontal line represents the solar \teff~measured by the Stefan-Boltzmann equation.
   Works that used theoretical models based on 1D atmosphere models are represented by circles, 
   and those that used 3D models by triangles. Gray circles represent works that used the theoretical model of BPO02, and 
   green circles represent a different/enhanced recipe. 
   Works that used KPNO solar atlases are labeled in blue. For them, for comparison purposes, our measurements from 
   corresponding KPNO spectra are included as red crosses in the same line.}
   \label{temperatures_literature}
   \end{figure*}

We used the 6 blaze-free coud\'{e} solar spectra listed in Table~\ref{proxies}
to determine the accuracy of H$\alpha$ profiles from 1D model atmospheres for the Sun. 
The profiles fitted are shown in Fig.~\ref{coude_fitted}, 
we obtain the average value $5744 \pm 7$ K.
Since we find good agreement between the determinations 
from coud\'{e}, MUSICOS, and HARPS spectra 
(Sec.~\ref{2DN} and \ref{Reliability}),
we determine the zero-point of the model by averaging the  
inferred \teff~values from all solar spectra,
resulting an offset of $-28 \pm 1$ K
with respect to the 5772 K \citep{Prsa2016,heiter2015}
measured by the Stefan-Boltzmann equation.

Our zero-point supports the temperature values initially found by 
BPO02 with their \mbox{MUSICOS} spectrum and the KPNO1984 atlas, 
and those found later by \citet{ram2011,cor2012} and \citet{ram2014}
with MIKE spectra. 
On the other hand, it disagrees with any value derived from 
KPNO solar atlases, including our own determinations. 
These values are presented in Table~\ref{zero-point} and Fig.~\ref{temperatures_literature}
along with those derived by other authors using 
enhanced theories from BPO02 on. 

Fig.~\ref{temperatures_literature} shows that none of the models recovered the solar \teff, 
included the most sophisticated ones, i.e. \citet{pereira2013} based on
3D models and \citet{amarsi2018} based on 3D models and NLTE conditions.
The plot also shows that the determinations from KPNO spectra are systematically cooler than those from other spectra,
except for the first one of BPO02. 
Notice that this determination disagrees with that of \citet{onehag2014} although they were 
obtained with the same version of KPNO atlas and 
the same broadening recipe. Which is explained by synthetic profiles computed from different versions of MARCS model atmospheres
that use distinct mixing-length parameters.

It is not satisfactory that such dispersion remains
for the Sun, our reference star from which spectra of supreme
quality are not difficult to obtain.
Thus, in the attempt of identifying the origin of the problem, we fitted KPNO atlases with the theoretical profiles 
of BPO02 (fittings with no further normalization). 
From these fits, we firstly computed the temperature difference that other models of 
H$\alpha$ produce with respect to that of BPO02 for the Sun, 
they are provided in Table~\ref{zero-point}. 
Secondly, we compared these fits with those of coud\'{e}/HARPS/MUSICOS to
analyze the goodness of their normalizations. 
The fits are shown in Fig.~\ref{kurucz_fittings}, they are very precise in the inner profile regions 
thanks to their high temperature sensitivity and to the high spectral quality in S/N and sampling.
However, when the outer regions are scrutinized, evident departures appear, see Fig.~\ref{transition_regions}.
We observed similar departures, after the first iteration in our normalization procedure, 
i.e. the custom normalization by polynomial interpolation (see Fig.~\ref{gross_norm}),
whose causes were explained in Sect.~\ref{norm}.

From KPNO2005 we obtain a 30 K 
cooler value than what we obtain with coud\'{e}/HARPS/MUSICOS spectra. 
This atlas version was normalized by polynomial fitting of the observed spectral fluxes, 
considering also the presence of broad $\mathrm{O_{3}}$ and $[\mathrm{O_{2}}]_{2}$ atmospheric 
features produced by synthetic spectra.  
The differences between the temperature values derived by us and the two authors 
that used profiles from 1D models are
entirely explained by the different physics of the models. 
H$\alpha$ profiles of \citet{cayrel2011} were synthesized by \mbox{ATLAS9}, \mbox{BALMER9} codes \citep{castelli2004} and the 
impact-broadening of \citet{allard2008} that includes more transitions than the self-broadening of BPO02.
The profiles of \citet{pereira2013} were synthesized also with a slight different input physics and 
an updated atmosphere model than that in BPO02.

From KPNO2011 we obtain a similar value to that obtained with KPNO2005, 
meaning that the relative flux of both spectra in the innermost regions of the profile agree.
On the other hand, significant differences are observed in the outer wings, see Fig.~\ref{transition_regions}.
No information is provided about the normalization method of this atlas,
but we suspect that the custom method was applied because 
we observe significant flux disparities 
around the continuum regions [6500.25, 6500.50], [6504.50, 6505.00] and [6625.60, 6625.80], 
see Fig.~\ref{normalization_errors}.
If their flux excess of $\sim\!0.2\%$ was constant through all the wavelength range, 
it would imply a temperature underestimate of at least \mbox{20}.

This analysis show that the systematic low temperatures 
from solar spectra in Table~\ref{zero-point} are associated to 
disparities with the synthetic spectra and/or the continuum, 
which may indicate minute normalization errors.
We show that when a special care is taken in the continuum placement and in fitting the outermost
profile regions, consistent results are obtained. 
These results are further supported by the agreement with all other \teffa~measurements 
from spectra other than KPNO, as Fig.~\ref{temperatures_literature} shows.

The temperature differences listed in the last column of Table~\ref{zero-point}
are computed subtracting the diagnostics by the BPO02 model 
to those by the H$\alpha$ models of the authors listed in the first column,
both obtained from the same solar spectra listed in second column.
Hence, they give the zero-points of the H$\alpha$ models relative to that of BPO02 \mbox{($-28 \pm 1$ K)},
so the two quantities added give the zero-point of the model.
Remarkably, we find that the two models using 3D atmospheric models improve 
the agreement with the actual solar \teff, and
\citet{amarsi2018} that also consider NLTE 
reproduce almost exactly the solar \teff.

\begin{table}
\caption{Third column lists $T_{\mathrm{eff}}$ values derived for the Sun 
with H$\alpha$ profiles from 1D model atmospheres (top table) and 
from 3D model atmospheres (bottom table).
Forth column lists the temperature differences given by different models of H$\alpha$ 
with respect to BPO02-grid based analysis for the same solar spectrum.
Fits of the spectra are shown in the appendix.}             
\label{zero-point}      
\centering  
\small
\begin{tabular}{c c c c}        
\hline\hline                 
Author & spectrum & $T_{\mathrm{eff}}$ (K) & $\Delta$\teff~(K) \\    
\hline
   BPO02 & KPNO1984 & $5733$ & ---\\
   BPO02 & MUSICOS & $5743$ & ---\\
   \citet{ram2011} & MIKE & $5732 \pm 32$ & ---\\
   \citet{cayrel2011} & KPNO2005 & $5678 \pm 5$ & $-29$\\
   \citet{cor2012} & MIKE & $5752 \pm 16$ & ---\\
   \citet{ram2014} & MIKE & $5731 \pm 21$ & ---\\
   \citet{pereira2013} & KPNO2005 & $5674$ & $-33$\\
   \citet{onehag2014} & KPNO1984 & $5670$ & ---\\
   \citet{amarsi2018} & KPNO2011 & $5681 \pm 40$ & $-14$\\
   This work & coud\'{e} & $5744 \pm 7$ & ---\\
   This work & HARPS & $5744 \pm 10$ & ---\\
   This work & MUSICOS & $5745 \pm 16$ & ---\\
   This work & KPNO2005 & $5707 \pm 6$ & ---\\
   This work & KPNO2011 & $5695 \pm 18$ & ---\\
   \hline
   \citet{pereira2013} & KPNO2005 & $5722$ & 15\\
   \citet{amarsi2018} & KPNO2011 & $5721 \pm 40$ &26\\
\hline
\end{tabular}
\end{table}

   \subsection{Accuracy for non solar stars}
   \label{interferomety_section}
\begin{figure}
   \centering
   {\includegraphics[width=.24\textwidth]{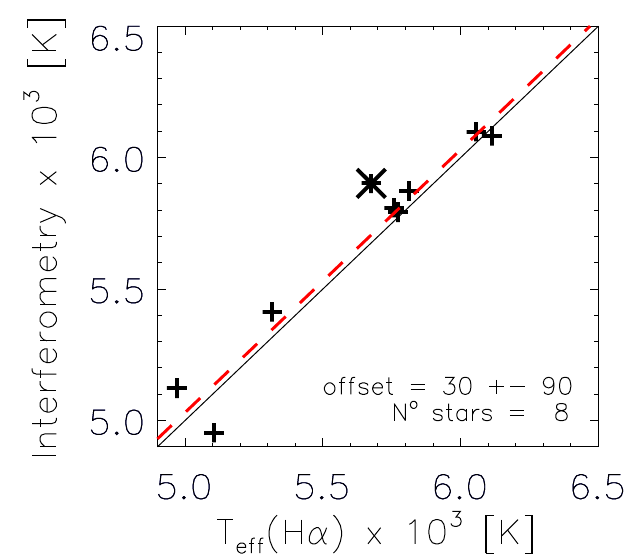}} 
   {\includegraphics[width=.24\textwidth]{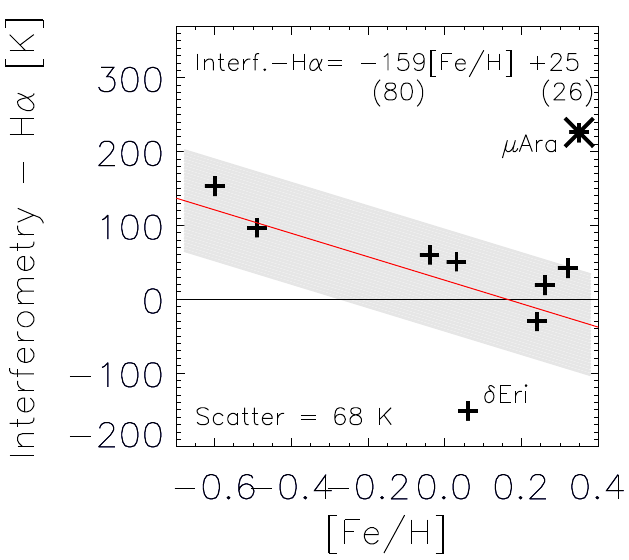}} 
   \caption{\textit{Left panel: }Comparison of $T_{\mathrm{eff}}^{H\alpha}$  with \teff~from interferometry of 
   the Gaia Benchmark stars (Heiter15). The red dashed line represent the offset.
   \textit{Right panel: }Relative temperatures in function of [Fe/H]. The red line and the shade represent 
   the trend and its scatter.
   The corresponding function and the errors of its coefficients (in brackets) are shown in the legends.
   The cross symbol in both plots point \mbox{$\mu$ Ara}'s \mbox{(HD 160691)}
   considered as outlier.}
   \label{interferometry}
   \end{figure}

   Atmospheric parameters of 34  Gaia Benchmark stars with a wide range of temperature and metallicity were published by Heiter15. 
   Their $T_{\mathrm{eff}}$'s were derived by measuring angular diameters with interferometry, 
   that is the least model-dependent technique. 
   We acquired coud\'{e} spectra of 9 Gaia Benchmark stars and 
   \teffa~were derived for them using the [Fe/H] and \logg~values given by the authors.
   The plot in Fig.~\ref{interferometry} shows the comparison of
   \teffa~with \teff~from interferometry.
   We find a constant offset of 30 K between the two scales, that confirms
   the $-28$ K zero-point found with the solar spectra in Sect.~\ref{zero-p}.
   No temperature dependence is found with \logg~but a trend is present with metallicity.
   The right panel of the figure shows that H$\alpha$ underestimates 
   \teff~by \mbox{$\sim\!100$ K} at [Fe/H] = $-0.5$.
   In the plots, the temperature values of \mbox{$\mu$ Ara} \mbox{(HD 160691)} appear highly discrepant
   and were ignored to compute the trend. 
   Its interferometric $T_{\mathrm{eff}}$ is flagged by the authors 
   as not reliable because its angular diameter is not directly measured (see Sect. 3.2 in paper);
   also its mass measurements derived by evolutionary and seismic techniques disagree. On the other hand, 
   we find its \teffa~to be consistent with IRFM and all 
   the spectroscopic values in following sections.
   The other star with a high discrepancy is $\delta$ Eri (HD 23249). 
   It appears also discrepant in the comparison with IRFM in Sect.~\ref{IRFM_section},
   and even our temperatures from coud\'{e} and HARPS disagree.
   However, its values in the plots of Fig.~\ref{interferometry} 
   were not ignored at computing the trends, 
   in order to do an homogeneous comparison with the trends in Fig.~\ref{IRFM}.
  
  Having determined with high precision the offset of \teffa~with respect to \teff~at solar 
  parameters in the previous subsection, 
  the \teffa~accuracy with respect to [Fe/H] over the metallicity range analyzed, is improved from the 
  relation in the plot on right panel of 
  Fig.~\ref{interferometry} to 
  \mbox{$T_{\mathrm{eff}} = T_{\mathrm{eff}}^{H\alpha}$ $-159(\pm80)$[Fe/H] $+ 28(\pm1)$ K} 
  (68 K scatter).
  
\section{Consistency with other \teff~scales}
\label{consistency}

We used 10 catalogs from  literature to determine the consistency of the 
H$\alpha$ profile diagnostics with other techniques.
Among them, Sousa08, Ghezzi10, Tsantaki13, Besnby14, Ramirez14a, Ramirez14b, and Maldonado15 
determine  spectroscopic $T_{\mathrm{eff}}$'s,
while Ramirez13 has $T_{\mathrm{eff}}$'s derived by photometric calibrations from IRFM.

In this section, as well in Sect.~\ref{interferomety_section}, \teffa's were 
derived for comparison purposes using as stellar imput \logg~and [Fe/H] parameters provided by each author,
so that the comparisons are consistent as far as the stellar parameters are concerned.
In the next subsections \teffa~determinations are separately compared with the results
obtained with each method.

   \subsection{IRFM effective temperatures} 
   \label{IRFM_section}
   
   \begin{figure*}
   \centering
   {\includegraphics[width=.24\textwidth]{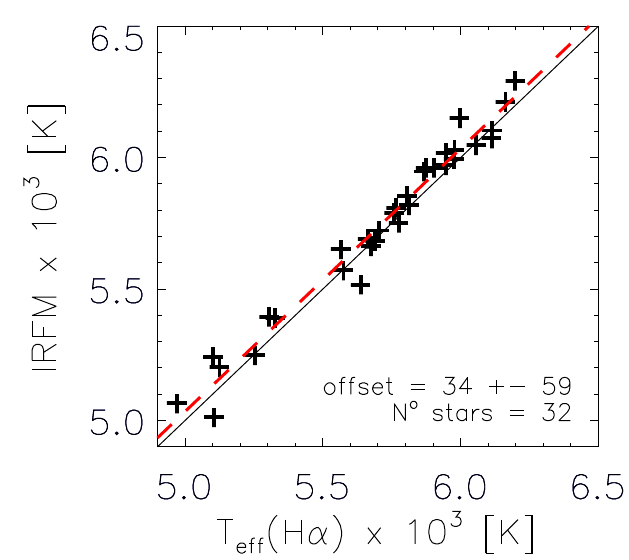}}
   {\includegraphics[width=.24\textwidth]{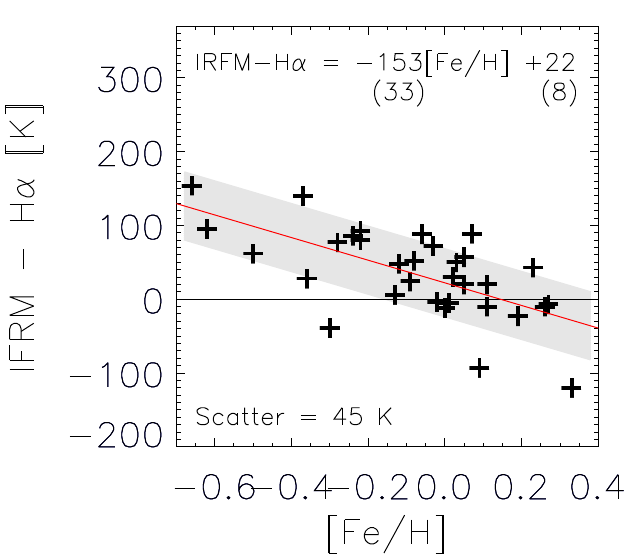}}  
   {\includegraphics[width=.24\textwidth]{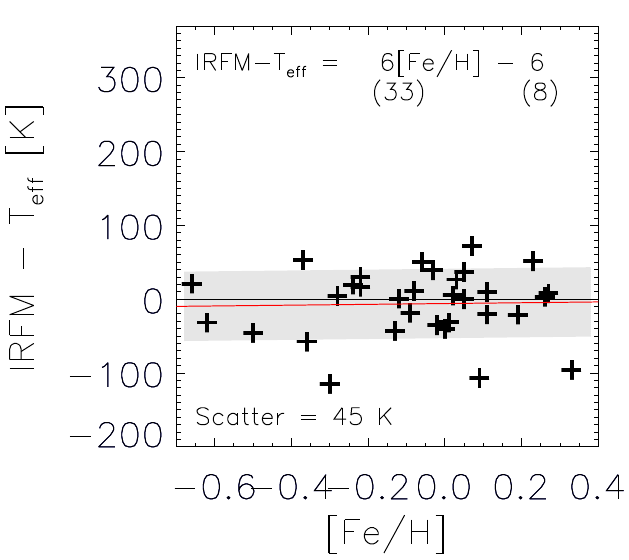}} 
   \caption{\textit{Left and middle panels:} Same as in Fig.~\ref{interferometry} for the $T_{\mathrm{eff}}^{IRFM}$'s of Ramirez13.
   \textit{Right panel:} Relative temperatures in function of [Fe/H] after 
   applying the correction relation given in~\ref{interferomety_section} to \teffa.}
   \label{IRFM}
   \end{figure*}  
   
   The comparison with IRFM is performed with the temperatures of Ramirez13, 
   that were derived by the metallicity-dependent color--$T_{\mathrm{eff}}$ 
   calibrations of \citet{casagrande2010} using the \mbox{Johnson-Cousins}, 2MASS, Tycho2 
   and Str\"{o}mgreen available photometry.
   To obtain these temperatures, represented by $T_{\mathrm{eff}}^{IRFM}$, 
   the authors used an homogeneous set of metallicity derived from Fe lines, 
   where \teff~is not obtained simultaneously with the other parameters 
   but fixed from photometric calibrations. In this way, both techniques are combined iteratively minimizing the
   \teff--[Fe/H] degeneracy.
 
   The plot in Fig.~\ref{IRFM} shows the comparison between $T_{\mathrm{eff}}^{IRFM}$
   and our coud\'{e} \teffa.
   There is a constant offset of +34 K between the two scales 
   with a \mbox{59 K} scatter. 
   Their difference show a trend with metallicity according to the equation displayed in the plot on the middle panel. 
   This trend is practically the same found in the comparison with interferometric measurements, 
   asserting the equivalence of the two scales \citep{casagrande2014}.
   After applying the relation given in Sect.~\ref{interferomety_section} to 
   \teffa, the trend is indeed fully removed, as shown in the right panel of the figure.
   The remaining \mbox{45 K} scatter is close to the average formal errors of 
   $T_{\mathrm{eff}}^{IRFM}$ of the stars compared \mbox{(52 K)},
   which implies that it is dominated by the uncertainties 
   of the color measurements. Therefore the contribution of random errors of \teffa~related 
   to the normalization is negligible, supporting the precision of our method.
   
   \subsection{Spectroscopic effective temperatures} 
   \label{spectroscopic}
   
      \begin{figure*}
   \centering
   {\includegraphics[width=.24\textwidth]{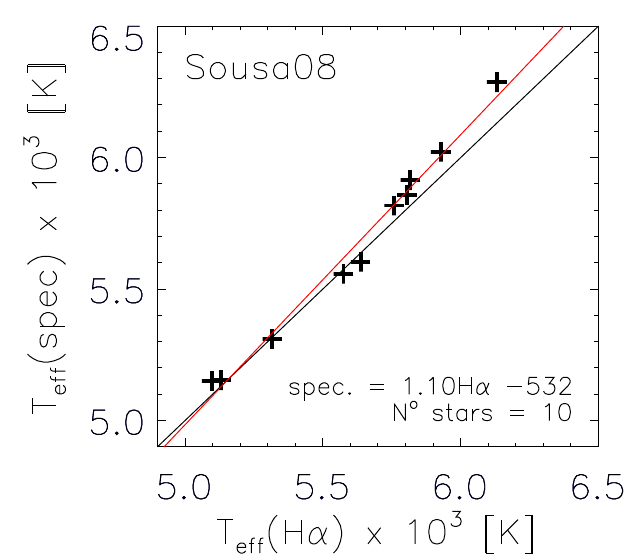}}
   {\includegraphics[width=.24\textwidth]{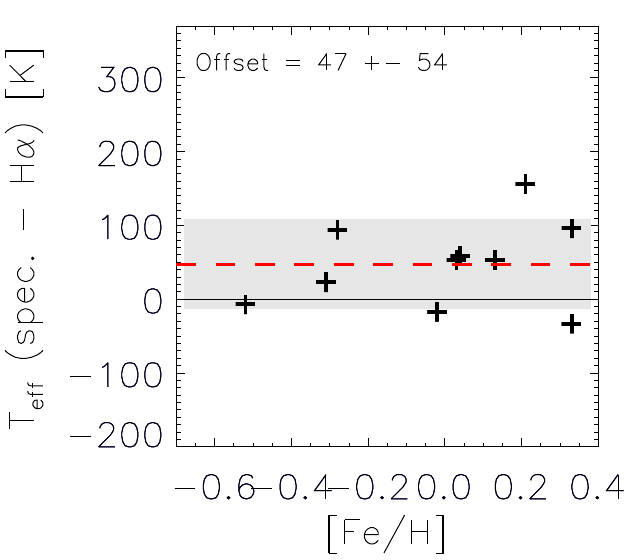}}
   {\includegraphics[width=.24\textwidth]{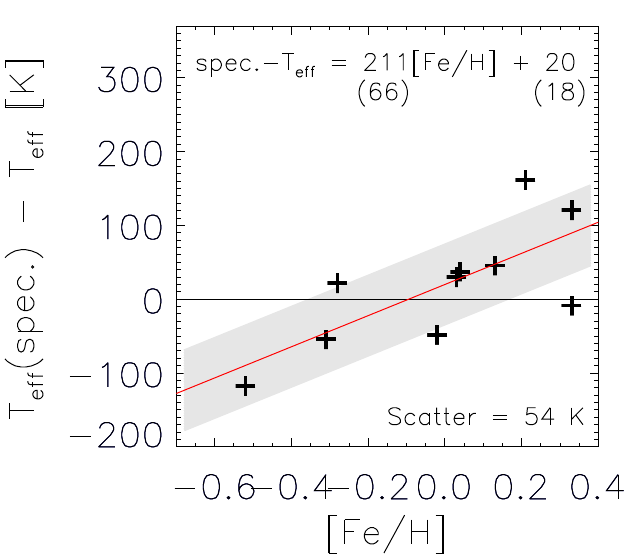}}\\
   {\includegraphics[width=.24\textwidth]{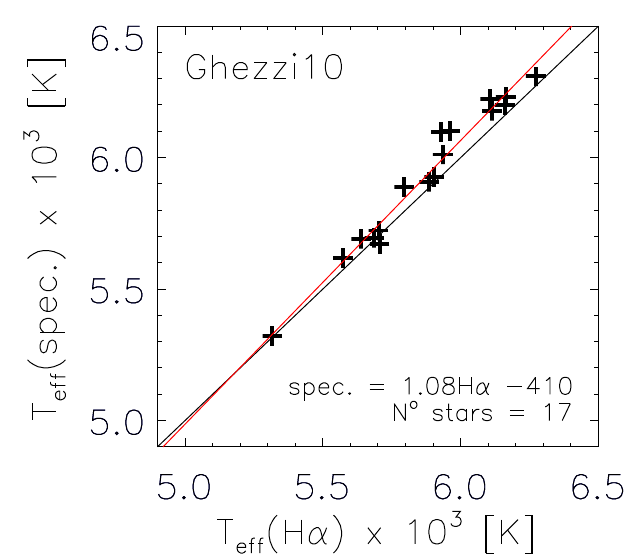}}
   {\includegraphics[width=.24\textwidth]{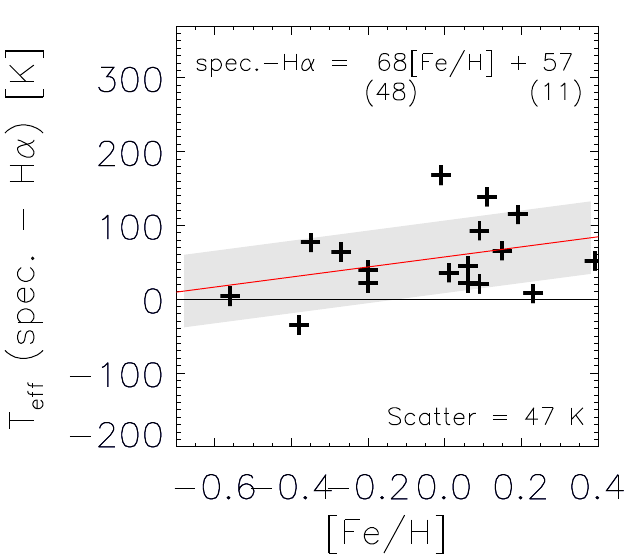}}
   {\includegraphics[width=.24\textwidth]{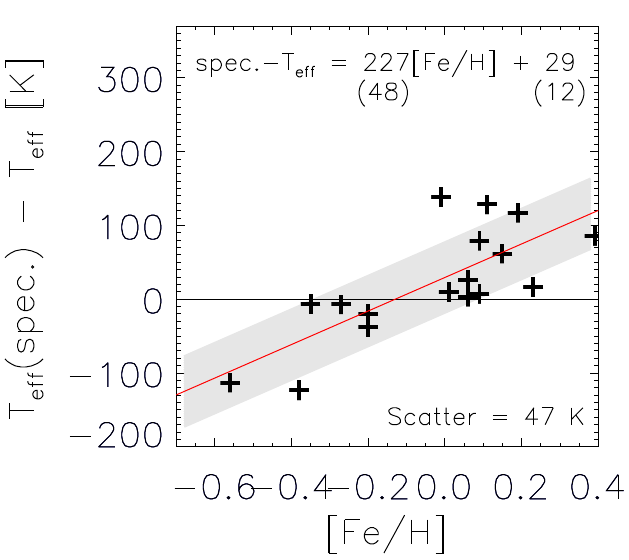}}\\   
   {\includegraphics[width=.24\textwidth]{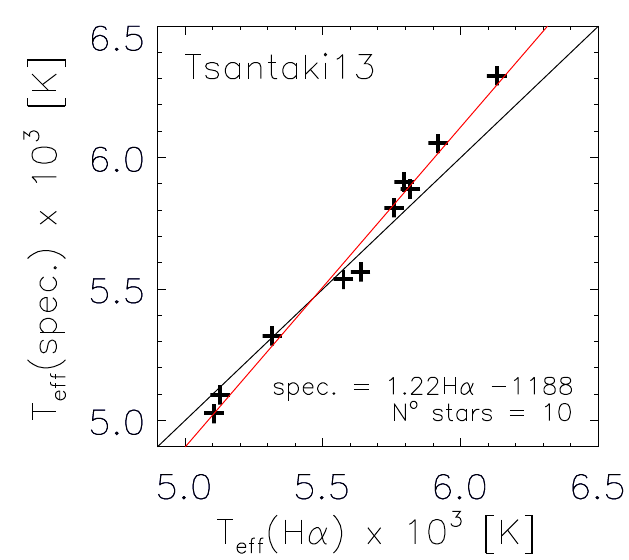}}
   {\includegraphics[width=.24\textwidth]{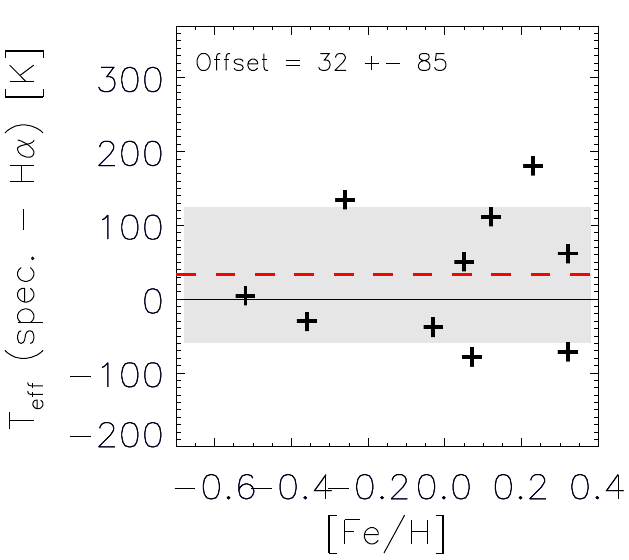}}
   {\includegraphics[width=.24\textwidth]{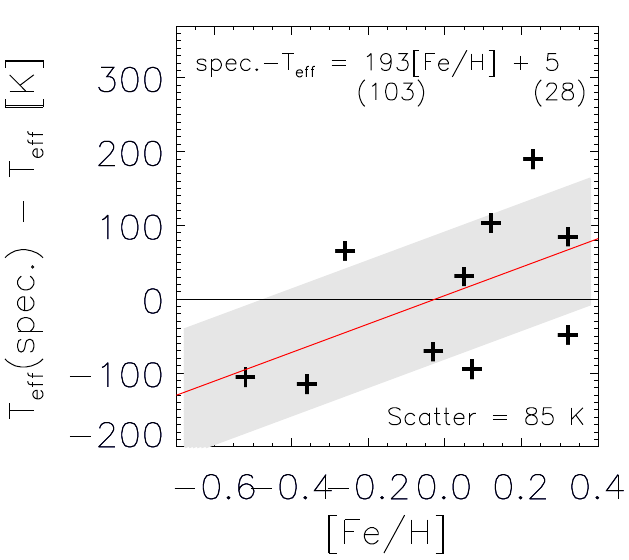}}\\ 
   {\includegraphics[width=.24\textwidth]{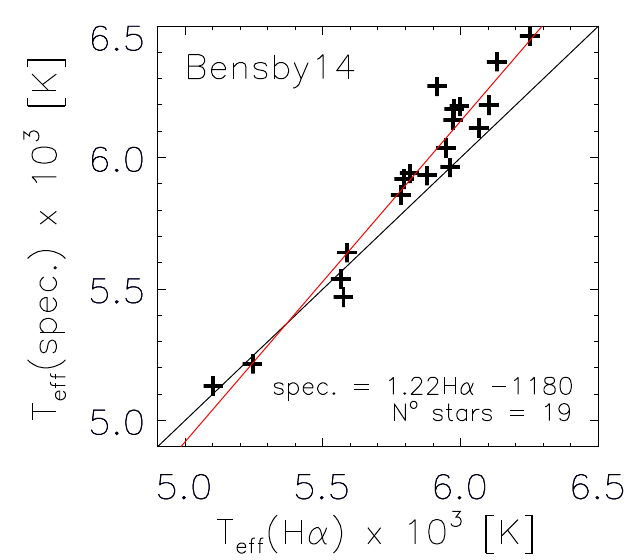}}
   {\includegraphics[width=.24\textwidth]{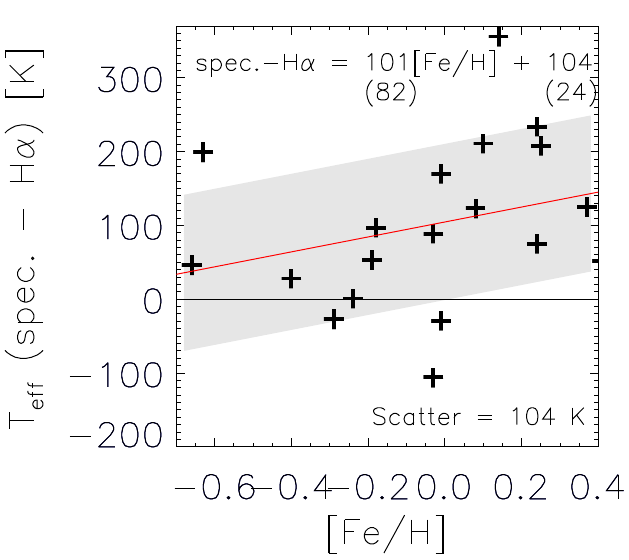}}
   {\includegraphics[width=.24\textwidth]{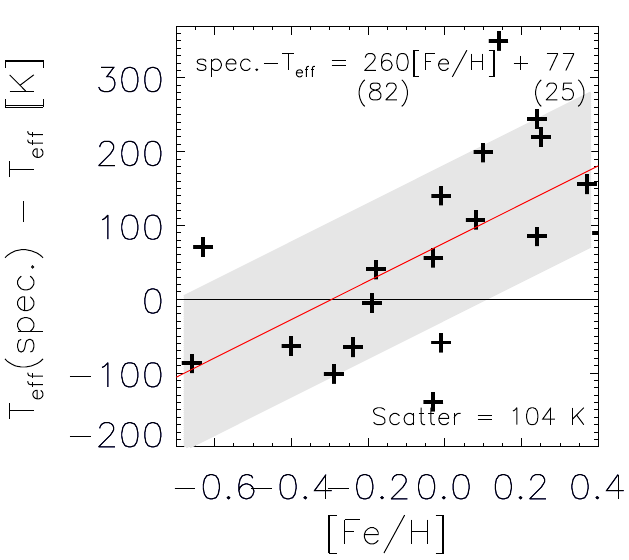}}\\
   {\includegraphics[width=.24\textwidth]{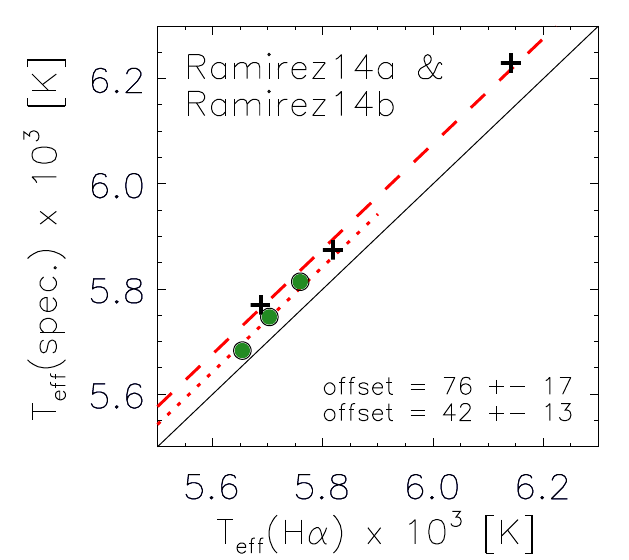}}
   {\includegraphics[width=.24\textwidth]{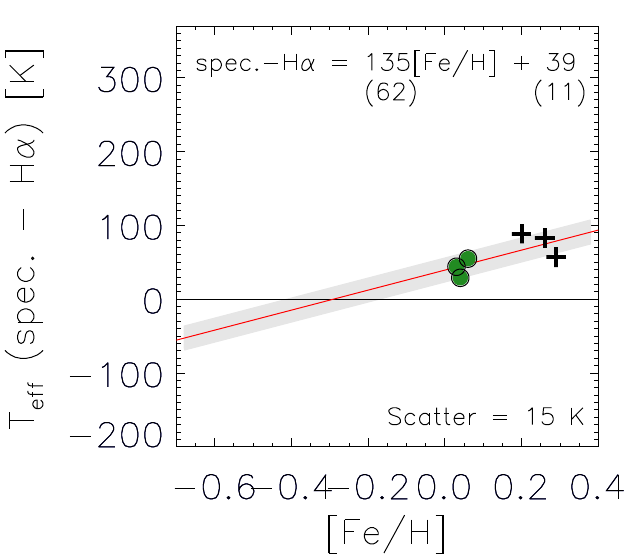}}
   {\includegraphics[width=.24\textwidth]{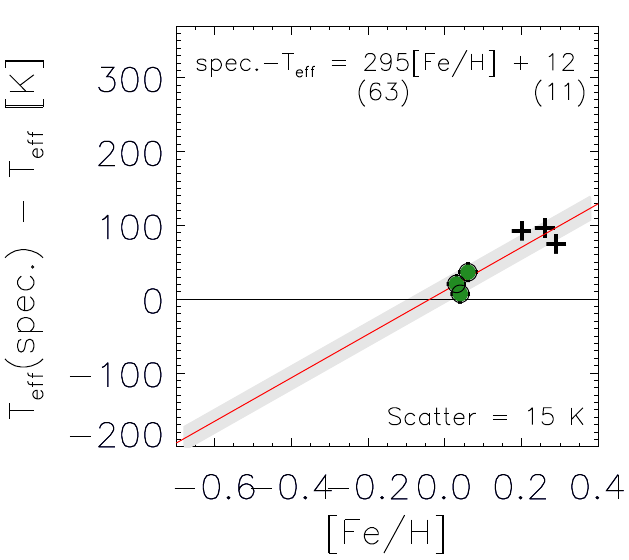}}\\
   {\includegraphics[width=.24\textwidth]{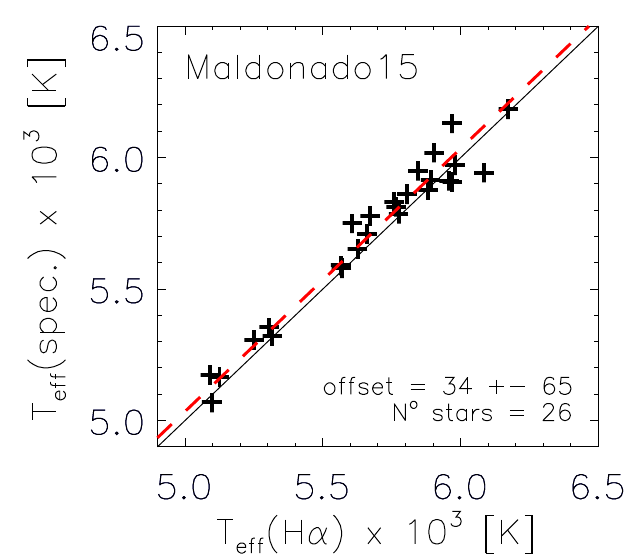}} 
   {\includegraphics[width=.24\textwidth]{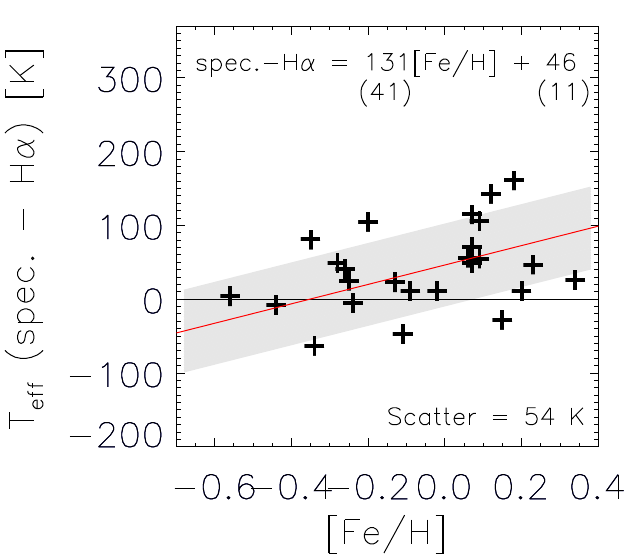}}
   {\includegraphics[width=.24\textwidth]{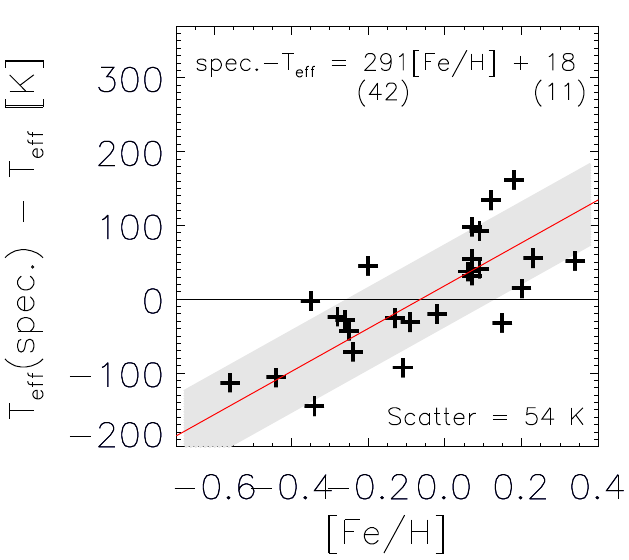}}
   \caption{Same as in Fig.~\ref{IRFM} for spectroscopic $T_{\mathrm{eff}}$'s.
   The authors are indicated in the plots on the left panels. 
   In all plots, the black lines represent the prefect agreement and the red lines the trends. 
   When the trends are found not significant, the offsets are drown with dashed red lines.
   $T_{\mathrm{eff}}$'s from Ramirez14a (plus symbols) and Ramirez14b (green circles), 
   derived with the same method, 
   are compared in the same plots.}
   \label{temperature-scales}
   \end{figure*}   
   
      \begin{figure*}
   \centering
   {\includegraphics[width=.24\textwidth]{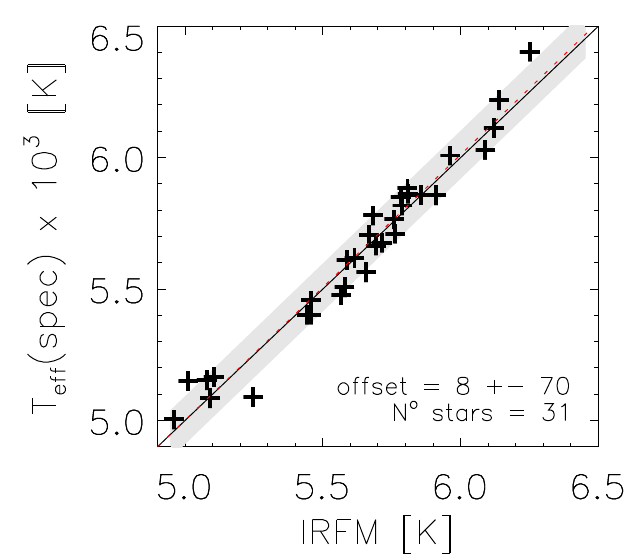}}
   {\includegraphics[width=.24\textwidth]{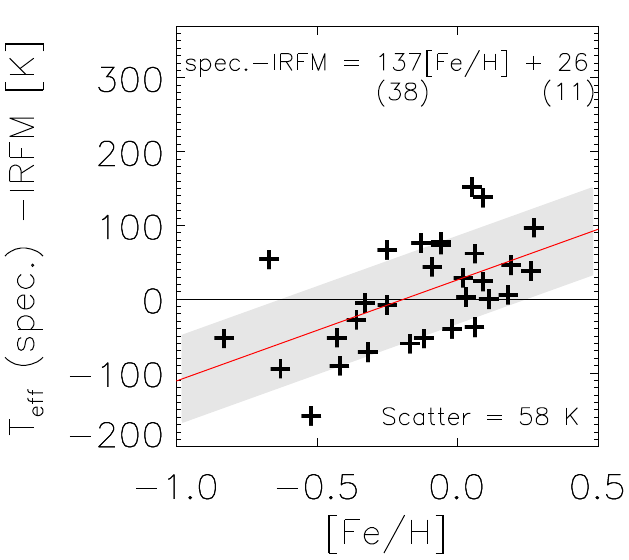}}  
   {\includegraphics[width=.24\textwidth]{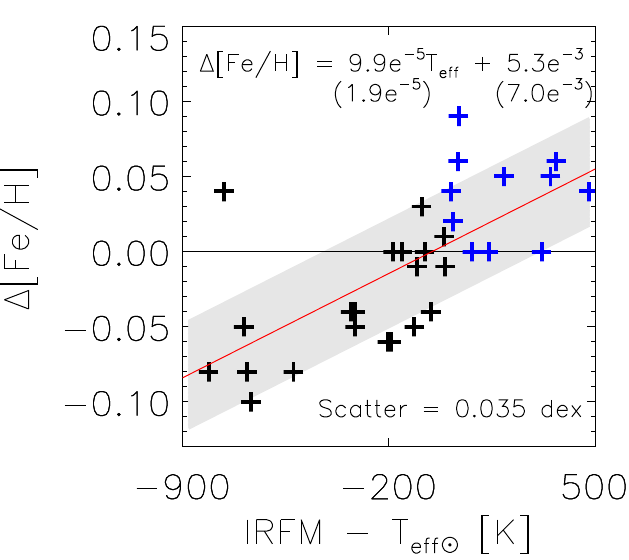}} 
   \caption{\textit{Left and middle panels:} Similar to Fig.~\ref{interferometry} 
   for the $T_{\mathrm{eff}}^{IRFM}$'s of Ramirez13 against 
   the spectroscopic \teff's of Sousa08.
   \textit{Right panel:} $\Delta$[Fe/H] represent the metallicity values of Sousa08 with respect to those of Ramirez13.
   The blue symbols are the stars with over-solar \teff's.}
   \label{IRFM_vs_spec}
   \end{figure*}  
   
   The need of deriving accurate stellar atmospheric parameters got more attention
   with the discovery of exoplanets, because their characterization depends directly on 
   how accurately and precisely the physical parameters of the host stars are known.
   Other studies also require a refined determination of $T_{\mathrm{eff}}$, 
   for instance, finding the nature of the connection between stellar metallicity and planetary presence 
   \citep[e.g.][]{santos2003,fischer2005,sousa2008,gh2010}, 
   the detection of diffusion effects in the stellar atmospheres \citep[e.g.][]{korn2006,korn2007} and
   the search for chemical signatures of planetary formation \citep[e.g.][]{Melendez2009,ram2009}.
   Some of them deal with a large amount of stars, for which automatic spectroscopic 
   procedures have been developed, that provide results with high internal precision. 
   However, as shown by \citet{ryab2015} in their Fig.~1, when results from different spectroscopic procedures
   are compared, significant discrepancies may appear.
   
   In this work we considered for comparison catalogs with small internal errors.
   Among them Ramirez14a and Ramirez14b are the most precise with \mbox{$\sim\!10$ K}. They are followed by 
   Sousa08, Tsantaki13 and Maldonado15 with \mbox{$\sim\!20$} K, a bit further Ghezzi10 and Heiter15 with \mbox{$\sim\!30$} K and 
   Bensby14 \mbox{$\sim\!70$} K.
   The plots in Fig.~\ref{temperature-scales} show the comparison of our temperature 
   determination from coud\'{e} with those derived by the different sources.

   Sousa08, Ghezzi10 and Tsantaki13:
   all derive \teff~assuming LTE and 1D geometry by the Kurucz Atlas 9 \citep{kurucz1993} model atmospheres.
   They used the 2002 version of  MOOG \citep{sneden1973}
   and the ARES code for automatic measurement of equivalent widths \citep{sousa2007}.
   They differ in the line-lists used and in the atomic data adopted.
   Tsantaki13's line-list is an upgrade of the Sousa08's list selected with HARPS, 
   where ``bad'' lines were suppressed to correct \teff~overestimate in cooler stars. 
   Both works computed log \textit{gf} values from an inverted solar analysis using equivalent widths measured 
   in solar spectra.
   Ghezzi10's list is short in comparison with those of Sousa08 and Tsantaki13, it was selected for 
   the FEROS spectrograph \citep{1999Msngr..95....8K} at lower resolution;
   the log \textit{gf} they used are obtained in laboratory.
   The  comparison with these three works show a trend with \teffa:
   the larger \teff, the larger is the discrepancy. 
   For Ghezzi10, the comparison between our measurements and theirs show a 
   positive trend with [Fe/H], while for Sousa08 and Tsantaki13
   no trend with [Fe/H] is found, but offsets of 48 and 33 K, respectively.

   Bensby14: derived \teff~considering NLTE corrections on spectral lines measured manually. 
   The 1D MARCS model atmospheres \citep{asplund1997} were used with 
   an own code of convergence of atmospheric parameters. 
   They used a large line-list and spectra from different instruments of medium and high resolution, 
   with log \textit{gf} values obtained in laboratory.
   The comparison of their \teff~scale against \teffa~is similar to those of Sousa08 and Tsantaki13.
   Indeed, Sousa08 find their scale to be compatible to an offset of $+18$ K respect Bensby14's
   (see Fig. 3 in paper).
   We find a slightly significant positive trend with [Fe/H].
   
   Ramirez14a and Ramirez14b: used a differential method \citep{Mel2006}
   with which the atmospheric parameters of high internal precision are 
   obtained. By means of the ``$\textit{q}^2$'' 
   package\footnote{The Python package ``$\textit{q}^2$'' \url{https://github.com/astroChasqui/q2}}
   both groups of authors used the 2013 version of MOOG and 1D-LTE model atmospheres grids.
   They, measured spectral lines manually and used atomic data from laboratory.
   There are two main differences between the procedures of Ramirez14a and Ramirez14b.
   Firstly, Ramirez14a used the ``odfnew'' version of Kurucz, while Ramirez14b used 
   MARCS atmosphere model \citep{gustafson2008}. 
   However, according to Ramirez14b the use of different models does not 
   affect significantly the parameters diagnostics because of the differential method applied.
   Secondly, the stars analyzed in both works differ in [Fe/H]: Ramirez14b analyzed 
   solar twins, while Ramirez14a more metal-rich stars, i.e. [Fe/H] $\gtrsim 0.2$. 
   Thus, Ramirez14b naturally used the Sun as standard for the solar twins, while in Ramirez14a
   the differential method was applied respect every star of the sample.
   For the Ramirez14b's scale of solar twins we find an offset of
   $+42 \pm 13$ K respect H$\alpha$, which agrees with
   the \mbox{$28 \pm 1$ K} needed to correct H$\alpha$ zero-point.  
   For the Ramirez14a's scale we find an offset of $+72 \pm 17$.
   Considering Ramirez14a and Ramirez14b a unique sample, we find a positive trend with [Fe/H].

   Maldonado15: assumed LTE and 1D geometry by the Kurucz Atlas 9 model atmospheres 
   as Sousa08, Ghezzi10, and Tsantaki13, but they used the line-list from \citet{G_S1999} and 
   spectra from several sources including HARPS. 
   For the convergence of the atmospheric parameters they used 
   TGVIT \citep{takeda2005}. The comparison of their \teff~scale against H$\alpha$ does not show a 
   significant trend, but an offset of +34 K. 
   We found the same offset for IRFM against H$\alpha$ (Sect.~\ref{IRFM_section}), 
   which confirms the agreement\footnote{Maldonado et al. find an offset of 41 K, 
   which is not significant considering the $\sim100$ K error bar relative to their IRFM calculations.} 
   between this \teff~scale and IRFM reported by the authors. 
   On the other hand we find a positive trend with [Fe/H].
   
   The spectroscopic scales analyzed in this section show, in general, 
   agreement with H$\alpha$ up to $\sim\!\!5700$ K and hotter diagnostics for hotter \teff's. 
   The trends with [Fe/H] are opposite to that we observe with interferometry and IRFM.
   After applying the correction relation for metallicity of Sect.~\ref{interferomety_section} to \teffa, 
   the H$\alpha$ scale can be considered in the same frame of the interferometry scale, allowing to
   study the accuracy of the spectroscopic scales. 
   This is shown in the right panels of Fig.~\ref{temperature-scales},
   the common pattern shows that spectroscopic temperatures are underestimated by \mbox{100-200 K}
   at [Fe/H] = $-0.6$ dex and overestimated by 
   \mbox{$\sim\!\!100$ K} at [Fe/H] = $+0.4$ dex.
   The most accurate [Fe/H] range is around the solar value, say between $-0.3$ and $+0.1$ dex.
   
   The relations presented in the plots can be used to empirically correct spectroscopic scales.
   These corrections become important as \teff~depart from solar,
   to derive unbiased [Fe/H] values. 
   An example of the impact of the \teff~scale on [Fe/H] 
   is provided in Fig.~\ref{IRFM_vs_spec}.
   The plots compare the temperature and metallicity scales of Sousa08 and Ramirez13.
   No offset between both temperature scales appears, 
   but their difference plotted against [Fe/H] replicate the trend 
   obtained in the top right panel of Fig.~\ref{temperature-scales}.
   The difference between metallicity scales also shows a trend with \teff,
   associating larger [Fe/H] discrepancies with \teff~farther from solar.
   
\section{Comparison with other H$\alpha$ scales}
\label{otherHa}
\begin{figure*}
   \centering
   {\includegraphics[width=.24\textwidth]{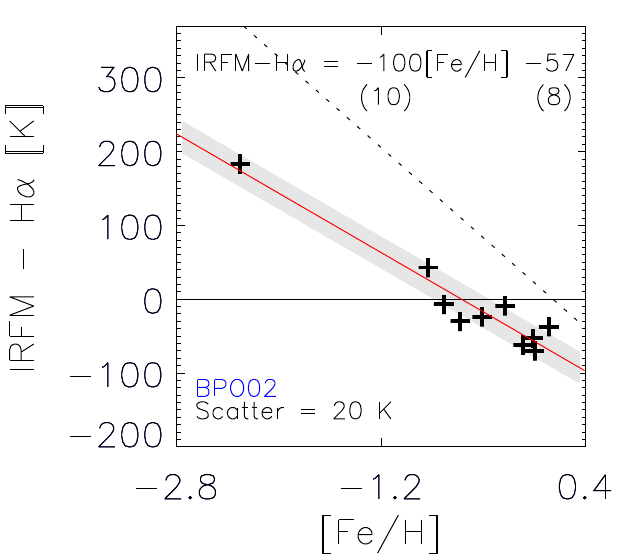}}
   {\includegraphics[width=.24\textwidth]{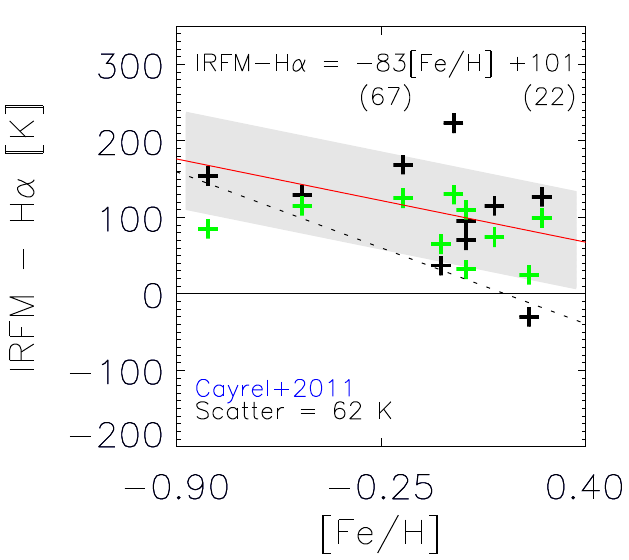}}  
   {\includegraphics[width=.24\textwidth]{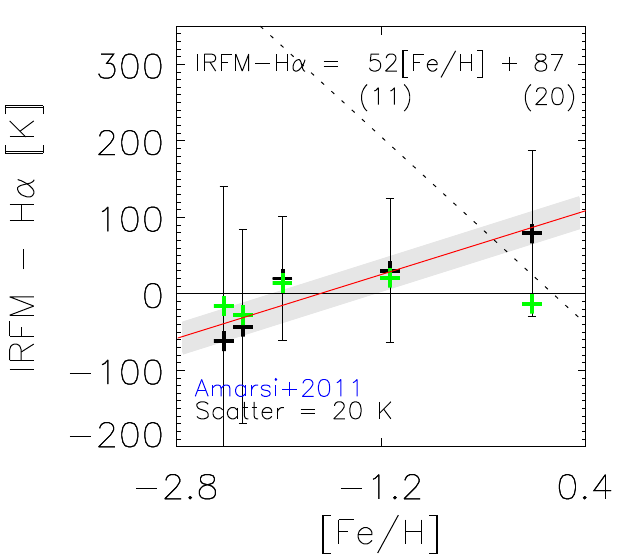}} 
   \caption{From left to right, analogous comparisons to the middle panel in Fig.~\ref{IRFM} for the H$\alpha$ scales of 
   BPO02, \citet{cayrel2011}, and \citet{amarsi2018}.
   In all plots, for a quick comparison, the trend with [Fe/H] of Fig.~\ref{IRFM} is represented by the dotted line.
   Green symbols represent interferometric \teff~replacing \teffI.}
   \label{Ha_scales}
   \end{figure*}  

In Sect.~\ref{zero-p} we determined \teffa~for the Sun and compared it with 
other authors that use the same diagnostic. 
In this section we compare not only zero-points but the temperature scales. 
We again discuss the possible sources of the differences between them
and how the enhanced models improve the results.
The works have stars in common with the IRFM catalog of Ramirez13, 
but they have a few or no stars in common with this work.
Accordingly, the comparisons are preformed with respect to IRFM in function of [Fe/H], 
as done in Sect.~\ref{IRFM_section} with our H$\alpha$ scale.
See the plots in Fig.~\ref{Ha_scales} to follow the discussions below. 

BPO02 scale: 
10 stars are in common with Ramirez13.
An analogous plot to that in Fig.~\ref{IRFM} show a similar 
slope shifted by \mbox{$\sim 70$ K} for the metallicity range we analyze. 
A probable cause for the shift is that the synthetic fitted spectra seem slightly biased towards lower relative fluxes,
see e.g. profiles of \mbox{HR 22879} and \mbox{HR 5914} at 6566-6568~\AA~in Fig. 6 in the paper. 
It may be due to the $\chi^2_{min}$ fitting method without sigma clipping
applied in low S/N spectra, 
e.g. Ramirez14b find systematic high \teffa~values for larger $\chi^2_{min}$.
It however deserves to be mentioned that BPO02's results are consistent with ours. 
Consider that quality of their spectra and their fitting method were not 
conceived to get the precision that this work attempts.

\textit{Cayrel et al. (2011) scale:} 
The comparison against \teffI~in function of [Fe/H] shows a slightly significant trend.
In the comparison against \teff~from interferometry the trend disappear remaining a flat offset of \mbox{$\sim100$ K} 
(check green symbols in the plot), as shown by the authors.
It appears that the H$\alpha$ model of \citet{allard2008} enhances the difference between the model of BPO02 
and interferometry close around the solar [Fe/H]. 
We obtain the same result in Sect.~\ref{zero-p} for the Sun, 
i.e. the zero-point of the model is nearly twice that of BPO02.

\textit{Ramirez14b scale:} Precise \teff~were derived for 88 solar analogs
\citep[i.e. stars that share the same atmospheric parameters with the Sun within 
an arbitrary narrow range of errors, according to the definition in][]{porto2014}
by the photometric calibrations of \citet{casagrande2010} (IFRM) and H$\alpha$ profiles using the model of BPO02, 
in addition to the spectroscopic technique described in Sect.~\ref{spectroscopic}.
In their Fig. 13, these authors compare their determinations from H$\alpha$ with spectroscopy and find, after a
zero-point correction, a small trend, as we did in Sect.~\ref{spectroscopic} 
comparing our H$\alpha$ scale with 
their spectroscopic scale and several others.
No comparison is presented against [Fe/H], which is to be expected, given that the range of their sample 
is very narrow around the solar metallicity ($\pm0.1$ dex). 

\textit{Amarsi et al. (2008) scale:} spectra of six templates were used to test the model.
Two of these stars, the Sun and Procyon, lie within the [Fe/H] range of our sample, 
while the other four with [Fe/H] between $-2.8$ and $-1.2$ dex 
exceed our range.
The comparison with \teffI~in function of [Fe/H] shows a trend, which disappears when interferometric 
\teff~is instead compared.
The change in slope is mainly given by the Procyon's interferometric measurement, that precisely agree with that from H$\alpha$.
The comparison with interferometry shows then a perfect agreement with this H$\alpha$ scale along the [Fe/H] range of analysis.
Further, we also estimated a perfect agreement for the Sun from a differential analysis in Sect.~\ref{zero-p}, 
i.e. the zero-point of the model is practically null.
  
\section{H$\alpha$ profiles from 3D models} 
\label{3D}
The previous sections have shown as the comparison with 
the accurate interferometric and IRFM scales 
is quite robust and free of biases or trends.
The only trend is a dependence on metallicity in both cases.
In order to further investigate such a trend, we have produced and analyzed  
eight H$\alpha$ profiles from 3D models, with which we expect to understand whether
the 1D approximation is indeed the main culprit. 
The eight 3D profiles are from the CIFIST grid of CO5BOLD models \citep{2009MmSAI..80..711L,freytag}, 
calculated using the spectral synthesis code Linfor3D (version 6.2.2) in LTE approximation.
Self-resonance broadening followed BPO02 and Stark broadening followed \citet{griem1967}. 
We chose the atmospheric parameters of four profiles to bracket a solar model \teff~and \logg. 
The four bracketing models were accompanied by four further models of sub-solar metallicity with [Fe/H]$=-0.5$ dex. 
The chemical composition follows \citet{grevesse1998} with the exception of the CNO
elements which were updated following \citet{asplund2005}. For the metal-depleted
models an $\alpha$-enhancement of $+0.2$ dex was assumed.
The variation of the continuum across the H$\alpha$ profile was modeled by
assuming a parabolic dependence of the continuum intensity on
wavelength. 
Doppler shifts stemming from the underlying velocity field were
fully taken into account -- albeit they have a minor effect on the overall
profile shape. The final flux profiles were horizontal and temporal averages
over typically 20 instants in time, the center-to-limb variation of the line
was calculated using three limb-angles.
   
\begin{figure}
   \centering
   {\includegraphics[width=.24\textwidth]{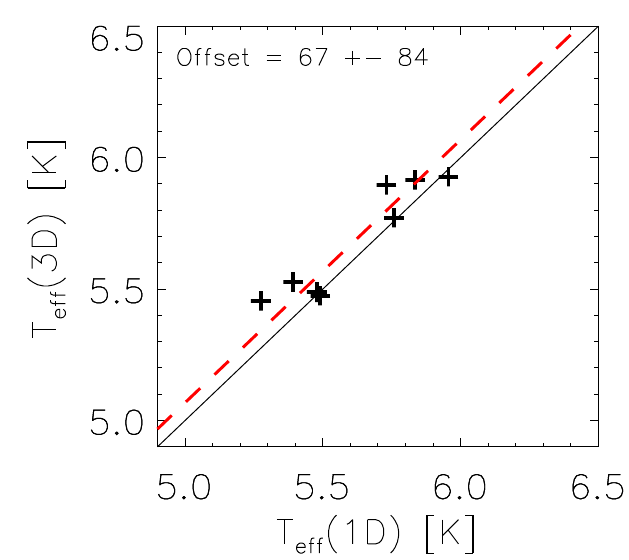}}
   {\includegraphics[width=.24\textwidth]{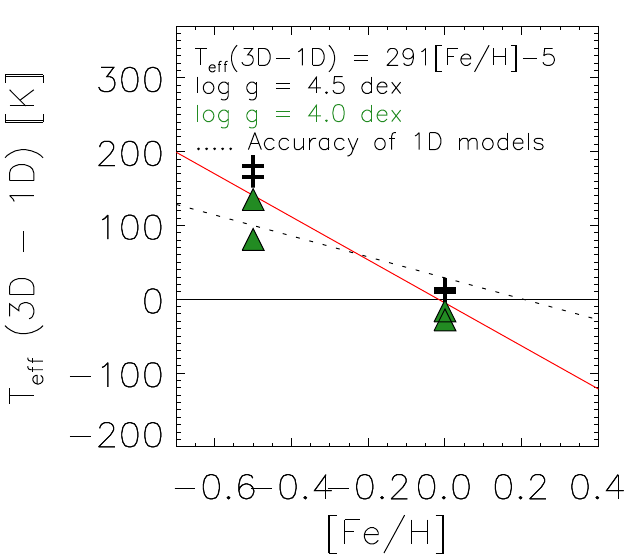}} 
   \caption{Same as in Fig.~\ref{interferometry} for 3D models.
   In the panel on the right different symbols and colors are used for the two \logg~values 
   according to the legends. The accuracy of 1D models 
   \mbox{$T_{\mathrm{eff}} = T_{\mathrm{eff}}^{H\alpha}$ $-159$[Fe/H] + 28} found in 
   Sect.~\ref{accuracy} is represented by the dotted line.}
   \label{3D_1D_}
   \end{figure}

   To estimate the effects of 3D models on \teff, we analyzed the synthetic H$\alpha$ profiles in the 
   same way as the observed ones. The synthetic profiles were resampled with the same pixel size of HARPS
    and 0.1\% of white noise was added. 
   The fits are shown in Fig.~\ref{3D_1D_fits} and the temperatures retrieved from 1D models 
   are compared with their nominal temperatures in Fig.~\ref{3D_1D_} 
   as done in Sect.~\ref{accuracy}. 
   In this figure, in the plot in function of [Fe/H], 
   the improvement given by the 3D models (continuous red line) can be estimated by 
   how similar the trend of Fig.~\ref{interferometry} (dotted line here)
   is reproduced.
   The comparison show that temperatures from 1D models are practically reproduced by 3D models at [Fe/H] = 0 dex,
   but at [Fe/H] = $-0.5$ dex 3D models produce 100-200 hotter temperatures depending on \logg.
   Hence, temperatures from 3D models are significantly closer to those from interferometry at [Fe/H] = $-0.5$ dex, 
   they particularly agree for low \logg~values.

   We therefore conclude that the most likely cause for the trend with metallicity 
   of our H$\alpha$ diagnostics with respect to interferometric and IRFM
   measurements is the use of 1D models. 
   We consider, on the other hand, and excellent approximation the use of 1D models, which are easily available, together
   with the correction for metallicity given in section 6.2.

\section{Suitability of HARPS} 
\label{Reliability}
\begin{table}[!]
\small
\caption{Solar proxies. The list is ordered by date of observation along with the
S/N of the spectra and the effective temperature
derived from H$\alpha$ profiles.}             
\label{proxies}      
\centering                          
\begin{tabular}{c c c c}        
\hline\hline                 
Date & object & S/N & \teffa~(K) \\
\hline \multicolumn{4}{c}{\textbf{coud\'{e}}}\\
\hline
2014/10 & Moon & 300 & $5741 \pm 32$\\
2017/07 & Moon & 400 & $5748 \pm 25$\\
2017/07 & Moon & 400 & $5746 \pm 28$\\
2017/07 & Moon & 400 & $5751 \pm 25$\\
2017/07 & Calisto & 350 & $5740 \pm 28$\\
2017/07 & Ganymede & 350 & $5732 \pm 35$\\
\hline \multicolumn{4}{c}{\textbf{MUSICOS}}\\
\hline
2017/11 & Ganymede & 300 & $5726\pm 28$ \\
2017/11 & Moon & 250 & $5756\pm 45$ \\
2017/11 & Moon & 250 & $5759\pm 41$ \\
2017/11 & Moon & 250 & $5753\pm 30$ \\
\hline \multicolumn{4}{c}{\textbf{HARPS}}\\
\hline
2007/04 & Ganymede & 174 & $5746\pm 52$ \\
2007/04 & Ganymede & 172 & $5750\pm 76$ \\
2007/04 & Ganymede & 171 & $5745\pm 88$ \\
2007/04 & Ganymede & 173 & $5745\pm 68$ \\
2007/04 & Ganymede & 174 & $5735\pm 99$ \\
2007/04 & Ganymede & 391 & $5747\pm 54$ \\
2009/03 & Moon & 532 & $5747 \pm 32$ \\
2010/10 & Moon & 263 & $5741 \pm 65$ \\
2010/10 & Moon & 307 & $5759 \pm 43$ \\
2010/10 & Moon & 288 & $5755 \pm 53$ \\
2010/10 & Moon & 299 & $5743 \pm 74$ \\
2010/10 & Moon & 308 & $5753 \pm 60$ \\
2010/10 & Moon & 304 & $5759 \pm 66$ \\
2010/12 & Moon & 578 & $5746 \pm 29$ \\
2010/12 & Moon & 408 & $5735 \pm 38$ \\
2010/12 & Moon & 412 & $5744 \pm 38$ \\
2010/12 & Moon & 494 & $5732 \pm 36$ \\
2012/06 & Moon & 479 & $5742 \pm 45$ \\
2012/06 & Moon & 478 & $5737 \pm 48$ \\
2012/06 & Moon & 488 & $5746 \pm 38$ \\
2012/06 & Moon & 487 & $5742 \pm 43$ \\
2012/06 & Moon & 485 & $5735 \pm 44$ \\
2012/06 & Moon & 486 & $5735 \pm 42$ \\
2012/06 & Moon & 488 & $5739 \pm 39$ \\
2012/06 & Moon & 490 & $5742 \pm 33$ \\
2012/06 & Moon & 478 & $5734 \pm 33$ \\
2012/06 & Moon & 476 & $5753 \pm 35$ \\
2014/02 & Ganymede & 119 & $5765\pm 98$ \\
2014/02 & Ganymede & 107 & $5750\pm 103$ \\
2014/02 & Ganymede & 117 & $5760\pm 105$ \\
2014/02 & Ganymede & 118 & $5750\pm 107$ \\
2014/02 & Ganymede & 109 & $5767\pm 97$ \\
2014/02 & Ganymede & 117 & $5757\pm 108$ \\
2014/02 & Ganymede & 116 & $5744\pm 139$ \\
2014/02 & Ganymede & 109 & $5757\pm 138$ \\
2014/02 & Ganymede & 109 & $5759\pm 118$ \\
2014/02 & Ganymede & 122 & $5760\pm 98$ \\
2015/07 & Ceres & 89 & $5754\pm 134$ \\
2015/07 & Ceres & 87 & $5748\pm 140$ \\
2015/07 & Ceres & 88 & $5751\pm 111$ \\
2015/07 & Ceres & 89 & $5745\pm 145$ \\
2015/07 & Ceres & 91 & $5755\pm 137$ \\
2015/07 & Ceres & 103 & $5753\pm 143$ \\
2015/07 & Ceres & 87 & $5751\pm 126$ \\
2015/07 & Ceres & 100 & $5753\pm 110$ \\
2015/07 & Ceres & 115 & $5754\pm 116$ \\
2015/07 & Ceres & 128 & $5746\pm 92$ \\
\hline                               
\end{tabular}
\end{table}

\begin{figure}
   \centering
   {\includegraphics{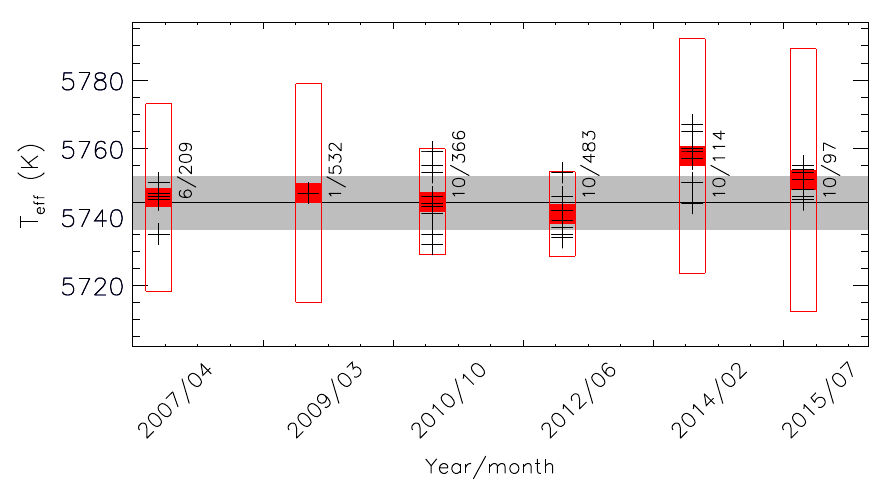}}
   {\includegraphics{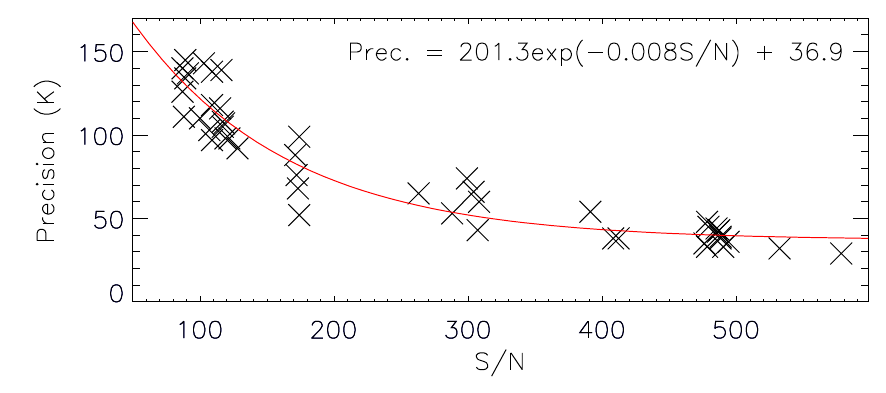}}
   \caption{\textit{Top panel:} Temperatures of the HARPS solar proxies in Table~\ref{proxies} 
   plotted versus date.
   Daily values are represented by plus symbols and 
   weighted means and errors for each month are drown in red. 
   {The weighted mean and error of all the measurements are represented by the continuous 
   line and the shade on 5744 $\pm 10$ K. 
   Next to the bars, the number of spectra analyzed and the mean S/N are noted.
   \textit{Bottom panel:} The errors of individual measurements in the top panel are plotted versus 
   S/N. The exponential curve given by the equation in the plot is the best fit to the points.}}
   \label{temporal}
   \end{figure}

Having shown the suitability of the method with the coud\'{e} spectra, 
we apply it to HARPS \citep{Mayor2003}. 
HARPS has been chosen because, in order to achieve high radial velocity precision,
the instrument has a very stable field and pupil injection. 
It is also thermally stable and in vacuum. 
In addition, HARPS archive contains a lot of observations of solar type stars, 
including a rich set of solar spectra taken by observing solar system bodies for many years.
All these characteristics make of HARPS the ideal instrument to investigate 
the precision of the H$\alpha$ method we have developed.
The fact that the solar siblings observations have been repeated  
for several years, allows us to also investigate the stability of this instrument in time, 
and to determine to which extent the HARPS H$\alpha$ profile has remained constant in time.
The test is performed with all solar spectra 
set out in Table~\ref{proxies},
for which \teffa's were derived. 
The plot in the top panel of Fig.~\ref{temporal} visually summarizes the results displayed in the table.
For each date, \teffa~values are represented by plus symbols. Their 
weighted mean and corresponding spread values are drawn with bars.
Next to them, the number of spectra used and their average S/N ratio are noted
to show the precision reached when measurements from several spectra are combined.
The weighted mean and spread of all measurements 
are represented by the horizontal line and the shade at \mbox{$5744 \pm 10$ K}. 
Evidently, there is no trend with time and the scatter is very low, which
confirms the blaze stability of HARPS.
This value is in perfect agreement with that of coud\'{e} (see values in Table~\ref{zero-point}),
which implies that not only the blaze is stable but it is also 
fully removed through the flat-field procedure.

In the bottom panel of Fig.~\ref{temporal} we plot 
the precision obtained from individual spectra in function of S/N.
It is observed that $\sim\!\!\!40$ K can be obtained from spectra of  
\mbox{S/N = 400-500}.

\begin{figure}
   \centering
   {\includegraphics[width=.45\textwidth]{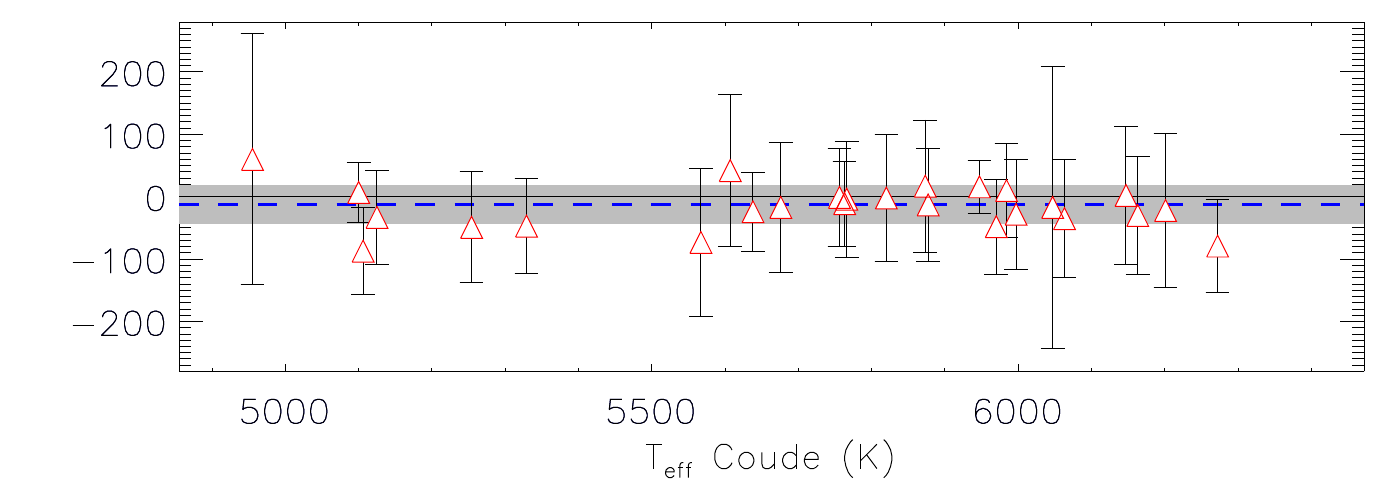}}
   {\includegraphics[width=.45\textwidth]{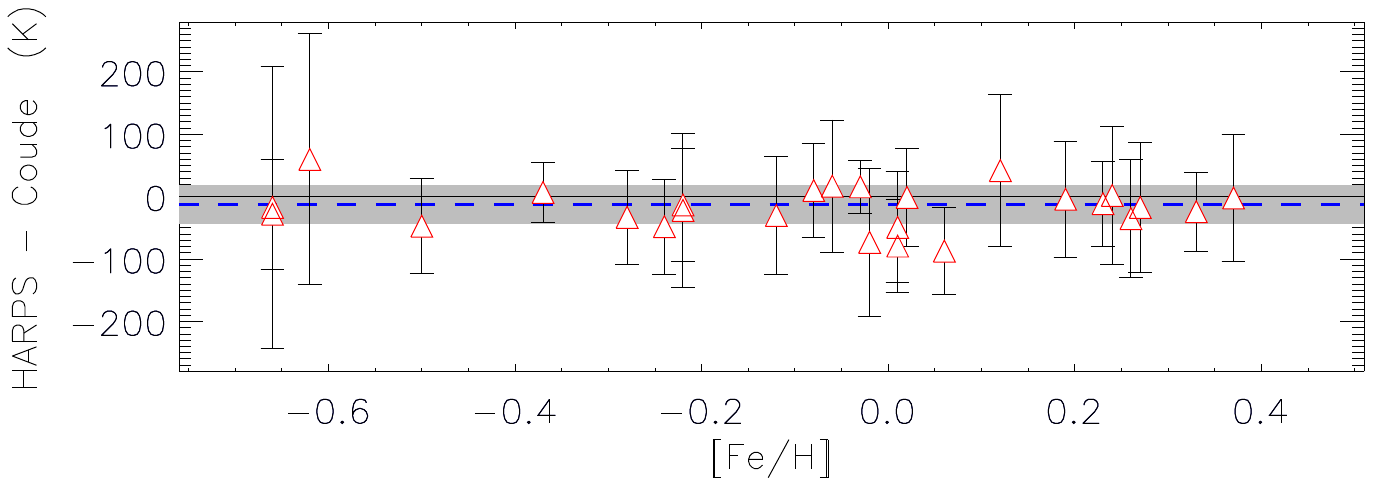}}
   {\includegraphics[width=.45\textwidth]{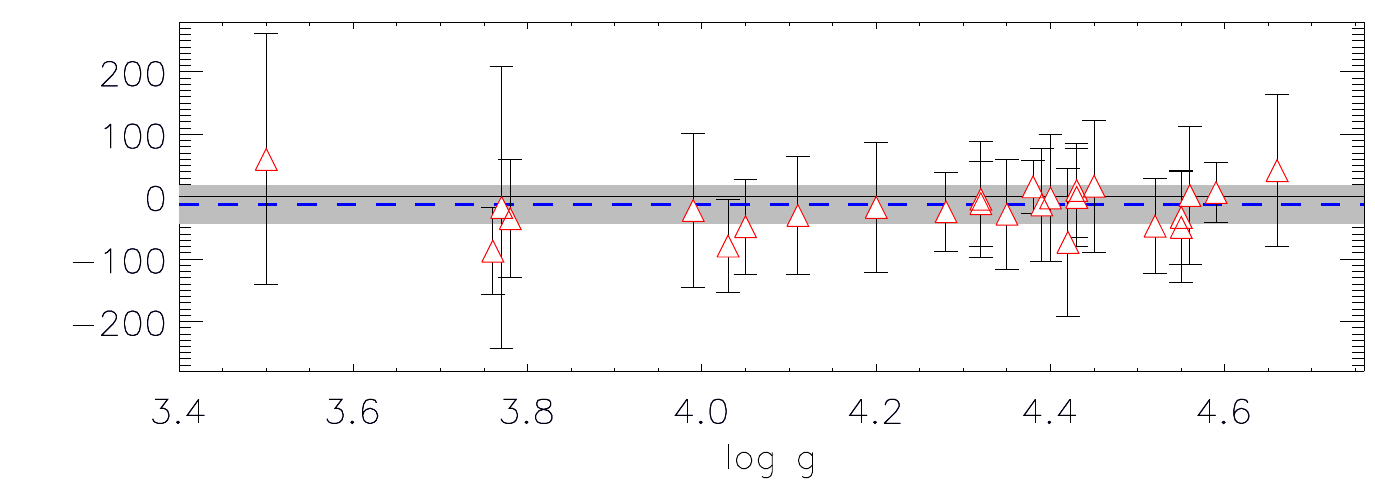}}
   \caption{Temperature diagnostics from HARPS respect those of coud\'{e} vs.
   atmospheric parameters.
   [Fe/H] and \logg~values from Table~\ref{objects} were used here.
   The \mbox{$-13$ K} offset and its 34 K scatter are represented by the dashed lines and the shades, respectively.}
   \label{coude_harps}
   \end{figure}

Finally, we compare the temperatures derived from HARPS 
with those derived from coud\'{e} spectra for the other stars in common.
The comparison is shown in Fig.~\ref{coude_harps} against the three main stellar parameters. 
It shows an excellent agreement with a negligible offset between the two samples of \mbox{$-13 \pm 34$ K} with no trends.
The temperatures of all stars agree within 1$\sigma$ errors,
with the exception of two ($\delta$ Eri and HD 184985)
that agree within 2$\sigma$.

\section{Summary and conclusions}
\label{resume}
With the aim of better understanding and minimizing the errors that affect H$\alpha$
measurements of effective temperature, we have developed a new method to analyze the spectra and
tested it extensively. The results are quite consistent, and they allow us also to test the accuracy 
of the temperature diagnostics with 
H$\alpha$ profiles from 1D model atmospheres in LTE conditions \citep{BPO2002}. 

The core of this work is the special effort adopted in recovering realistic H$\alpha$ profiles 
free from instrumental signatures. Namely, the blaze function of the echelle spectrographs 
and those induced by random errors of normalization.
We eliminated the blaze by using the single-order coud\'{e} instrument at 
do Pico dos Dias Observatory. With it, spectra of 44 F, G, and K stars, including the Sun, 
with a wide parameter range
\mbox{\teff--[Fe/H]--\logg} (see Fig.~\ref{sample_space}) were acquired.
We minimized the errors of normalization of H$\alpha$ profiles, 
by integrating normalization  and fit  into an iterative procedure, with which we derive precise \teffa's.
This procedure, additionally uses synthetic spectra of telluric features of PWV 
to optimize the continuum location. PWV features may be very small and nearly omnipresent 
around H$\alpha$, so they can be easily confused with spectral noise
and shift the continuum to lower flux values.

The accuracy of H$\alpha$ lines from 1D model atmospheres  
is found to follow the relation \mbox{$T_{\mathrm{eff}} = T_{\mathrm{eff}}^{H\alpha}$ $-159$[Fe/H] + 28}
within the metallicity range $-0.7$ to $+0.45$ dex.
It was determined at solar parameters by 
\teffa's from 57 coud\'{e}/HARPS/MUSICOS solar spectra (Table~\ref{proxies})
compared with the reference solar \teff~= 5772 K \citep{Prsa2016,heiter2015},
and at non solar parameters comparing \teffa's of 10 \textit{Gaia Benchmark Stars} 
\citep{heiter2015} with their \teff's from interferometric measurements.

The  consistency of our results with effective temperature scales from IRFM and excitation and 
ionization equilibrium of Fe lines  
was also investigated.
The comparison with IRFM using the photometric calibrations of \citet{casagrande2010} 
show exactly the same trend as the interferometric one of \citet{heiter2015} 
(compare Fig.~\ref{IRFM} with Fig.~\ref{interferometry}),
asserting the equivalence of the two scales. 
As far spectroscopic measurements, the  results vary slightly with the authors, but in general they show agreement 
with H$\alpha$ up to 5700 K.
A trend with metallicity is present and is opposite to that observed with interferometry and IRFM.
Implying that the spectroscopic scale, in general, 
underestimates/overestimates \teff~by \mbox{100 K} at [Fe/H] = $-0.6/$+0.4 dex 
with respect to interferometry and IRFM (see Fig.~\ref{temperature-scales}).

In order to investigate the observed trend with metallicity 
when comparing our measurements with the
interferometric and IRFM ones, we tested  3D model atmospheres. 
H$\alpha$ profiles from 3D models 
produce quite similar diagnostics to 
1D models at solar parameters (we obtain a \mbox{$-15$ K} zero point),
while at the metal-poor range [Fe/H] = $-0.5$ dex, 
they almost fully correct 1D models underestimates (see Fig.~\ref{3D_1D_}).
This therefore indicates that the
trend with metallicity is largely due to the use of 1D models.
The correction we provide by the equation above, however, brings
the three scales H$\alpha$(1D + LTE), interferometry and IRFM on the same base.

We further find that the systematic ``cool'' 
solar temperature determinations from H$\alpha$ models in the literature
are associated to normalization errors of the different versions of Kitt Peak National Observatory solar atlases.
We quantified the impact of the errors in \teff~and find that models 
enhanced by 3D atmosphere geometry and NLTE conditions do improve 
the accuracy of 1D + LTE models, leading to practically null differences with the
solar \teff~derived by Stefan-Boltzmann equation 5772 K.

We tested the suitability of HARPS for the temperature determination with H$\alpha$ profiles.
The tests were performed analyzing spectra of 26 stars in common with the coud\'{e} sample and
47 solar spectra from the period 2007-2015, The solar spectra  show consistent results, to better than $\pm$ 10 K,  
demonstrating the stability of the HARPS blaze and the goodness of the de-blazing process.
The very small ($-13$ K) offset resulting from  the comparison of the stars in common with the coud\'{e} sample,
confirms that the normalization-fitting integrated method minimizes random normalization errors.
Hence, when this method is applied, the internal errors of the H$\alpha$ profiles 
fitting are entirely due to the spectral noise.

Finally, in Table~\ref{final_teff} we list \teffa~as measured 
(by combining all measurements from coud\'{e}, HARPS, and MUSICOS spectra)
and our best \teff~estimate obtained applying the correction for 
metallicity.
The [Fe/H] and log \textit{g} values used for deriving $T_{\mathrm{eff}}^{H\alpha}$
follow the hierarchy Heiter15, Ramirez13, Ramirez14b, 
Ramirez14a, Maldonado15, Ghezzi10, Sousa08, Tsantaki13, Bensby14.

\begin{table}
\tiny
\centering
\caption{\small $T_{\mathrm{eff}}$ of the sample stars.
Column 4 lists the [Fe/H] values used to derive $T_{\mathrm{eff}}^{H\alpha}$ and their sources are 
shown in last column in the same way as in Table~\ref{objects}:
(1) \citet{sousa2008}, (2) \citet{gh2010},
(3) \citet{tsa2013}, (4) \citet{ram2013}, (5) \citet{bensby2014}, (6) \citet{ram_2014},
(7) \citet{ram2014}, (8) \citet{maldo2015}, (9) \citet{heiter2015}.
Column 5 lists the weighted mean of the temperatures derived with coud\'{e}, HARPS, and MUSICOS 
spectra.
Column 6 lists $T_{\mathrm{eff}}$ corrected from the H$\alpha$ diagnostics following the 
relation \mbox{$T_{\mathrm{eff}} = T_{\mathrm{eff}}^{H\alpha} -159$[Fe/H] + 28}. 
The errors presented are internal and are associated to the dispersion of the fit. 
These are the best estimates.
}
\label{final_teff}
\begin{tabular}{l c c c c c c}
\hline\hline \\
Name & HD & HIP & [Fe/H] &$T_{\mathrm{eff}}^{H\alpha}$ (K) & best $T_{\mathrm{eff}}$ (K) & ctlg\\
\hline \\
$\zeta$ Tuc & 1581	& 1599 & $-0.22$ & $5866 $  & $5930 \pm 17$  & 4\\ 
$\beta$ Hyi & 2151	& 2021 & $-0.04$ & $5813 $ & $5848 \pm 20$  & 9\\
& 3823	& 3170 & $-0.34$ & $5947$ & $6030 \pm 18$  & 8\\
$\tau$ Cet & 10700	& 8102 & $-0.49$ &  $5311 $ & $5417 \pm 22$  & 9\\
$\epsilon$ For & 18907	& 14086 & $-0.60$ &  $4984 $ & $5108 \pm 48$  & 9\\
$\alpha$ For & 20010	& 14879 & $-0.30$ &  $6112 $ & $6188 \pm 23$  & 4\\
$\kappa$ Cet & 20630	& 15457 & $~~~0.00$ &  $5675 $ & $5704 \pm 22$  & 4\\
$10$ Tau & 22484	& 16852 & $-0.09$ &  $5947 $ & $5990 \pm 25$  & 4\\
$\delta$ Eri & 23249	& 17378 & $+0.06$ &  $5090 $ & $5110 \pm 12$  & 9\\
40 Eri & 26965	& 19849 & $-0.28$ &  $5109 $ & $5182 \pm 33$  & 4\\
& 100623	& 56452 & $-0.37$ &  $5101 $ & $5188 \pm 17$  & 4\\
$\beta$ Vir & 102870	& 57757 & $+0.24$ &  $6096 $ & $6087\pm 18$  & 9\\
& 114174	& 64150 & $+0.05$ &  $5703 $ & $5724 \pm 32$  & 4\\
$59$ Vir & 115383	& 64792 & $+0.11$ &  $5975 $ & $5987 \pm 23$  & 4\\
$61$ Vir & 115617	& 64924 & $-0.02$ &  $5557 $ & $5589 \pm 18$  & 4\\
$\eta$ Boo & 121370	& 67927 & $+0.32$ &  $6042 $ & $6020 \pm 25$  & 9\\
& 126053	& 70319 & $-0.36$ &  $5663$ & $5749 \pm 58$  & 4\\
$\alpha$ Cen A & 128620	& 71683 & $+0.26$ &  $5765 $ & $5753 \pm 12$  & 9\\
$\psi$ Ser & 140538	& 77052 & $+0.12$ &  $5653 $ & $5663 \pm 21$  & 8\\
& 144585	& 78955 & $+0.29$ &  $5816 $ & $5799 \pm 27$  & 6\\
18 Sco & 146233	& 79672 & $+0.06$ &  $5760 $ & $5780 \pm 20$  & 9\\
& 147513	& 80337 & $+0.03$ &  $5805 $ & $5829 \pm 24$  & 4\\
$\zeta$ TrA & 147584	& 80686 & $-0.08$ &  $6012 $ & $6054 \pm 17$  & 4\\
12 Oph & 149661	& 81300 & $+0.01$ &  $5209 $ & $5236 \pm 34$  & 4\\
& 150177	& 81580 & $-0.66$ &  $6056 $ & $6189 \pm 60$  & 5\\
& 154417	& 83601 & $-0.03$ &  $5950 $ & $5984 \pm 12$  & 4\\
$\mu$ Ara & 160691	& 86796 & $+0.35$ &  $5690 $ & $5664 \pm 13$  & 9\\
70 Oph & 165341	& 88601 & $+0.07$ &  $5305 $ & $5323 \pm 33$  & 4\\
$\iota$ Pav & 165499	& 89042 & $-0.13$ &  $5891 $ & $5941 \pm 32$  & 8\\
& 172051	& 91438 & $-0.24$ &  $5565 $ & $5632 \pm 71$  & 4\\
& 179949	& 94645 & $+0.2$ &  $6134 $ & $6131 \pm 32$  & 6\\
31 Aql & 182572	& 95447 & $+0.41$ &  $5581 $ & $5545 \pm 14$  & 5\\
& 184985	& 96536 & $+0.01$ &  $6255 $ & $6282 \pm 21$  & 2\\
$\delta$ Pav & 190248	& 99240 & $+0.33$ &  $5633$ & $5610 \pm 14$  & 4\\
15 Sge& 190406	& 98819 & $+0.05$ &  $5904 $ & $5925 \pm 16$  & 4\\
$\phi^2$ Pav & 196378	& 101983 & $-0.44$ &  $5979$ & $6078 \pm 28$  & 8\\
$\gamma$ Pav & 203608	& 105858 & $-0.66$ &  $5991$ & $6124 \pm 31$  & 4\\
& 206860	& 107350 & $-0.06$ &  $5878 $ & $5916 \pm 27$  & 4\\
$\xi$ Peg & 215648	& 112447 & $-0.27$ &  $6125$ & $6197 \pm 21$  & 2\\
49 Peg & 216385	& 112935 & $-0.22$ &  $6193 $ & $6257 \pm 35$  & 4\\
51 Peg & 217014	& 113357 & $+0.19$ &  $5785 $ & $5784 \pm 15$  & 4\\
$\iota$ Psc & 222368	& 116771 & $-0.12$ &  $6150$ & $6198 \pm 25$  & 4\\
\hline \\
\end{tabular}
\end{table}

\begin{acknowledgements}
R.E.G. acknowledges a ESO PhD studentship. 
R.E.G. and M.L.U.M. acknowledge CAPES studentships.
G.F.P.M. acknowledges grant 474972/2009-7 from CNPq/Brazil. 
D.L.O. acknowledges the support from FAPESP (2016/20667-8).
S.U. Acknowledges the support of the Funda\c{c}\~ao para a Ci\^{e}ncia e Tecnologia (FCT) through national funds 
and of the FEDER through COMPETE2020 by these grants UID/FIS/04434/2013 \& POCI-01-01-145-FEDER-007672 and PTDC/FIS-AST/1526/2014
\& POCI-01-0145-FEDER-016886.
H.G.L. acknowledges financial support by the Sonderforschungsbereich SFB\,881
``The Milky Way System'' (subprojects A4) of the German Research Foundation (DFG).
We thank the staff of the OPD/LNA for considerable support in the observing runs needed to complete this project. 
Use was made of the Simbad database, operated at the CDS, Strasbourg, France, 
and of NASA Astrophysics Data System Bibliographic Services.
\end{acknowledgements}

\bibliographystyle{aa} 
\bibliography{riano} 

\onecolumn
\begin{appendix} 
\section{H$\alpha$ profile fits}

\begin{figure*}[h]
   \centering
    \includegraphics[width=14cm]{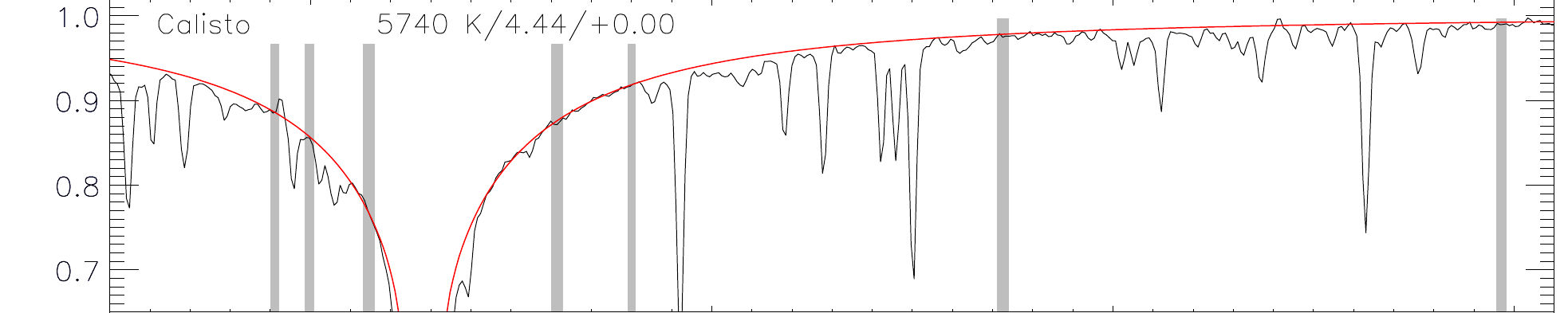}
    \includegraphics[width=3.5cm]{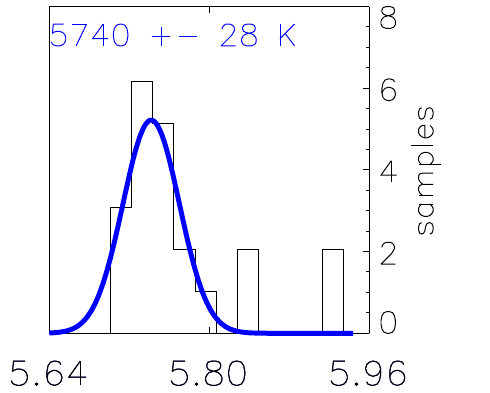}\\
    \includegraphics[width=14cm]{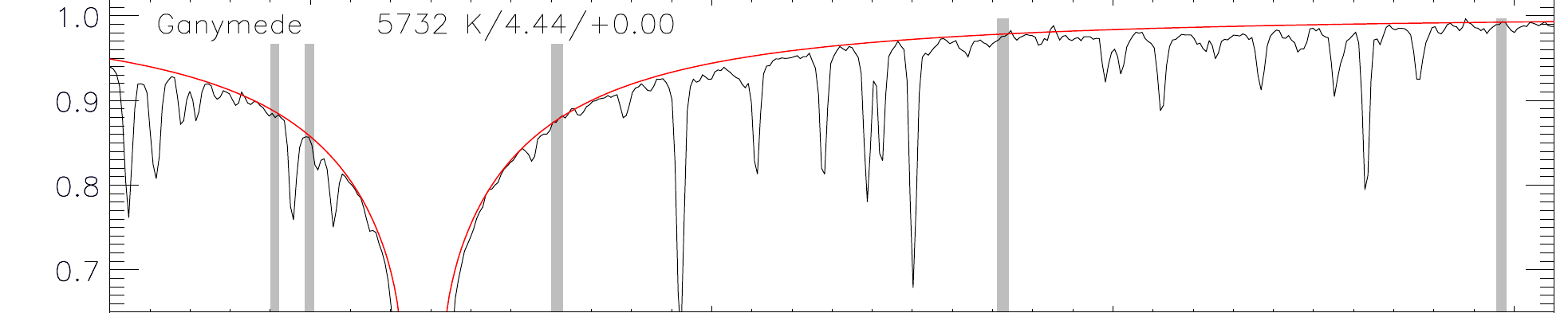}
    \includegraphics[width=3.5cm]{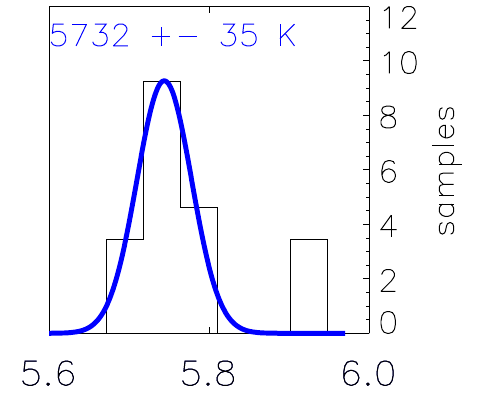}\\
    \includegraphics[width=14cm]{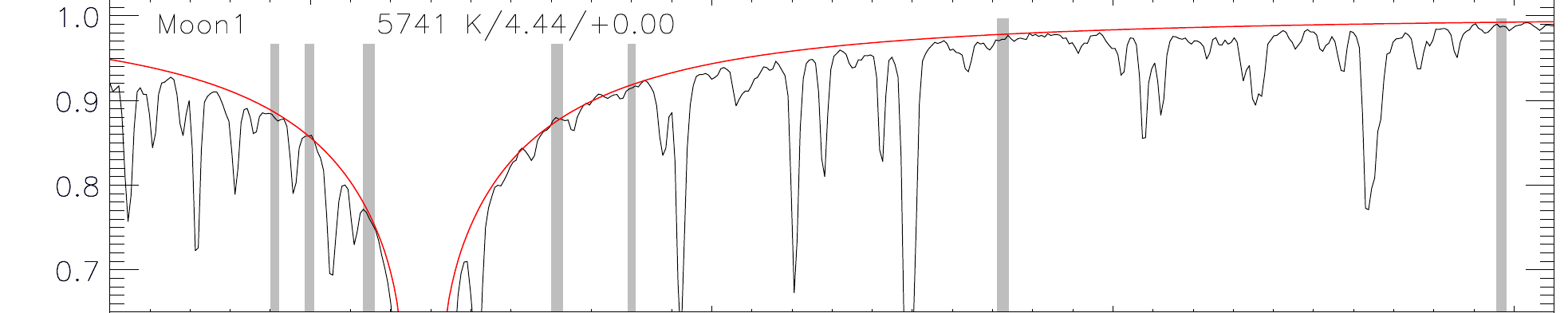}
    \includegraphics[width=3.5cm]{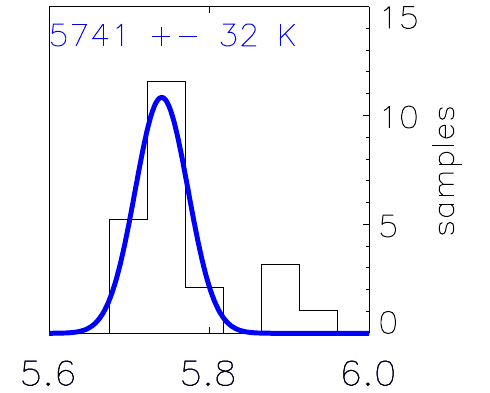}\\
    \includegraphics[width=14cm]{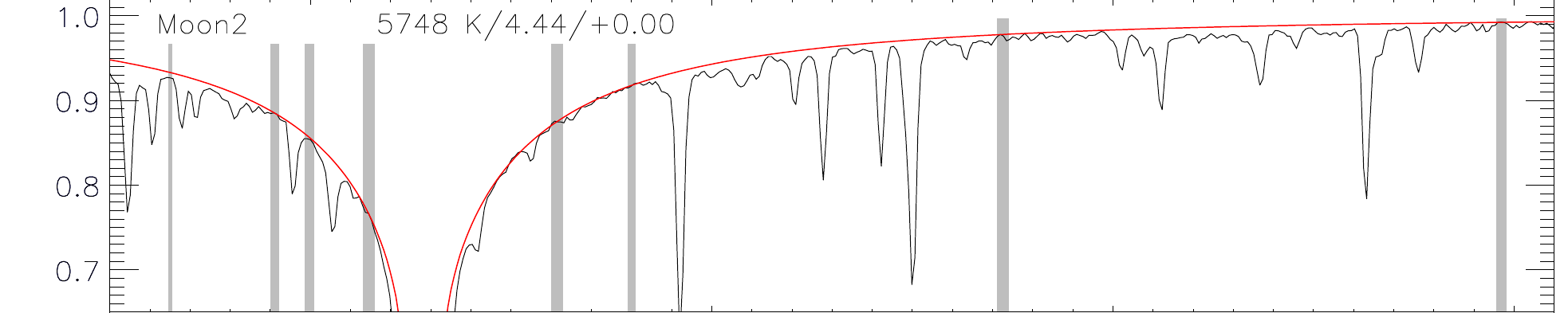}
    \includegraphics[width=3.5cm]{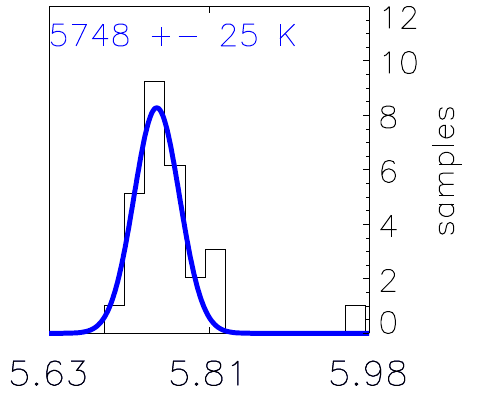}\\
    \includegraphics[width=14cm]{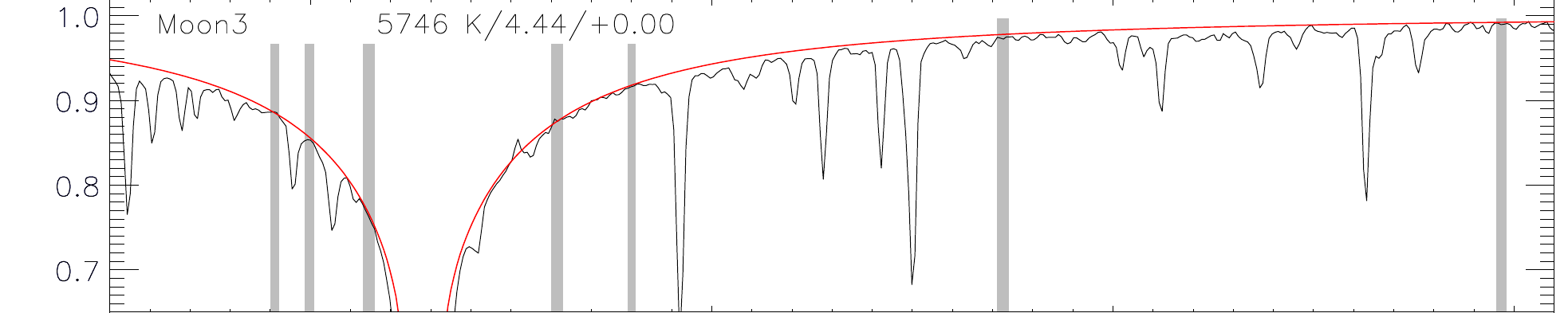}
    \includegraphics[width=3.5cm]{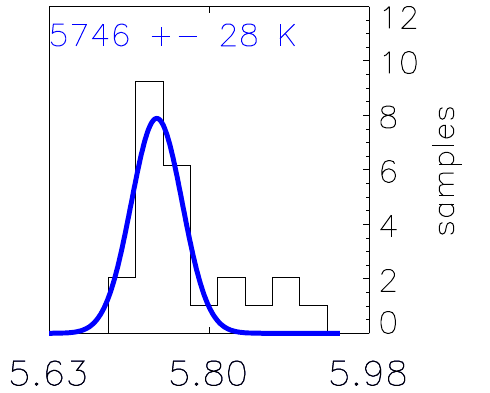}\\
    \includegraphics[width=14cm]{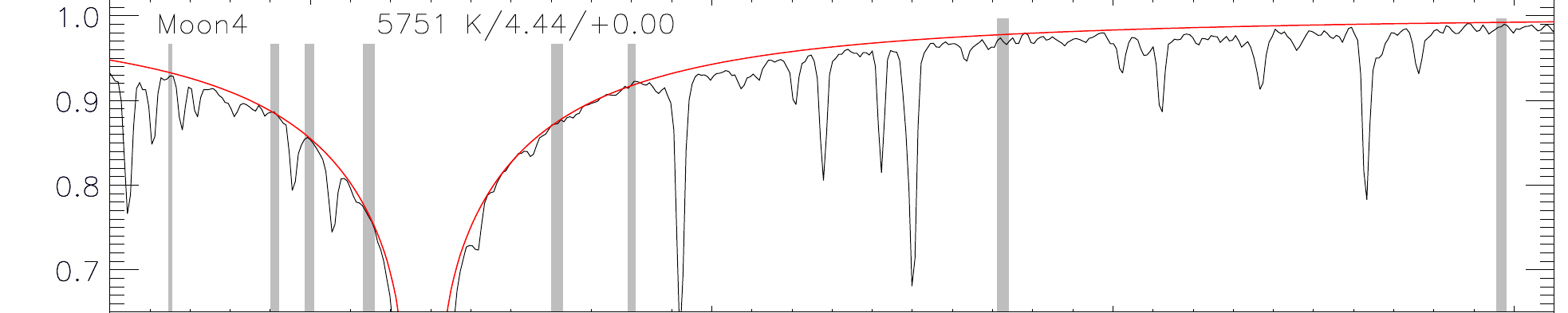}
    \includegraphics[width=3.5cm]{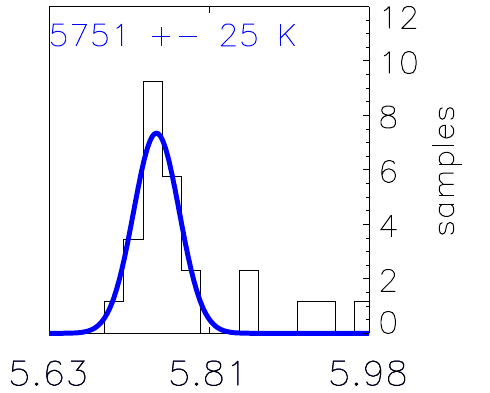}\\
    \includegraphics[width=14cm]{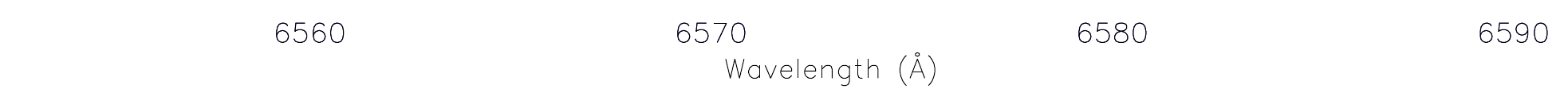}
    \includegraphics[width=3.5cm]{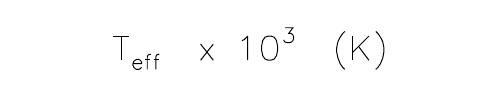}\\
\caption{Profile fits of coud\'{e} solar spectra.}
\label{coude_fitted}
\end{figure*}

\begin{figure*}
\centering
\includegraphics[width=14cm]{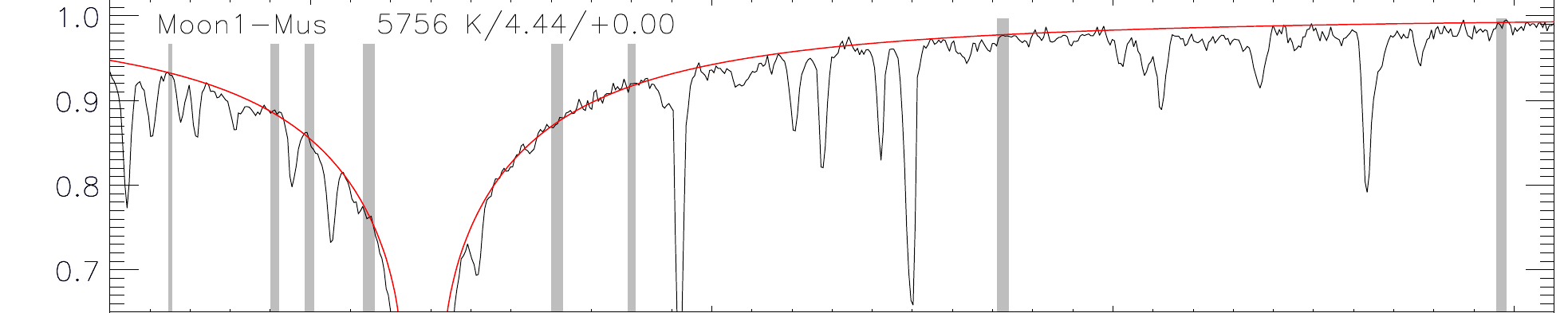}
\includegraphics[width=3.5cm]{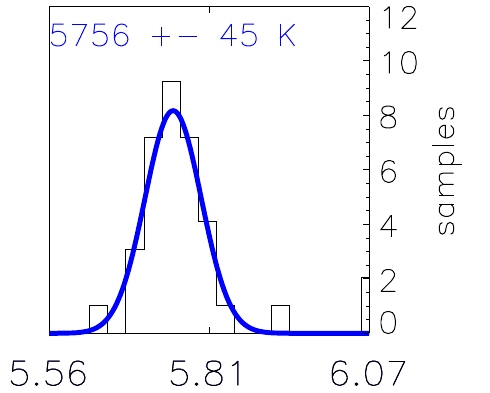}
\includegraphics[width=14cm]{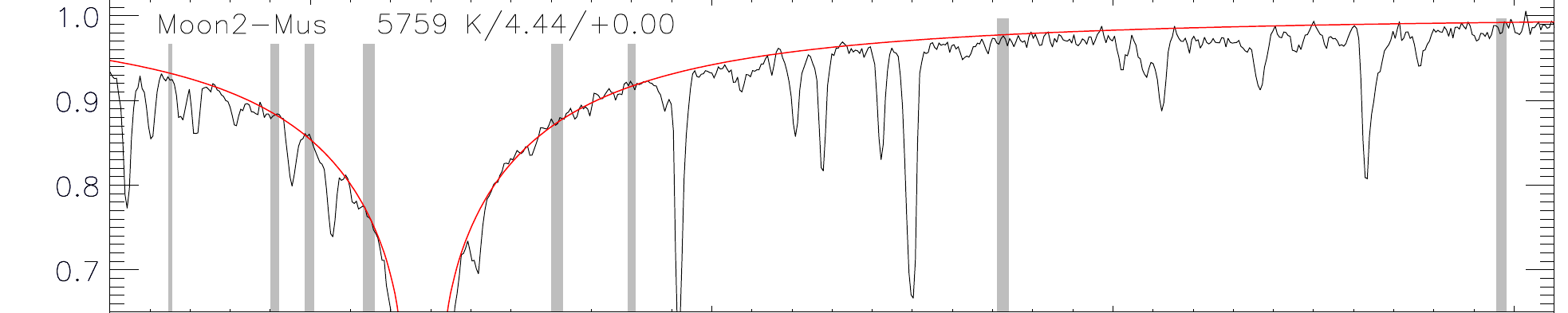}
\includegraphics[width=3.5cm]{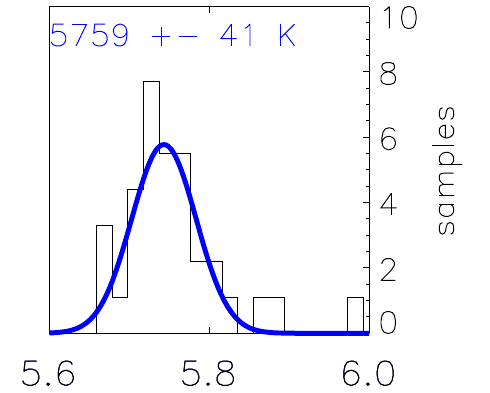}
\includegraphics[width=14cm]{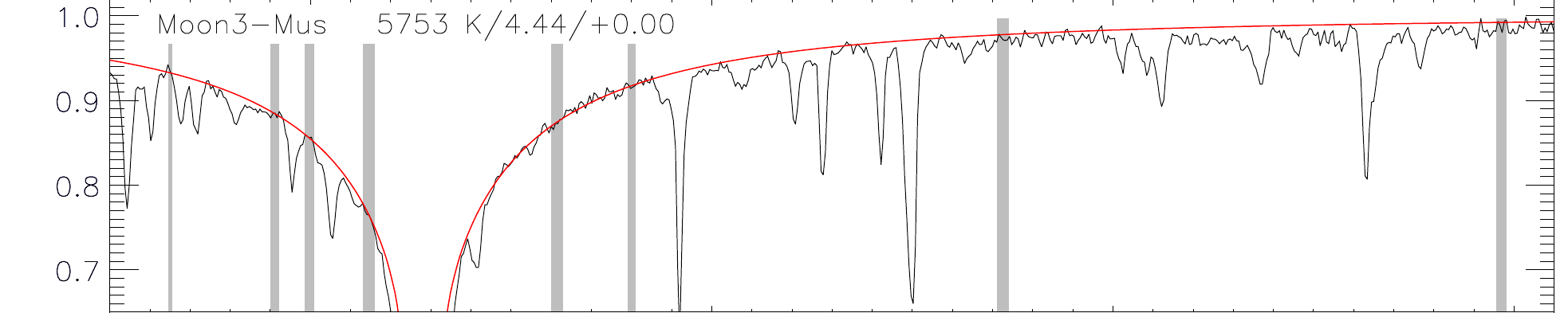}
\includegraphics[width=3.5cm]{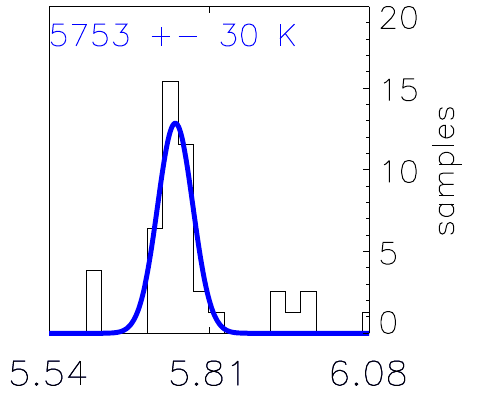}
\includegraphics[width=14cm]{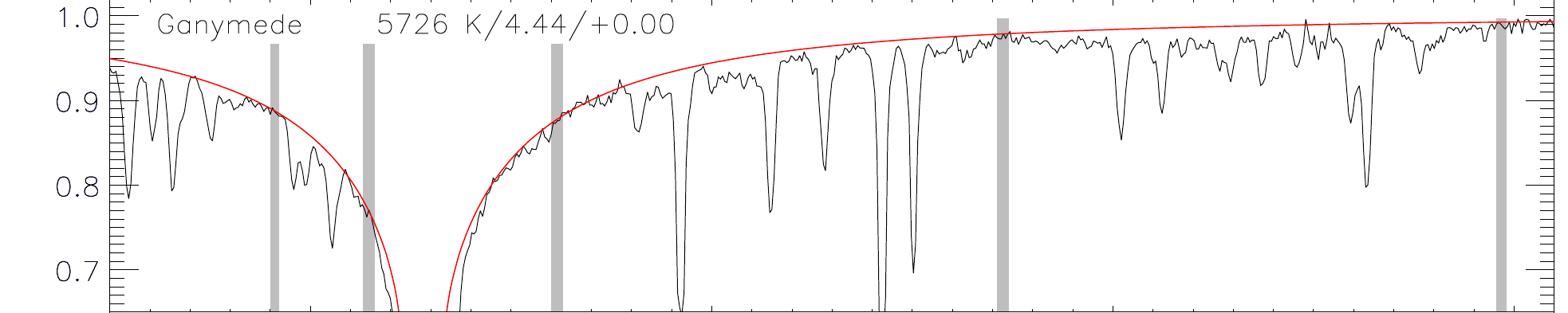}
\includegraphics[width=3.5cm]{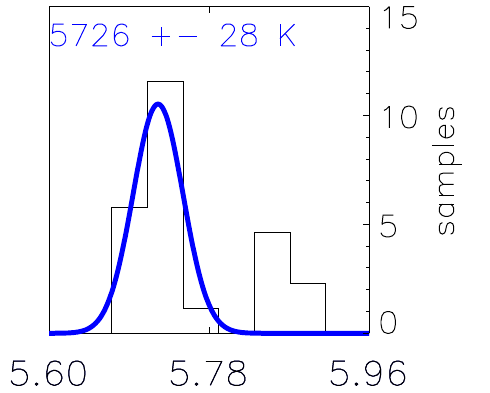}
\includegraphics[width=14cm]{fittingXtitle_moons-eps-converted-to}
\includegraphics[width=3.5cm]{distributionXtitle-eps-converted-to}\\
\caption{Profile fits of MUSICOS solar spectra.}
\label{MUSICOS_fittings}
\end{figure*}

\begin{figure*}
\centering
    \includegraphics[width=14cm]{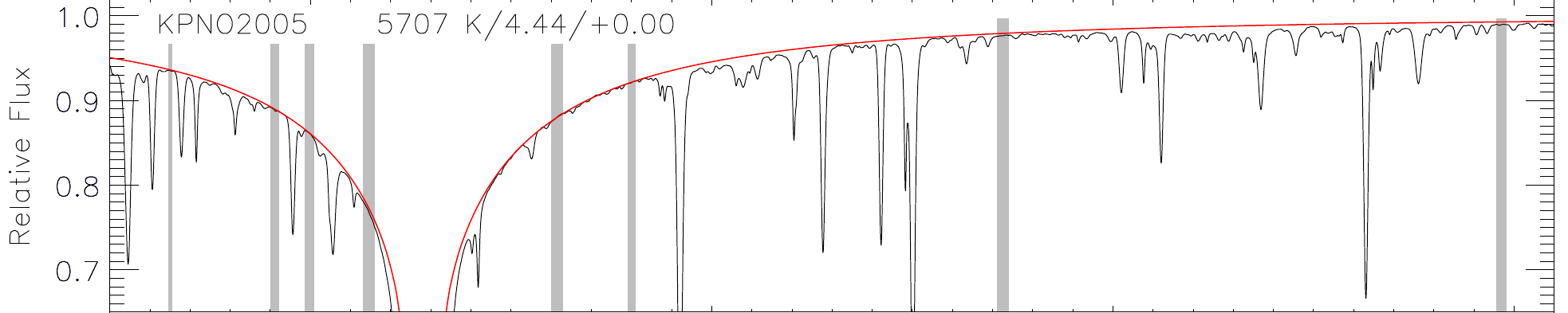}
    \includegraphics[width=3.5cm]{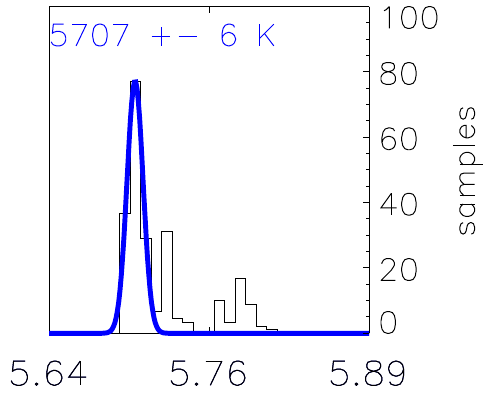}
     \includegraphics[width=14cm]{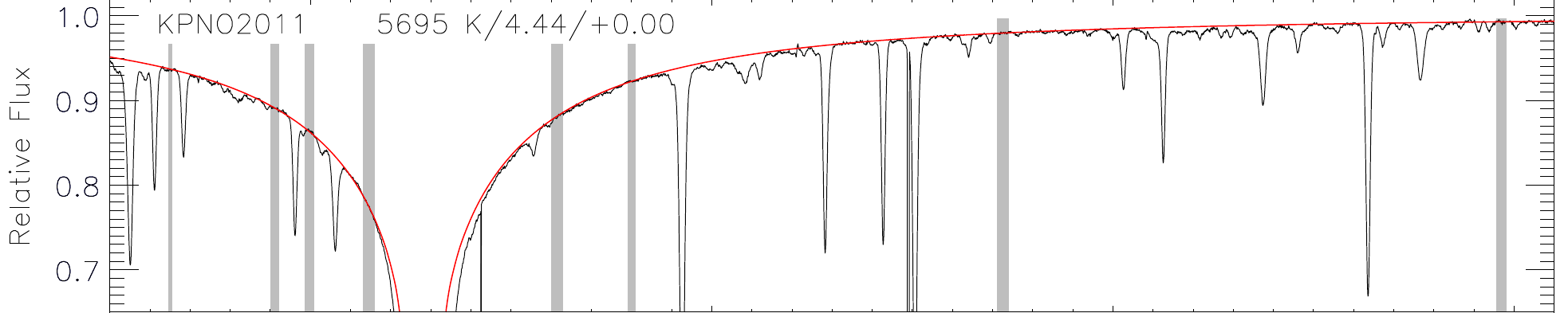}
    \includegraphics[width=3.5cm]{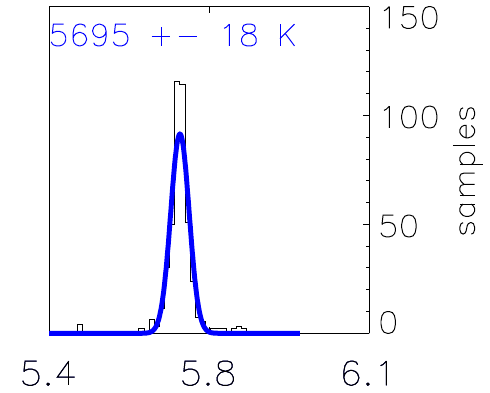}   
    \includegraphics[width=14cm]{fittingXtitle_moons-eps-converted-to}
    \includegraphics[width=3.5cm]{distributionXtitle-eps-converted-to}\\
    \caption{Profile fits of KPNO2005 (top), and KPNO2011 (bottom) spectra.}
    \label{kurucz_fittings}
\end{figure*}

\begin{figure*}
\centering
    \includegraphics[width=.8\textwidth]{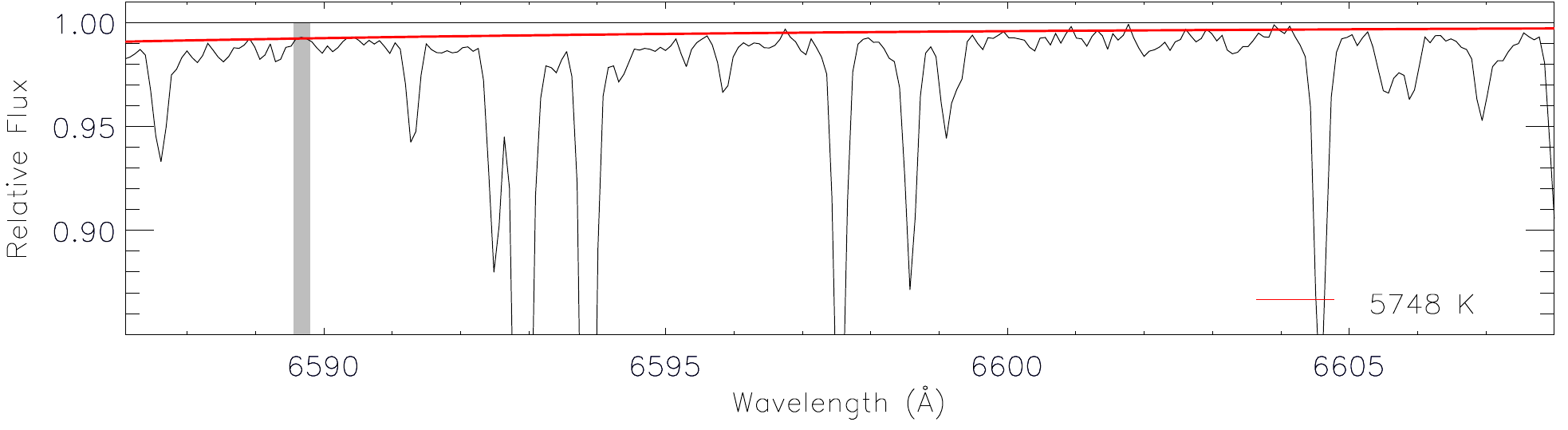}\\
    \includegraphics[width=.8\textwidth]{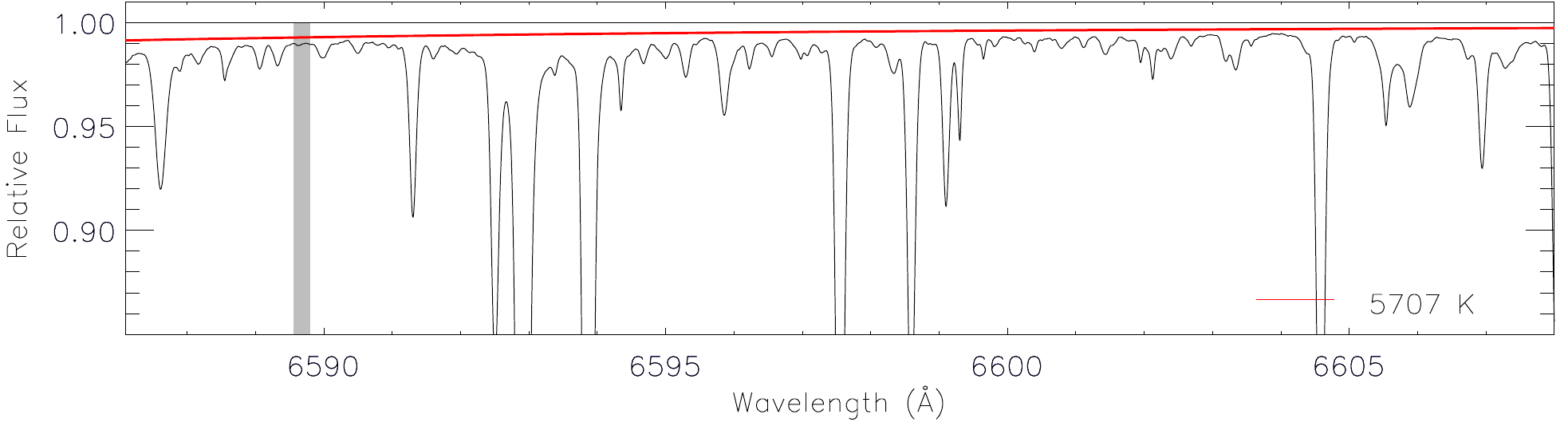}\\
    \includegraphics[width=.8\textwidth]{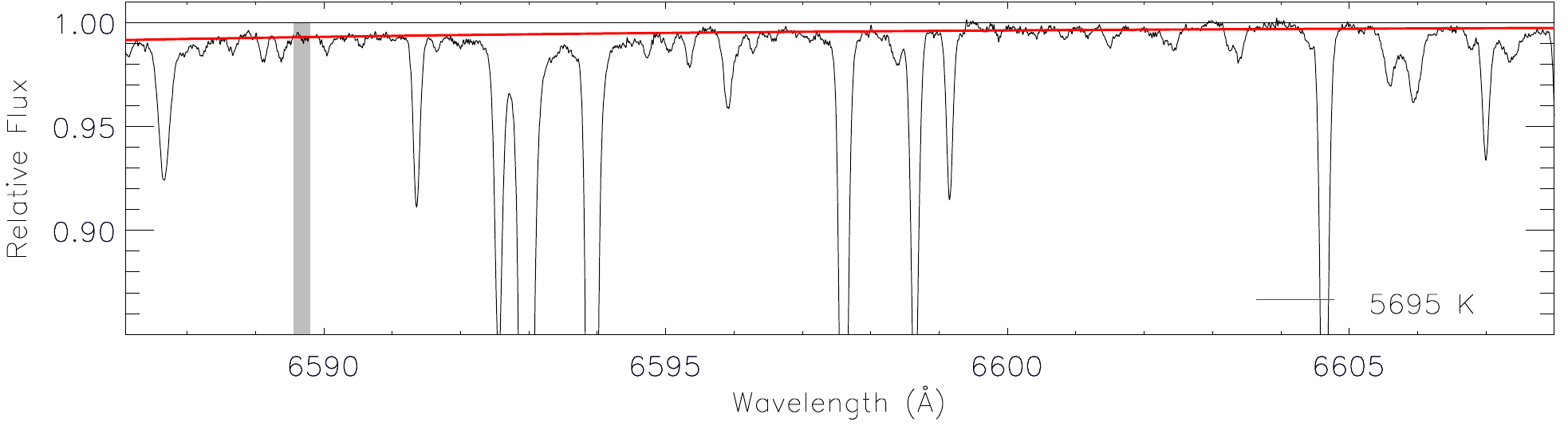}\\
    \caption{From the top to the bottom, transition regions at the red wing in fitted coud\'{e}, KPNO2005, and KPNO2011 spectra. 
    The panel in the top is related to the spectrum in Fig.~\ref{sun_normal}, 
    and the two panels below are related to the spectra in Fig.~\ref{kurucz_fittings}.}
    \label{transition_regions}
\end{figure*}

\begin{figure*}
\centering
    \includegraphics[width=.8\textwidth]{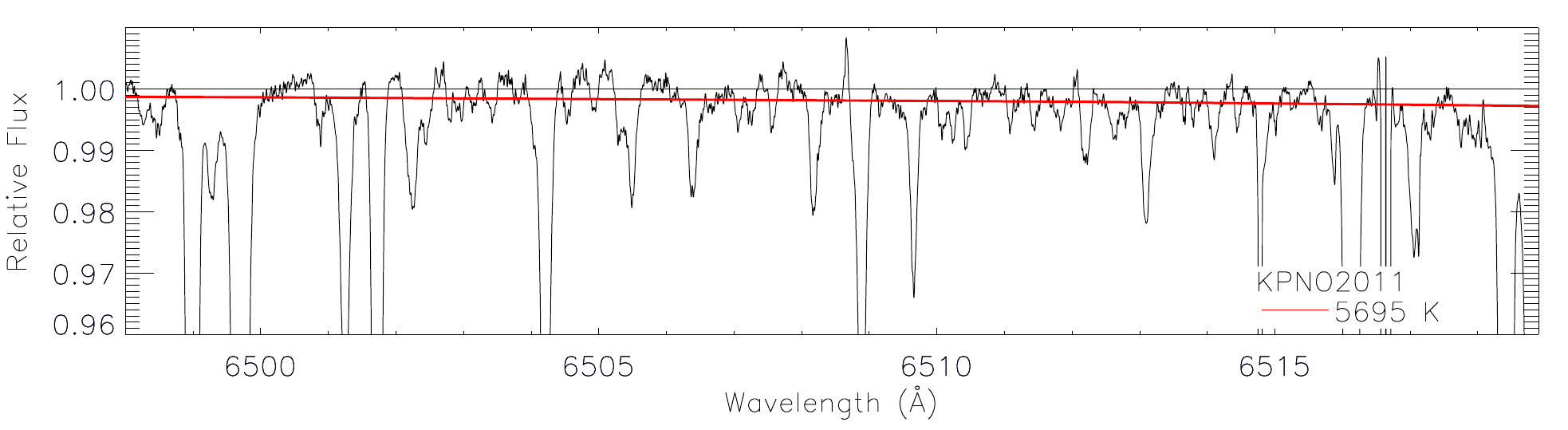}\\
    \includegraphics[width=.8\textwidth]{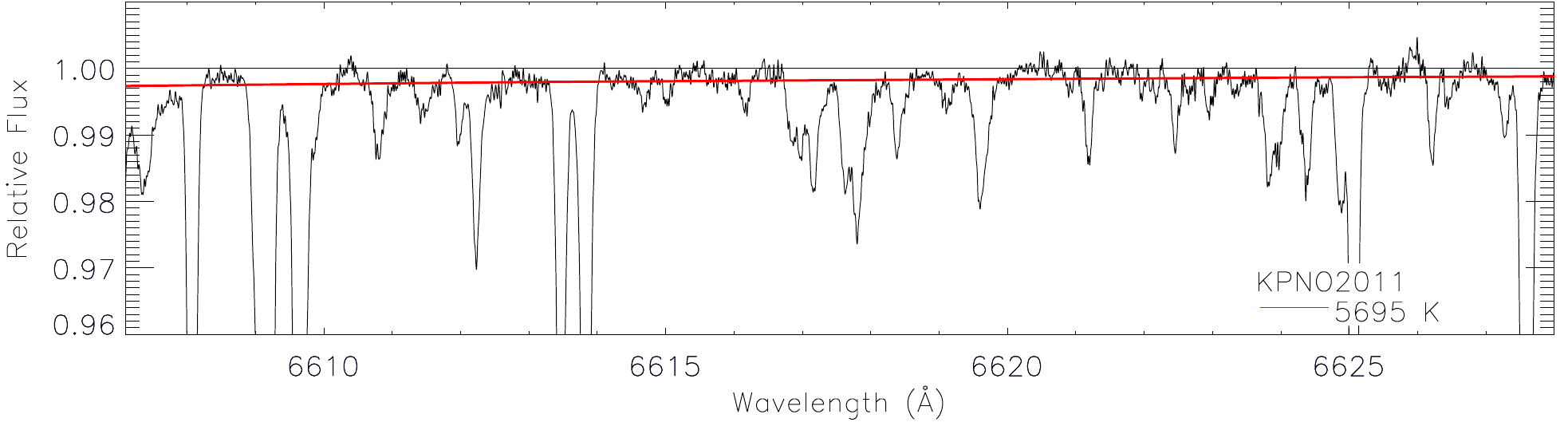}\\
    \caption{For the KPNO2011 atlas, spectral regions that contain the continuum windows [6500.25, 6500.50] and [6504.50, 6505.00] (top), 
    and [6619.70, 6620.50], [6625.60,6625.80] 
    and [6626.50, 6626.80] (bottom).}
    \label{normalization_errors}
\end{figure*}

\begin{figure*}
\centering
    \includegraphics[width=.8\textwidth]{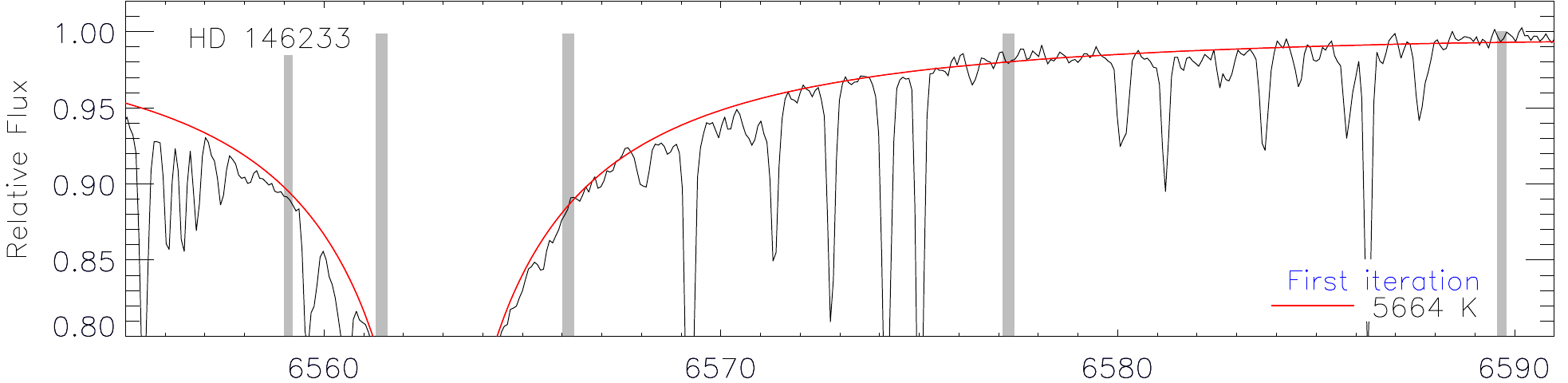}\\
    \includegraphics[width=.8\textwidth]{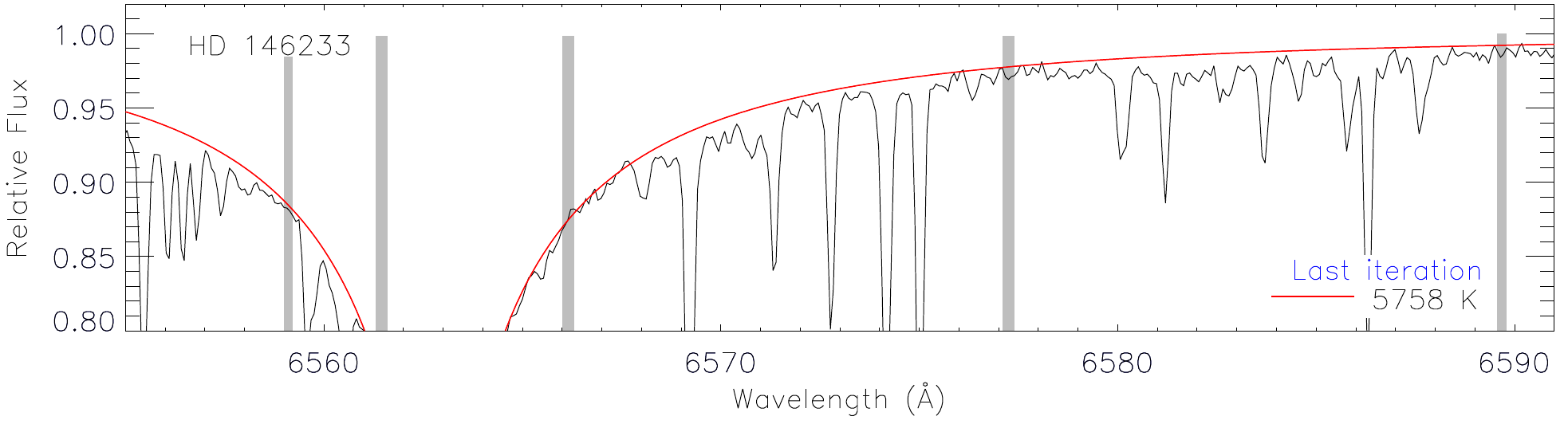}
    \caption{Results from the iterative H$\alpha$ normalization-fitting of one of the two coud\'{e} 
    spectra of 18Sco (HD 146233) following the procedure described in Sect.~\ref{coude}.}
    \label{gross_norm}
\end{figure*}

\begin{figure*}
\centering
    \includegraphics[width=14cm]{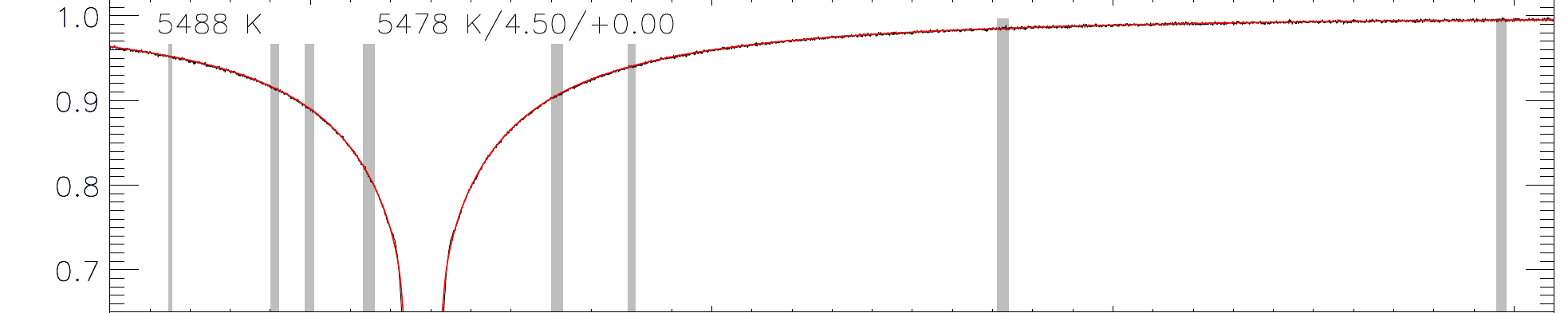}
    \includegraphics[width=3.5cm]{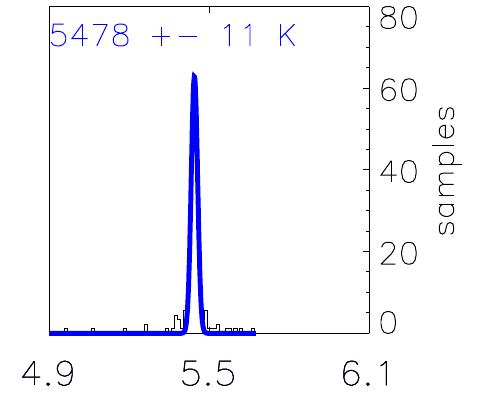}\\
    \includegraphics[width=14cm]{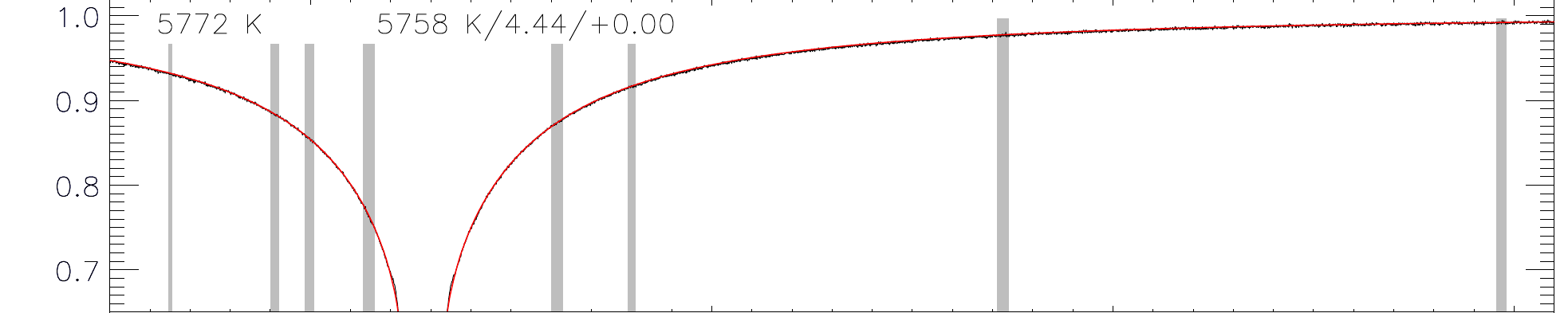}
    \includegraphics[width=3.5cm]{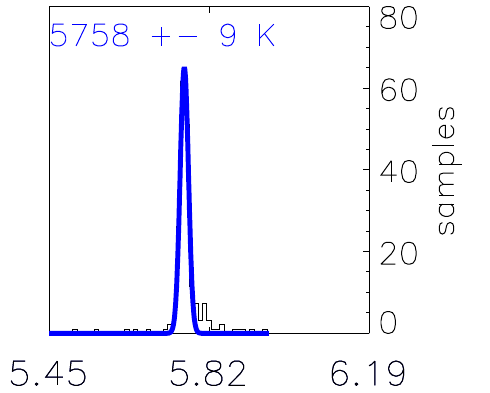}\\
    \includegraphics[width=14cm]{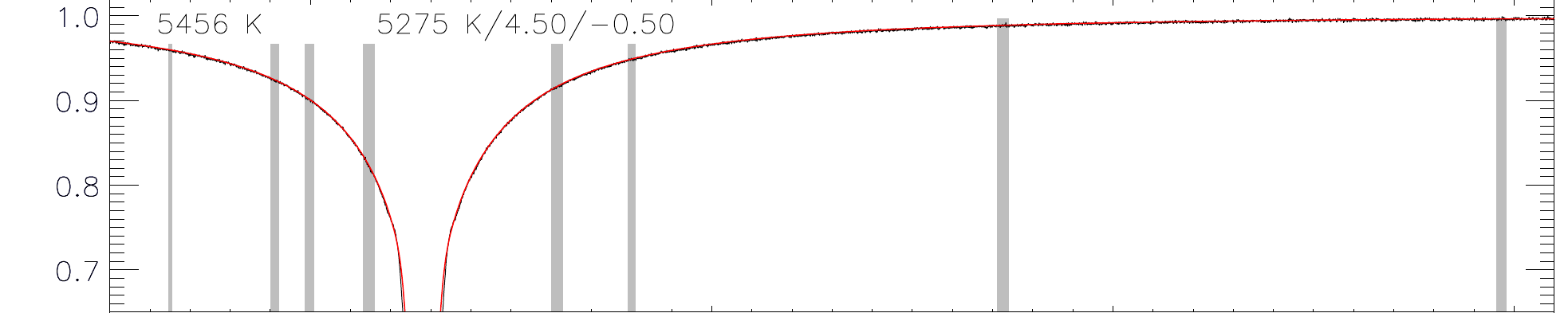}
    \includegraphics[width=3.5cm]{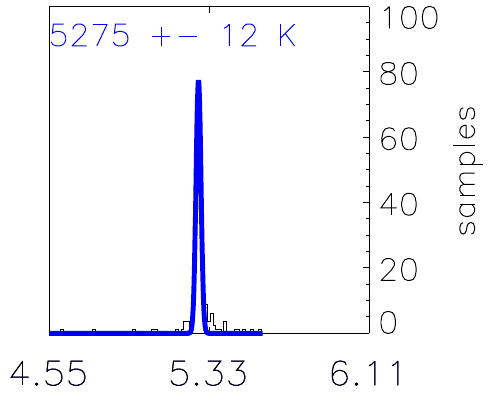}\\
    \includegraphics[width=14cm]{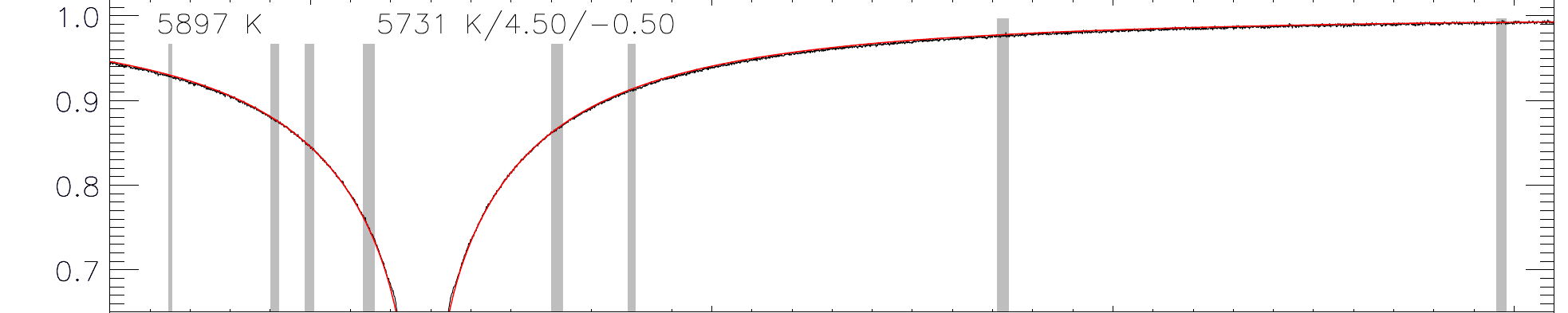}
    \includegraphics[width=3.5cm]{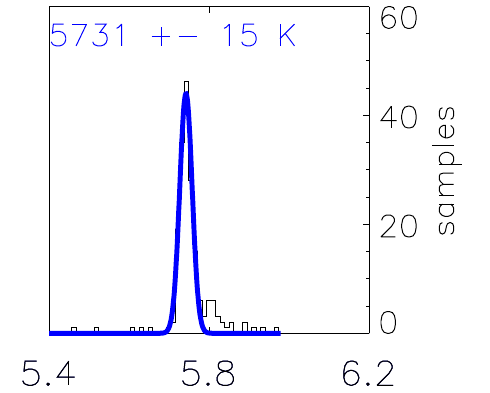}\\
    \includegraphics[width=14cm]{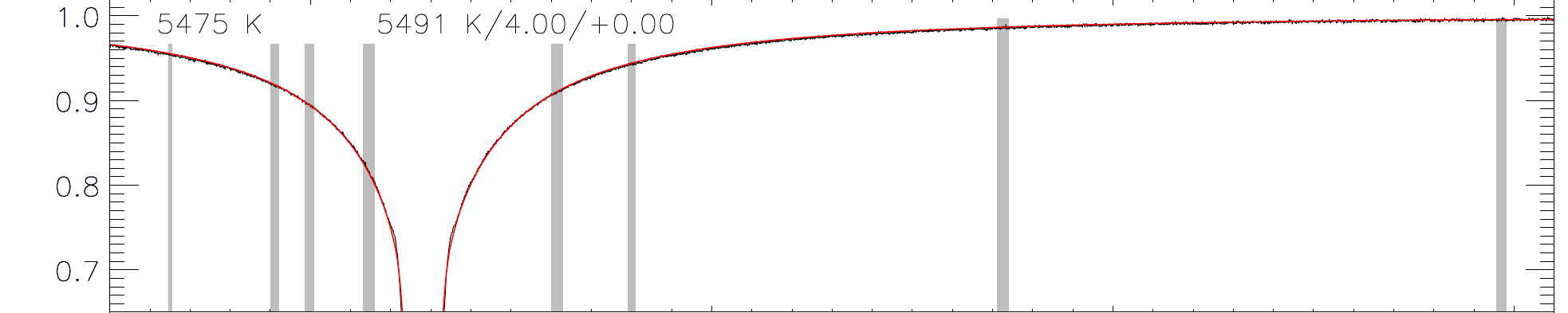}
    \includegraphics[width=3.5cm]{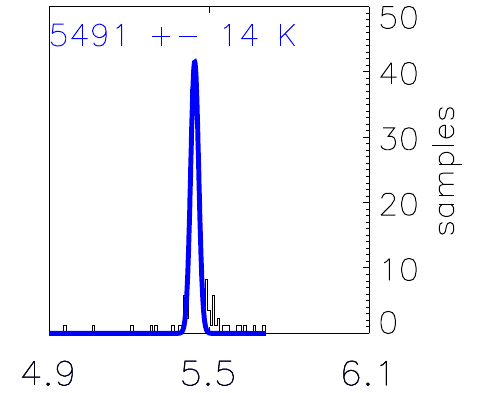}\\
    \includegraphics[width=14cm]{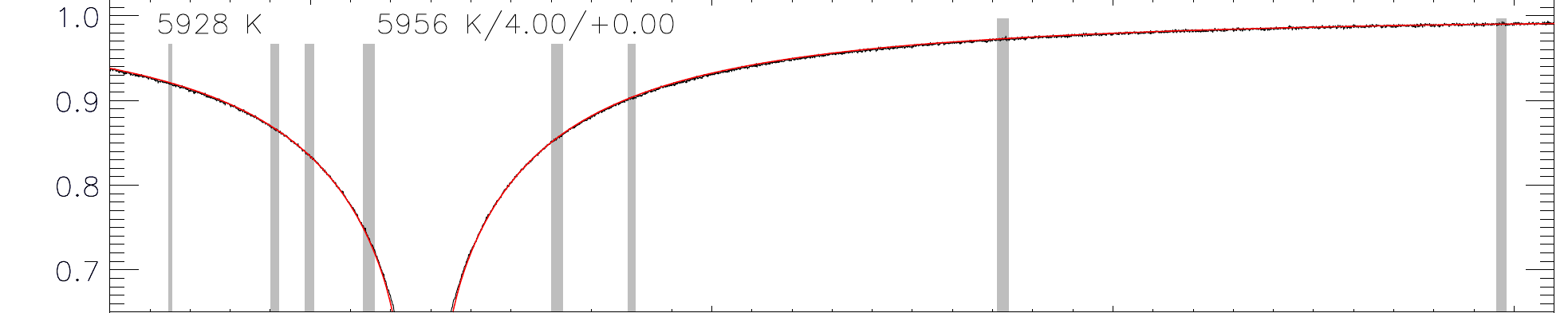}
    \includegraphics[width=3.5cm]{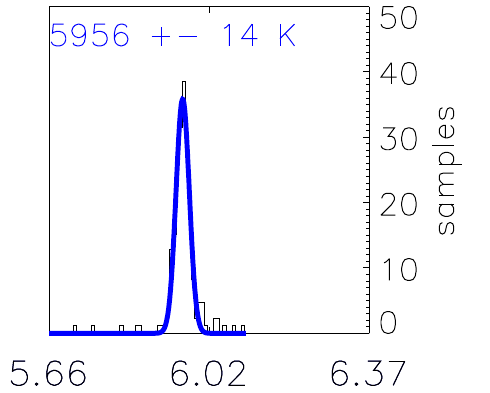}\\
    \includegraphics[width=14cm]{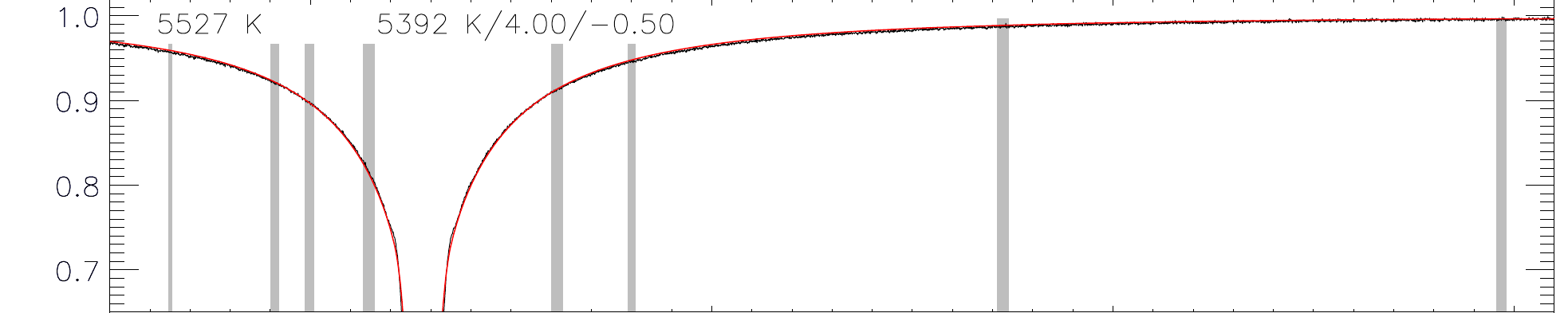}
    \includegraphics[width=3.5cm]{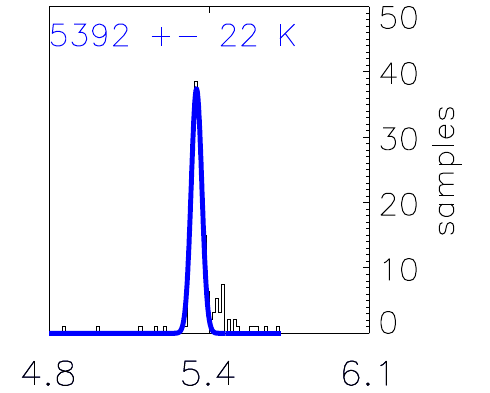}\\
    \includegraphics[width=14cm]{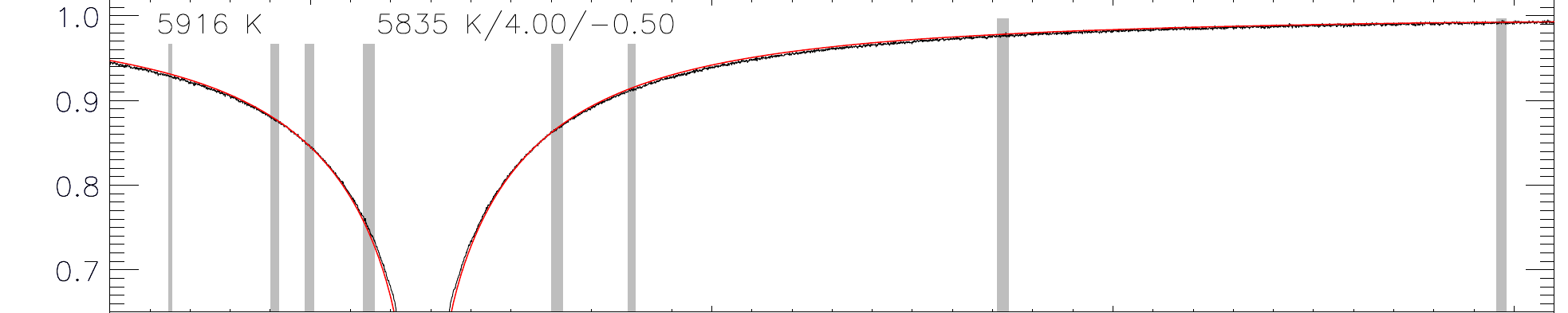}
    \includegraphics[width=3.5cm]{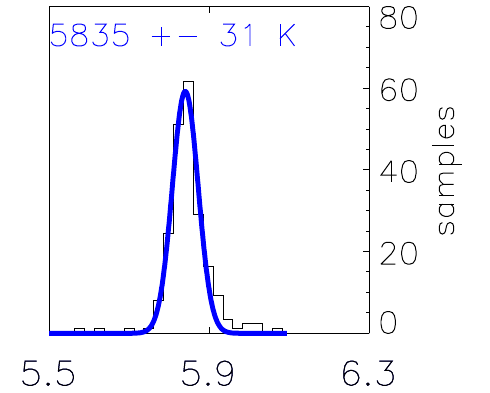}\\
    \includegraphics[width=14cm]{fittingXtitle_moons-eps-converted-to}
    \includegraphics[width=3.5cm]{distributionXtitle-eps-converted-to}\\
    \caption{Fits of 3D profiles (black) with 1D profiles (red). 
    The nominal temperature values of the 3D profiles are noted in the left, 
    while the parameters of the 1D profiles are at the right side.}
    \label{3D_1D_fits}
\end{figure*}

\end{appendix}
\end{document}